\documentclass[preprint,prd,showpacs,superscriptaddress,amssymb,amsmath,preprintnumbers,nofootinbib,fleqn,floatfix]{revtex4}

\usepackage{graphicx}
\usepackage{dcolumn}
\usepackage{axodraw}
\usepackage{longtable}
\usepackage{color}
\usepackage{hyperref}

\newcommand{\del}{\partial}

\begin{document}   

\preprint{KANAZAWA-13-09, HUPD-1308, KEK-CP-290}

\title{
 Lattice Study on quantum-mechanical dynamics of
two-color QCD with six light flavors} 

\author{M.~Hayakawa}
\affiliation{Department of Physics, Nagoya University, 
Nagoya 464-8602, Japan}

\author{K.-I.~Ishikawa}
\affiliation{Department of Physics, Hiroshima University, Higashi-Hiroshima, 
739-8526, Japan}

\author{S.~Takeda}
\affiliation{School of Mathematics and Physics,
College of Science and Engineering, Kanazawa University, Kakuma-machi,
Kanazawa, Ishikawa 920-1192, Japan}

\author{M.~Tomii}
\affiliation{Department of Physics, Nagoya University, 
Nagoya 464-8602, Japan}

\author{N.~Yamada}
\affiliation{
KEK Theory Center, Institute of Particle and Nuclear Studies,
High Energy Accelerator Research Organization (KEK),
Tsukuba 305-0801, Japan
}
\affiliation{
School of High Energy Accelerator Science,
Graduate University for Advanced Studies (Sokendai), Tsukuba 305-0801, Japan
}

\date{\today}

\begin{abstract}
 We investigate
the chiral properties of ${\rm SU(2)_C}$ gauge theory
with six flavors, i.e.~six light Dirac fermions
in the fundamental representations by lattice simulation,
and point out that the spontaneous breakdown of 
chiral symmetry does not occur in this system.
 The quark mass dependence of the mesonic spectrum
provides an evidence for such a possibility.
 The decay constant tends to be increased 
by the finite size effect, which is opposite to the behavior predicted
by chiral perturbation theory and 
indicates that the long distance dynamics in the six-flavor theory
could be different from the theory with chiral symmetry breaking.
 The subtracted chiral condensate, 
whose utility is demonstrated by the simulation of two-flavor theory, 
is shown to vanish in the chiral limit within the precision of available data.
\end{abstract}


\maketitle 

\section{Introduction}
\label{sec:intro}

 The standard model contains 
a Higgs doublet field which plays the role
to trigger the electroweak symmetry breakdown
and to generate the masses of all known elementary particles.
 The standard model, however, is incapable of predicting
observed flavor structure; 
all of fermion masses as well as mixing angles require 
corresponding number of input parameters.
 The technicolor \cite{Susskind:1978ms,Weinberg:1975gm}
and its extension \cite{Dimopoulos:1979es,Raby:1979my} 
are such an attempt to attribute flavor structure
and breakdown of electroweak symmetry
to quantum-mechanical dynamics of some gauge theory.
 The components absorbed into W and Z bosons eventually
resolve into their constituent fermions at shorter distance.
 Therefore, stability of electroweak scale against radiative correction
is guaranteed.

 However, the technicolor dynamics must be quite different from QCD
to resolve inconsistency with the size of flavor changing neutral current
(See, for example, Ref.~\cite{Farhi:1980xs})
and the constraints on the nature of electroweak symmetry breakdown
\cite{Peskin:1990zt,Peskin:1991sw}.
 The former problem can be solved if technicolor dynamics
give rise to relatively large anomalous
dimension for the chiral condensate
over wide range of energy scale (walking technicolor)
\cite{Holdom:1981rm,Yamawaki:1985zg,Akiba:1985rr,Appelquist:1986an}.
 It needs, however, explicit nonperturbative calculation
in order to access to the question 
whether such dynamics settle the second problem.
 The simulation formulated on lattice gauge theory
is expected to play a vital role to answer to this question
\cite{Shintani:2008qe,Boyle:2009xi,Appelquist:2010xv}, 
once we successfully find out the gauge system with walking dynamics.
  
 Fairly many works have been done to search
a candidate for walking technicolor by means of lattice simulation 
after the seminal work \cite{Appelquist:2007hu,Appelquist:2009ty},
which gave an evidence for the conformality
in the infrared (IR) limit of 
${\rm SU}(3)_{\rm C}$ gauge theory with twelve light Dirac fermions
by calculating the running gauge coupling constant defined 
in the Schr\"{o}dinger functional scheme (SF)
\cite{Luscher:1992an,Luscher:1992zx,Sint:1993un}.
 Other than SF coupling constant calculation, 
various methods have been proposed and attempted
to approach the issue;
analysis of phase structure of lattice systems
\cite{Iwasaki:2003de,DeGrand:2008kx},
finite size scaling test \cite{DeGrand:2009hu},
large scale spectroscopy study \cite{Fodor:2011tu}, 
Monte-Carlo renormalization group approach
\cite{Hasenfratz:2009ea,Hasenfratz:2010fi,Hasenfratz:2011xn}, 
and test of hyperscaling relations in the attractive
basin of infrared fixed point (IRFP)
\cite{DelDebbio:2010hu,DelDebbio:2010ze},
the calculation of coupling constant
defined via twisted Polyakov loop 
\cite{Bilgici:2009nm,Itou:2012qn}, 
and the search of modified asymptotic behavior 
of the correlation function \cite{Ishikawa:2013wf}.

 Thus far, ${\rm SU}(3)_{\rm C}$ gauge theories have been 
most often investigated as a candidate gauge system, 
in particular, with fermions in the fundamental representations, 
called flavors.
 In a series of such $N_F$-flavor QCD, 
the work \cite{Appelquist:2012nz} showed 
that ten-flavor QCD may have large mass anomalous dimension $\gamma_m \sim 1$
at the IRFP, the existence of which was demonstrated 
by the SF coupling calculation \cite{Hayakawa:2010yn}.

 We focus here on a series of ${\rm SU(2)_C}$ gauge theories
to search a candidate gauge theory with walking technicolor dynamics.
 It is probable that the quantum dynamics of ${\rm SU(2)_C}$ gauge theories
differ from those of ${\rm SU(3)_C}$ gauge theories.
 Firstly, in pure ${\rm SU(2)_C}$ Yang-Mills theory, 
the finite temperature transition from confinement to deconfinement
is second-order \cite{Kuti:1980gh,Engels:1980ty} 
and belongs to the universality class of three-dimensional Ising model
\cite{Svetitsky:1982gs},
while it is first-order in pure ${\rm SU(3)_C}$ Yang-Mills theory.
 If ${\rm SU(2)_C}$ gauge theory involves fermions coupled to the gauge field,
it should not be 
regarded as one of ${\rm SU({\it N_C})_C}$ gauge theories;
rather it is one of ${\rm Sp(2{\it N})_C}$ gauge theories, 
in which the fundamental representation is pseudo-real.
 As a consequence,
${\rm SU(2)_C}$ gauge theory with
$N_F$ Dirac fermions in the fundamental representation,
often referred to as
two-color QCD with $N_F$-flavors of ``quarks'',
has the chiral symmetry ${\rm SU(2{\it N_F})}$
enhanced from the familiar
${\rm SU}(N_F)_{\rm L} \times {\rm SU}(N_F)_{\rm R} \times {\rm U}(1)_{\rm B}$.
(Appendix \ref{sec:EWSB} summarizes this point.)
 It is thus anticipated that
the chiral dynamics of two-color QCD
can differ significantly from those of three-color QCD, 
in particular, at the critical number $N_F^{\rm crtl}$ of fermions,
above which gauge system becomes conformal in the infrared limit, 
called {\it IR-conformal}.

 Moreover, from the point of view of the application to the dynamical 
realization of the electroweak symmetry breaking,
the content of the effective Higgs sector is quite
different in ${\rm Sp}(2 N)_{\rm C}$ gauge theories 
from that in ${\rm SU}(N_C)_{\rm C}$ ($N_C \ge 3$), 
as described in Appendix \ref{sec:EWSB}.
 This fact also motivates us to carry out lattice simulation
to grasp its properties of nonperturbative dynamics
such as spectra of bound states.

  Study of two-color QCD intended to search
the conformal window has been done by means of lattice simulation
so far
\cite{Kogut:1987ai,Iwasaki:2003de,Bursa:2010xn,Ohki:2010sr,
Karavirta:2011zg,Voronov:2011zz,Voronov:2012la}.
 The perturbatively calculated $\beta$ function 
\cite{vanRitbergen:1997va} suggests
that $6 \le N_F^{\rm crtl} \le 8$.
 Ref.~\cite{Iwasaki:2003de} investigated the phase structure of Wilson fermions
and indicated that $N_F^{\rm crtl} = 2$ and that the massless theory
is not in the confinement phase for $N_F \ge 3$.
 Ref.~\cite{Bursa:2010xn} is the first trial to compute
the SF running gauge coupling constant
for the six-flavor theory, and pointed out the existence of 
infrared fixed point (IRFP) 
in the region $4 \lesssim g^2_\star \lesssim 6$.
 The authors in Ref.~\cite{Karavirta:2011zg} use
$O(a)$-improved fermionic actions (clover actions) and find no evidence of IRFP
at least in the region $g_\star^2 \lesssim 9$.
 Recently, Ref.~\cite{Voronov:2012la} suggests 
the absence of IRFP through the calculation of 
the SF running coupling with smeared link fields, 
where the continuum limit has not been taken yet.
 Our latest result for the SF coupling constant
with perturbative improvement indicates that
IR fixed point $g_\star^2$ exists in the range 
\cite{Hayakawa:2013yfa}
\begin{equation}
 0.06 < \frac{1}{g_\star^2} < 0.15\,.
 \label{eq:ourSF:IRFP}
\end{equation} 

 The purpose of this paper is to report our result for 
chiral properties of ${\rm SU(2)_C}$ gauge theory with six
Dirac fermions in the fundamental representation
by the lattice simulation.
 Overall, our strategy here is complementary to 
the calculation of the SF running coupling constant 
reported in a separate paper \cite{Hayakawa:2013yfa}.
 The SF running coupling constant is obtained 
by taking the explicit continuum limit of the data
for the massless quark.
 Instead, we focus here on 
the dependence on the quark mass 
of the various observables, 
such as mesonic spectra, 
decay constant, etc.
 In the report \cite{Hayakawa:2012gf} on the preliminary result,
the gluonic observables are analyzed to investigate
their fate in the chiral limit, 
but we leave them for the future study until further 
accumulation of statistics is accomplished.
 In this paper, we concentrate on the investigation
of the chiral properties of two-color QCD
with the six-flavors, which has not been done in Ref.~\cite{Hayakawa:2012gf}.

 Here we attempt to get a new insight on the possible role 
of the finite size effect so that it can provide valuable knowledge
concerning with the long distance properties of quantum-mechanical dynamics
(Sec.~\ref{subsec:FV}).
 We are inspired by the result for the mesonic and gluonic spectra 
in the ${\rm SU}(2)_{\rm C}$ gauge theory with 
two fermions in the adjoint representation
obtained in Ref.~\cite{DelDebbio:2010hu}.
 In three-color QCD with two-flavors, 
the finite size effect is known to increase the mass of, say, the 
lightest pseudoscalar meson, pion \cite{Gasser:1986vb}.
 However, in the ${\rm SU(2)_C}$ gauge theory with two adjoint fermions, 
the pion mass is decreased by the finite size effect.
 Another notable result found in Ref.~\cite{DelDebbio:2010hu} 
is the fact that
the $0^{++}$ glueball is {\it lighter} than the pion 
in the measured quark mass range, 
which is in striking contrast to the usual QCD.
 This tendency is actually compatible
with the theory of the finite size effect 
on the meson masses
\cite{Luscher:1983rk,Luscher:1985dn,Koma:2004wz}. 
 In Sec.~\ref{subsec:FV}, we focus on the decay constant, 
and list possible tendency for the finite size effects
in order to compare them with our data. 
 We find that $f_P$ is decreased by the finite size effect
in the six-flavor theory, which is opposite to
the behavior 
predicted by the chiral perturbation theory.

 Our results for the meson masses
and the decay constant $f_P$ in the six-flavor system 
have a similarity with those found for ${\rm SU(3)}_{\rm C}$ gauge theory 
with two Dirac fermions in the symmetric representation \cite{DeGrand:2008kx}.
 A look at the formula to calculate $f_P$, Eq.~(\ref{eq:fP}), 
indicates that $f_P$ approaches to zero 
in the chiral limit once the pion mass is bounded below.
 In Sec.~\ref{sec:two-flavor}, 
we make a preliminary study of finite size effects
in the theory with chiral symmetry breaking at very weak coupling,
as it may provide a benchmark 
to decide if the observed behavior 
is caused by the genuine dynamics of the IR-conformal theory
or just due to the finite size effect.


 The paper is organized as follows.
 Section~\ref{sec:simulation} describes the method of calculation.
 There, we also discuss the possible role of 
finite size effects to discriminate between 
the theory with chiral symmetry breaking and the IR-conformal theory.
 Section.~\ref{sec:two-flavor} has two purposes.
 One is to demonstrate the utility of the subtracted chiral condensate
defined through the pseudoscalar correlator
(See Eq.~(\ref{eq:def:scond})).
 Another is to get the knowledge on the finite size effects
on the various observables in the theory
with chiral symmetry breaking, but at such weak coupling 
that the chiral perturbation theory at finite volume is not applicable.
 Those knowledges will be confronted with the data 
to be obtained for our target system, $N_F = 6$ theory.
 After explaining the choice for the value of the bare coupling constant
in Sec.~\ref{sec:phaseStructure},
we present our results in Sec.~\ref{sec:results}.
 Section \ref{sec:summary} is devoted
to the summary of this paper and discussion.

\section{Lattice simulation}
\label{sec:simulation}

\subsection{generation of gauge configurations}
\label{subsec:gg}

 To study ${\rm SU(2)_C}$ gauge theory with 
six dynamical Dirac fermions, it is necessary to start with
generating a certain finite number of gauge configurations of this system.
 They form a sample to estimate
the ensemble average corresponding to  
the vacuum expectation value (VEV) $\left<{\cal A}\right>$ of an operator, 
${\cal A}$, with respect to some lattice-regularized Euclidean action 
\begin{equation}
 S_{\rm lat} = S_G + S_F\,,
 \label{eq:latticeAction}
\end{equation}
or the secondary quantities.

 In the lattice action (\ref{eq:latticeAction}),
$S_G$ is given solely by the gauge field, so called link variable
$U(n,\,\mu)$ obeying periodic boundary condition.
 In this paper, we employ the plaquette gauge action for 
\begin{eqnarray} 
 S_{\rm G}[U]
 &=& 
 \frac{\beta}{2 N_C}
 \sum_{n \in \Gamma_4} \sum_{0 \le \mu < \nu \le 3}
 2\,{\rm Re}\,{\rm tr}\left(1 - U(n;\,\mu,\,\nu)\right) \,,\nonumber\\
 U(n;\,\mu,\nu) &\equiv& 
 U(n,\,\mu) U(x + \widehat{\mu},\,\nu) 
 U(n + \widehat{\nu},\,\mu)^{-1} 
 U(n,\,\nu)^{-1} \, , 
 \label{eq:plaquetteGaugeAction}
\end{eqnarray}  
where $\widehat{\mu}$ denotes the unit vector along the $\mu$-direction, 
and the sum is taken over the whole lattice points (sites) $\Gamma_4$.

 $S_F$ in Eq.~(\ref{eq:latticeAction}) is the part containing 
the coupling of the ``quarks'' $\psi_j$, $\overline{\psi}_j$
to the gauge field.
 In this paper, we employ the Wilson fermion action for $S_F$
\begin{eqnarray}
 S_F[\psi,\,\overline{\psi},\,U] &=& 
 \sum_{n \in \Gamma_4} \sum_{j=1}^{N_F}
 \overline{\psi}_j(n) (D_W[U] \psi_j)(n) \nonumber\\
 &=&
 \sum_{n \in \Gamma_4} \sum_{j=1}^{N_F}
 \left[
  \overline{\psi}_j(n) \psi_j(n)
  -
  \kappa 
  \sum_{\mu = 1}^4
  \left\{
   \overline{\psi}_j(n) \left(1 - \gamma_\mu\right) 
   U(n,\,\widehat{\mu})
   \psi_j(n + \widehat{\mu})
  \right.
 \right.\nonumber\\
 &&\qquad\qquad\qquad\qquad\qquad\qquad
 \left.
  \left.
   +
   \overline{\psi}_j(n + \widehat{\mu}) \left(1 + \gamma_\mu\right) 
   U(n,\,\widehat{\mu})^\dagger
   \psi_j(n)
  \right\}
 \right] \, . \label{eq:actionWilsonFermion}
\end{eqnarray} 
 The Wilson fermion action $S_F$ contains a parameter $\kappa$, 
called hopping parameter, representing
the strength of the nearest-neighboring fermionic variables, 
in place of the bare quark mass.
 In the simulation, we adopt the periodic boundary conditions 
for $\psi$, $\overline{\psi}$ along all directions.
 Actually, in the quantum ${\rm SU}(2)_C$ gauge theory, 
the anti-periodic boundary condition is easily seen to be equivalent 
to the periodic boundary condition
if all the representation of the matter fields 
are the $Z_2$-odd conjugacy classes, such as fundamental representation.

 The gauge configurations
corresponding to the action (\ref{eq:latticeAction}) are produced
in the standard manner.
 The important sampling is processed
according to 
the lattice action (\ref{eq:latticeAction})
by Hybrid Monte-Carlo method (HMC)
\cite{Duane:1987de}, 
which utilizes $N_F /2$ ($=3$ for $N_F = 6$) pseudo-fermions
to represent $N_F$ copies of functional determinants
of the Wilson Dirac operator $D_W[U]$.
  The evolution in the modular dynamics is numerically performed by 
the improved integrator \cite{Omelyan:2003}.
 The most computationally expensive part 
occupied in this evolution
and the calculation of change of stochastic Hamiltonian 
is solving the linear problem, $D_W[U] \phi = b$.
 This part can be accelerated 
by using $3$ GPU cards
and adopting the mixed precision solver
with flavor-parallelized single-precision preconditioning
\cite{Hayakawa:2010gm}. 

\subsection{meson masses}

 Throughout this paper, a ``meson'' means a color-singlet and flavor-nonsinglet 
state with vanishing baryon number.
 In the rest of this section, all quantities such as meson masses
are dimensionless.
 The masses of the lightest mesons in the channel 
$H \in \left\{{\rm P,\,V,\,S,\,A}\right\}$
are determined by measuring the connected-type contribution
to the two-point function 
\begin{eqnarray}
 f_{HH}(t \equiv n_0)
 \equiv
 \sum_{{\bf n}} 
 \left< O_H(n) O_H(0)^\dagger \right>\,.
 \label{eq:def:2-pt-corrFunc}
\end{eqnarray}
 $O_H(n)$ is a certain bilinear field with the quantum number
corresponding to $H$, for instance, the flavor-non-singlet 
and strictly local operator 
$O_P(n) = P(n) \equiv \overline{\psi}(n) \gamma_5 \psi^\prime(n)$
in the pseudoscalar channel $P$.
 Unless the fermions are massless,
the excited state gets non-zero energy gap from  
the ground state in every channel even
in the IR-conformal theory.
 The mass $M_H$ of the ground state is extracted 
from the asymptotic behavior
of the correlation function (\ref{eq:def:2-pt-corrFunc}) 
at large $t$ so that the contribution of the excited states 
all drop out and it is dominated by the ground state contribution
\begin{equation}
 f_{HH}(t) = A^{HH} e^{-M_H t} + \cdots \,.
 \label{eq:asymTwoPoint}
\end{equation}
 The one-pole dominance can be monitored by checking if the effective mass 
\begin{equation}
 M_{HH}(t) \equiv {\rm ln}\left(\frac{f_{HH}(t)}{f_{HH}(t+1)}\right)\,,
\end{equation}
exhibits a plateau.
 Actually, there are channels $H$ on the lattices which do not exhibit
plateau, probably due to short length along the time direction.
 We then give up finding $M_H$.
 If something like plateau is seen in a certain limited range of time,
the correlation function is measured with
the gauge-invariant smeared source as in Ref.~\cite{Allton:1993wc}
(Jacobi smearing) for the operator $O_H(0)$.
 The corresponding effective mass $M_{HH}(t)$ approaches to a plateau
from below, enabling to check
if its position is consistent with that observed using the point source.

\subsection{PCAC mass, decay constant}
\label{subsec:PCACmass_decayConstat}

 The Wilson fermion explicitly breaks chiral symmetry.
 Thus, the ``naive mass'' proportional to $1/ \kappa$
suffers additive and nonperturbative shift.
 Instead, mimicking the PCAC relation in the continuum theory,
(bare) PCAC mass $m_{\rm PCAC}$ derived from 
the plateau of the ratio of the correlation functions
summed over spatial volume
\begin{eqnarray}
 \widehat{m}_{\rm PCAC}(t) &\equiv& 
 \frac{\frac{1}{2}(\del_t + \del_t^*) f_{A_t P}(t)}{2 f_{PP}(t)}\,,\nonumber\\
 f_{A_t P}(t \equiv n_0) &\equiv&
 \sum_{\vec{n}}
 \left<A_t(n) P(0) \right>\,,
  \label{eq:def_PCACmass}
\end{eqnarray}
where $A_t(n) \equiv \overline{\psi}^\prime \gamma_t \gamma_5 \psi$ 
and $\del_t$ ($\del_t^*$) is the forward (backward) difference
operator, is used as a measure of bare quark mass.

 Similarly, the (bare) decay constant $f_P$ of the lightest pseudoscalar meson
is obtained as
\begin{equation}
  f_P = (2 \kappa) 2 m_{\rm PCAC} 
  \sqrt{\frac{2 A^{PP}}{M_P}}\,
  \frac{1}{\sinh(M_P)}\,.
  \label{eq:fP}
\end{equation}
 In the context of the application to the electroweak symmetry breaking,
the decay constant in the continuum theory serves the mass scale
to make qualitative predictions, 
and will be obtained by multiplying the above $f_P$ by
the constant $Z_A$ of the renormalization of the axial-vector current
and an appropriate matching factor.
 Since our main interest here is to explore the qualitative features of 
the dynamics of the system, the multiplication of all renormalization
factors, which are short-distance quantities and do not affect 
to the long distance dynamics, is omitted.

\subsection{subtracted chiral condensate}
\label{subsec:scond}

 The Wilson fermion breaks chiral symmetry explicitly through the Wilson term.
 As a consequence, the vacuum expectation value 
of the chiral condensate 
$\left< \overline{\psi}(0) \psi(0) \right>$
(no sum over flavors is understood throughout this paper
for this quantity) suffer hard ultra-violet divergence $\sim O(a^{-3})$, 
the singularity that does not vanish in the massless quark limit.
 The renormalization requires the subtraction of
such a hard component as the first step.
 We will call the resulting chiral condensate 
as the {\it subtracted} chiral condensate. 
 It is possible to calculate
\begin{equation}
 \left< \overline{\psi}(0) \psi(0) \right>
 = - \left< {\rm tr}^\prime D_W[U]^{-1}(0) \right>\,,
 \label{eq:condNiave}
\end{equation}
where $D_W[U]^{-1}$ is the inverse of the Wilson Dirac operator 
in Eq.~(\ref{eq:actionWilsonFermion}) 
and ${\rm tr}^\prime$ implies the sum over colors. 
 However, no one knows a practical method to subtract away
such an $O(a^{-3})$ component to get the subtracted chiral condensate.

 Even for the Wilson fermion,
the Ward-Takahashi identity 
with respect to the axial-vector current
can be written down 
\cite{Bochicchio:1985xa}
\begin{equation}
 \delta^{ab} \cdot 
 \left<\overline{\psi} \psi\right>_{\rm subt}\left(m_{\rm PCAC},\,L/a\right)
 =
 2 m_{\rm PCAC} \cdot \left(2 \kappa\right)^2
 \sum_n \left< P^a(n) P^b(0) \right>\,,
 \label{eq:def:scond}
\end{equation}
where $P^a(n) = \overline{\psi} T^a \gamma_5 \psi$
for ${\rm SU}(N_F)$ generators $T^a$.
 Through this identity,
the subtracted chiral condensate in the Wilson fermion is 
{\it calculated using the right hand side} of Eq.~(\ref{eq:def:scond}).
 The quantity in Eq.~(\ref{eq:def:scond})
actually requires the multiplicative renormalization constants 
to be confronted with the continuum physics.
 Those constants are short-distance quantities, 
and are thus considered not to affect to the 
observation on the essence of the long distance dynamics of the system.

 Recently, 
$\left<\overline{\psi} \psi\right>_{\rm subt}$ was used 
to monitor the phase structure of many-flavor Wilson fermions 
at the strong coupling limit ($\beta = 0$) \cite{Nagai:2009ip}, 
and to study the chiral phase transition at finite temperature
\cite{Umeda:2012nn}.
 However, as long as we know,
the properties of $\left<\overline{\psi} \psi\right>_{\rm subt}$ 
have not been investigated thus far.
 We will examine them in Sec.~\ref{sec:two-flavor}
before we use $\left<\overline{\psi} \psi\right>_{\rm subt}$
to study the occurence of chiral symmetry breaking
in $N_F = 6$ theory. 


\subsection{finite size effect}
\label{subsec:FV}

 The simulation must be done for the system put in a finite box. 
 All the quantities measured in the box 
thus receive more or less the effect due to this limitation.
 As is done in the statistical mechanics,
as long as the size of the effect 
is small enough that it can be treated as a correction
\footnote{
 We recall that the inverse of the lattice size, $1/l = a/L$, 
is one of relevant perturbations.
},
the information on the dependence of the quantities 
upon the system size and the boundary condition
may help to extract the long distance dynamics of the system.

 In this work, the finite size effect 
will actually play a crucial role to investigate
the dynamical features of the target gauge theory.
 The purpose of this subsection is to provide the materials 
that form the basis of the forthcoming analysis in regard to this point;
the summary of the known facts
on the finite size effects in the system with spontaneous breakdown
of chiral symmetry, simply abbreviated as
{\it ${\chi\hspace{-5pt}/}$-theory}, 
and the observation on those in the IR-conformal system.

 We consider first the pseudoscalar meson $P$ and the finite size effect 
on its mass $M_P$.
 If the linear size $L$ of the spatial volume becomes comparable 
to the Compton wavelength of $P$, 
the finite size correction $\Delta M_P(L) = M_P(L) - M_P$
due to the elastic scattering of $P$ is known to increase the mass, 
$\Delta M_P^{\rm sc}(L) > 0$
\cite{Luscher:1983rk,Luscher:1985dn}.
 For instance, 
in ${\rm SU(3)_C}$ gauge theory with two or three light quarks,
where low-energy dynamics can be approximately described by 
Nambu-Goldstone bosons $\pi$, 
the $\pi\pi$ scattering effect
dominates over the finite size effect on $M_\pi$ and thus increases it.
 
 One is often inclined to identify the Compton wavelength $\sim 1/ M(L)$ 
with the spatial correlation length $\xi$.
 As $\xi$ cannot exceed the system size $L$,
the mass is then bounded from below in proportion to $\sim 1/L$, 
suggesting that the finite size effect increases $M(L)$.

 However, the analysis in the framework of the quantum field theory
implies that this is not always the case.
 As illustrated in Ref.~\cite{Luscher:1983rk}, 
in pure Yang-Mills theory,
the finite size effect acts on 
the mass $m_G$ of the glueball $G$ in the $0^{++}$-channel
in such a way that
it {\it decreases} $m_G(L)$, i.e. $\Delta M_G^{\rm tri}(L) < 0$,
through the non-trivial trilinear coupling 
among glueballs with a dimensionful coefficient.

 Now, let us suppose that the dynamics of the theory admits
such a light $0^{++}$ state 
$\sigma $ that is active in the low-energy effective theory for $P$.
 In case $\sigma$ is lighter than $P$, 
we can apply the result in Ref.~\cite{Koma:2004wz}, 
which generalizes L\"{u}scher's analysis to the system of two species.
 The finite size effect decreases the mass of $P$;
$\Delta M_P(L) \simeq M_P^{\rm tri}(L) < 0$, 
provided that it is dominated by the $\sigma P P$ interaction. 

 An example of negative $\Delta M_{P}(L)$ 
seems to be realized
in the ${\rm SU(2)_C}$ gauge theory with two adjoint fermions
\cite{DelDebbio:2010hu}.
 See Fig.~2 in Ref.~\cite{DelDebbio:2010hu}.
 Indeed, Fig.~11 in Ref.~\cite{DelDebbio:2010hu} shows that 
the $0^{++}$-glueball
is lighter than the pseudoscalar meson $P$ in that theory.

 Those results illustrate that
the finite size effect reflects the dynamical features 
of the system and thus provides us a device to investigate them. 
 In this paper, we would like to point out that the decay constant $f_P$
of the pseudoscalar meson could be also interesting 
from such a viewpoint.

\begin{figure}[thp]
\includegraphics[width=11.0cm,clip]{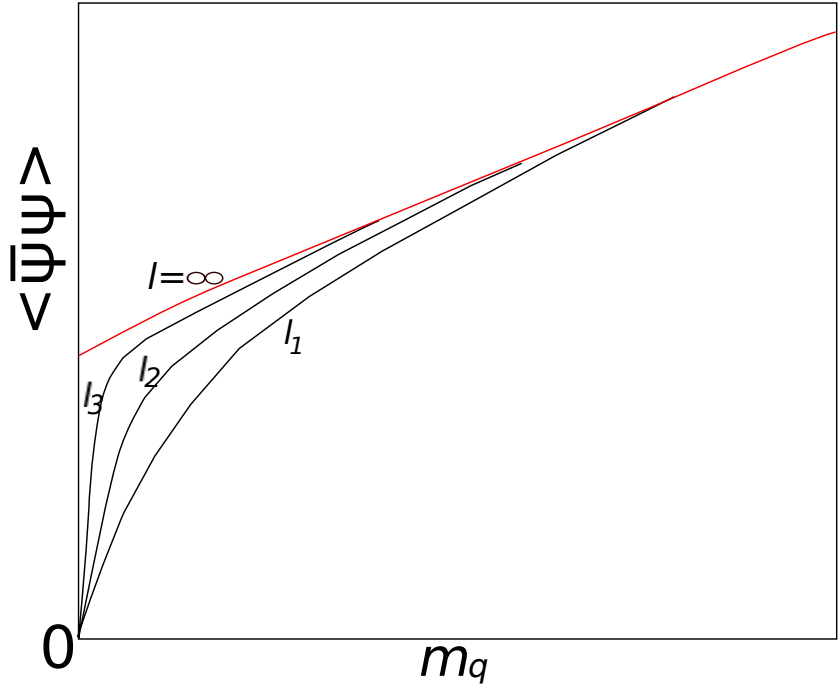}
\caption{
 $\left<\overline{\psi} \psi\right>$
versus the quark mass at finite volume
in the ${\chi\hspace{-5pt}/}$-theory.
 Here, the linear sizes  $l_1,\,l_2,\,l_3$ of three spatial volumes
are ordered such that $l_3 > l_2 > l_1$.
 The condensate in the limit of the vanishing quark mass
becomes nonzero only after the thermodynamic limit is taken first.
}
\label{fig:conds_finiteVolume}
\end{figure}

\begin{figure}[htp]
\includegraphics[width=11.0cm,clip]{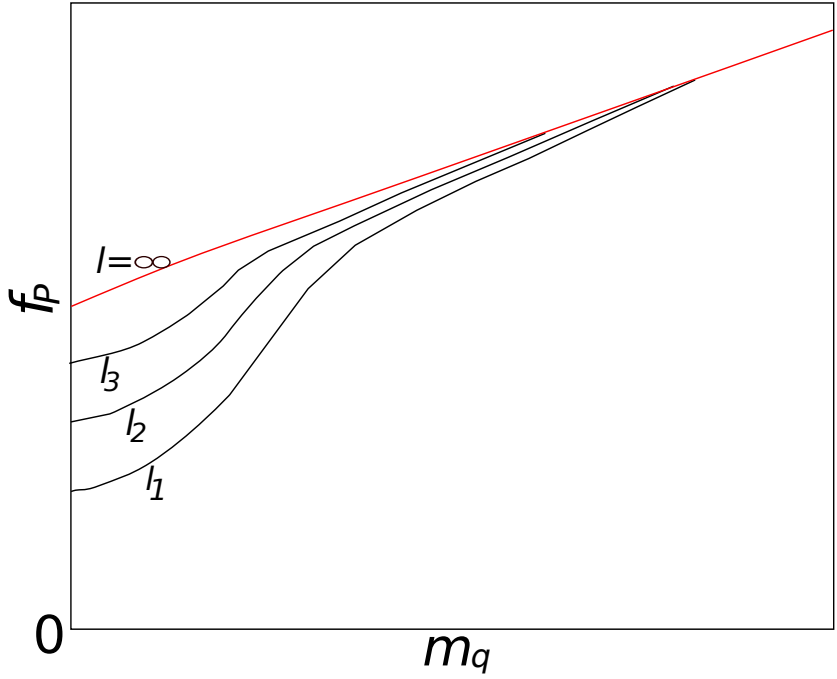}
\caption{
 The decay constant $f_P(m_q)$
versus the quark mass $m_q$ at finite volume
in the ${\chi\hspace{-5pt}/}$-theory.
}
\label{fig:fP_finiteVolume_chibreak}
\end{figure}

 We first recall that
the chiral condensate $\left<\overline{\psi} \psi\right>$
is an order parameter of the chiral symmetry.
 It thus inevitably vanishes in the chiral limit at finite volume.
 The situation is schematically represented as in 
Fig.~\ref{fig:conds_finiteVolume} for the ${\chi\hspace{-5pt}/}$-theory.
 Contrastingly, the decay constant $f_P$ is not 
the order parameter of the chiral symmetry
in QCD or the gauge theory with fermionic matters in general.
 At finite volume, the linear size $L$ of the system 
provides the infrared cut off $\sim 1/L$, 
and there is no symmetry reason that its chiral limit 
should vanish. 
 In the ${\chi\hspace{-5pt}/}$-theory, 
the chiral perturbation 
can be applied to the $p$-regime, $M_\pi \gg 1/L$, 
and it is found that
$f_P(m_q)$ at quark mass $m_q$ is reduced
by the finiteness of system size for any $N_F$ 
as long as chiral symmetry breaking occurs 
and the parameters are in the $p$-regime
\cite{Gasser:1986vb,Colangelo:2005gd}.
 Also in the $\epsilon$-regime, $M_\pi \ll 1/L$, 
the finiteness of system size acts to decrease $f_P(m_q)$, 
leaving a non-vanishing constant at $m_q = 0$ even at finite volume
\cite{Aoki:2011pza}.
 Figure \ref{fig:fP_finiteVolume_chibreak} shows 
the schematic dependence of $f_P(m_q)$ on quark mass $m_q$
and the linear size $L$ of the volume
in the ${\chi\hspace{-5pt}/}$-theory.

\begin{figure}[htp]
\begin{tabular}{cc}
\begin{minipage}{0.47\hsize}
\begin{center}
\includegraphics[width=\hsize,clip]{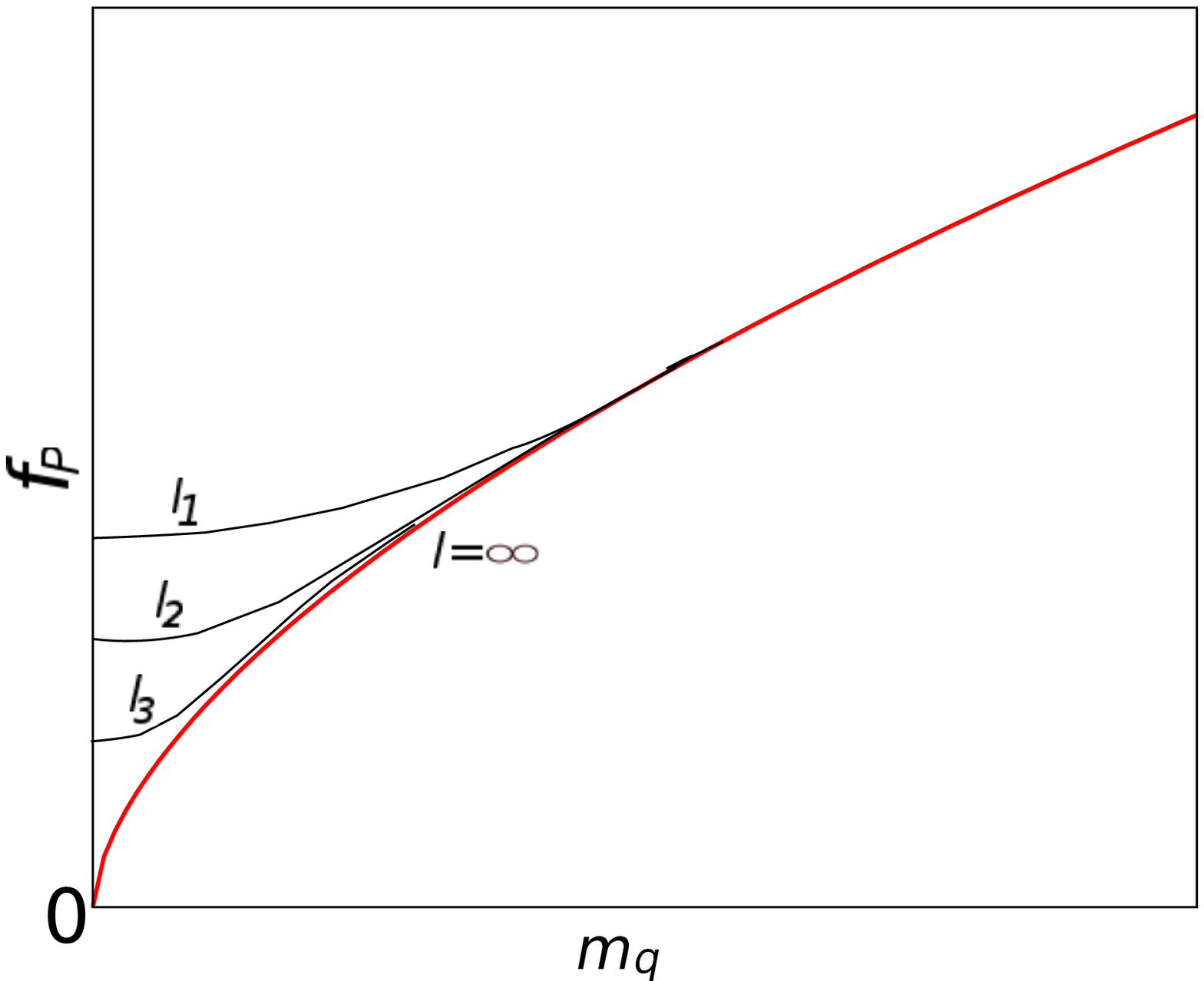}
\caption{
 Example $1$ of possibility
for finite size effect on the decay constant $f_P(m_q)$
versus the quark mass $m_q$ in the IR-conformal theory. 
 The linear sizes  $l_1,\,l_2,\,l_3$ of three spatial volumes
are ordered such that $l_3 > l_2 > l_1$.
}
\label{fig:fP_finiteVolume_IRconformal_1}
\end{center}
\vspace{0.2cm}
\end{minipage}
\quad 
\begin{minipage}{0.47\hsize}
\begin{center}
\includegraphics[width=\hsize,clip]{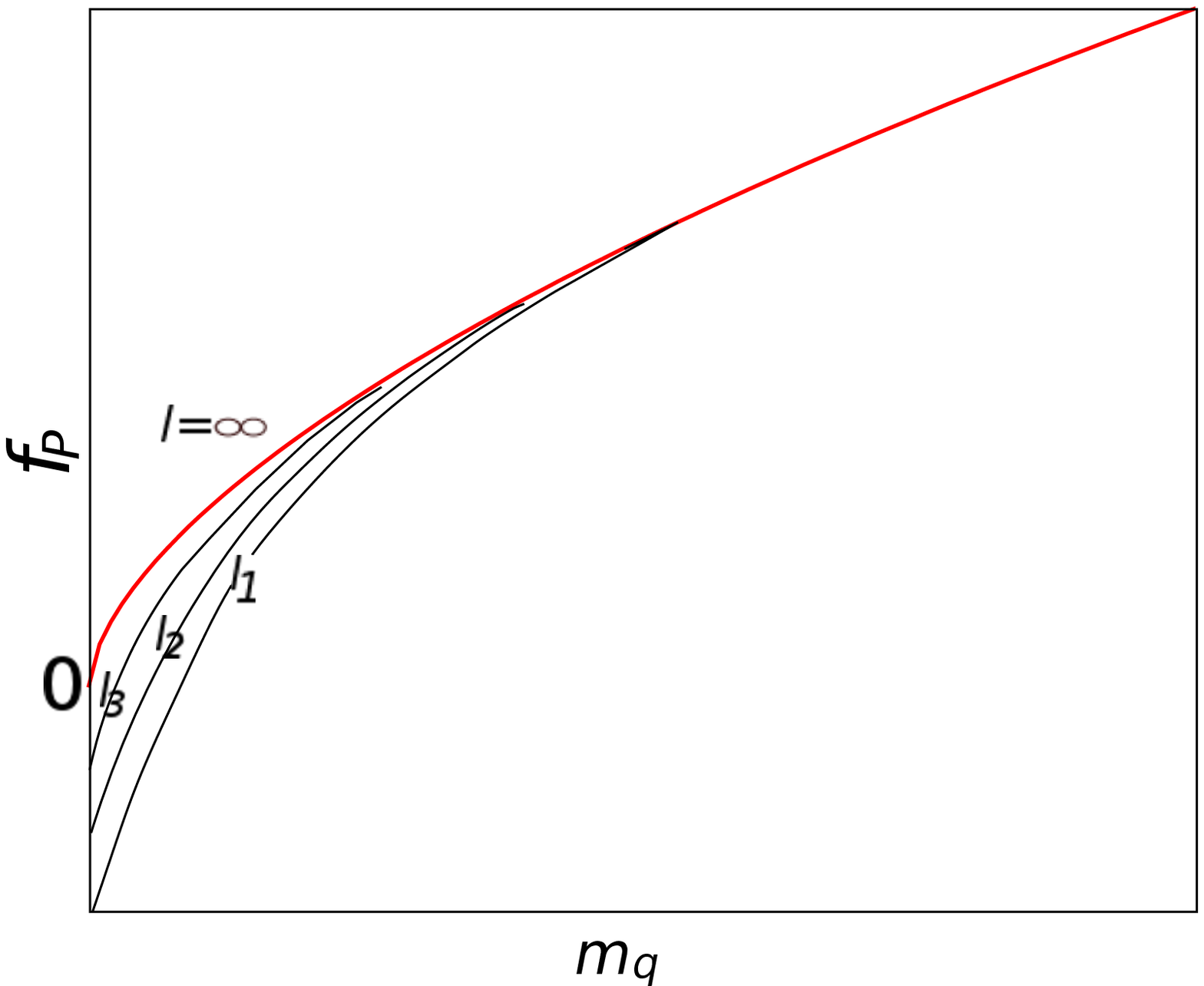}
\caption{
 Example $2$ of possibility
for finite size effect on the decay constant $f_P(m_q)$
versus the quark mass $m_q$ in the IR-conformal theory, 
in which $f_P$ is reduced by the finite size effect 
and approaches to a negative value in the chiral limit.
}
\label{fig:fP_finiteVolume_IRconformal_2}
\end{center}
\end{minipage}
\end{tabular}
\end{figure}
\begin{figure}[htp]
\begin{tabular}{cc}
\begin{minipage}{0.47\hsize}
\begin{center}
\includegraphics[width=\hsize,clip]{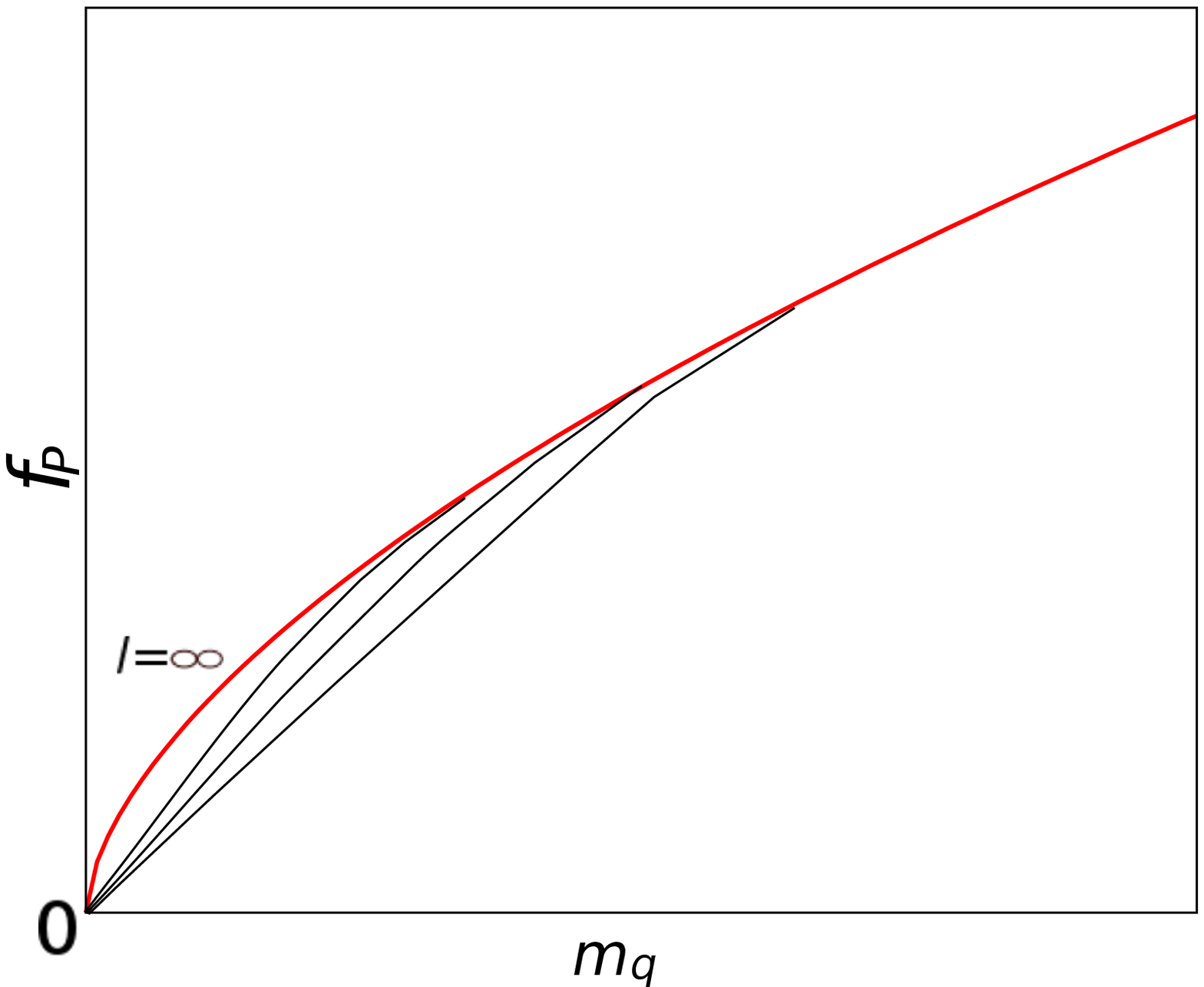}
\caption{
 Example $3$ of possibility
for finite size effect on the decay constant $f_P(m_q)$
versus the quark mass $m_q$ in the IR-conformal theory, 
in which $f_P$ becomes zero in the chiral limit. 
}
\label{fig:fP_finiteVolume_IRconformal_3}
\end{center}
\end{minipage}
\quad 
\begin{minipage}{0.47\hsize}
\begin{center}
\includegraphics[width=\hsize,clip]{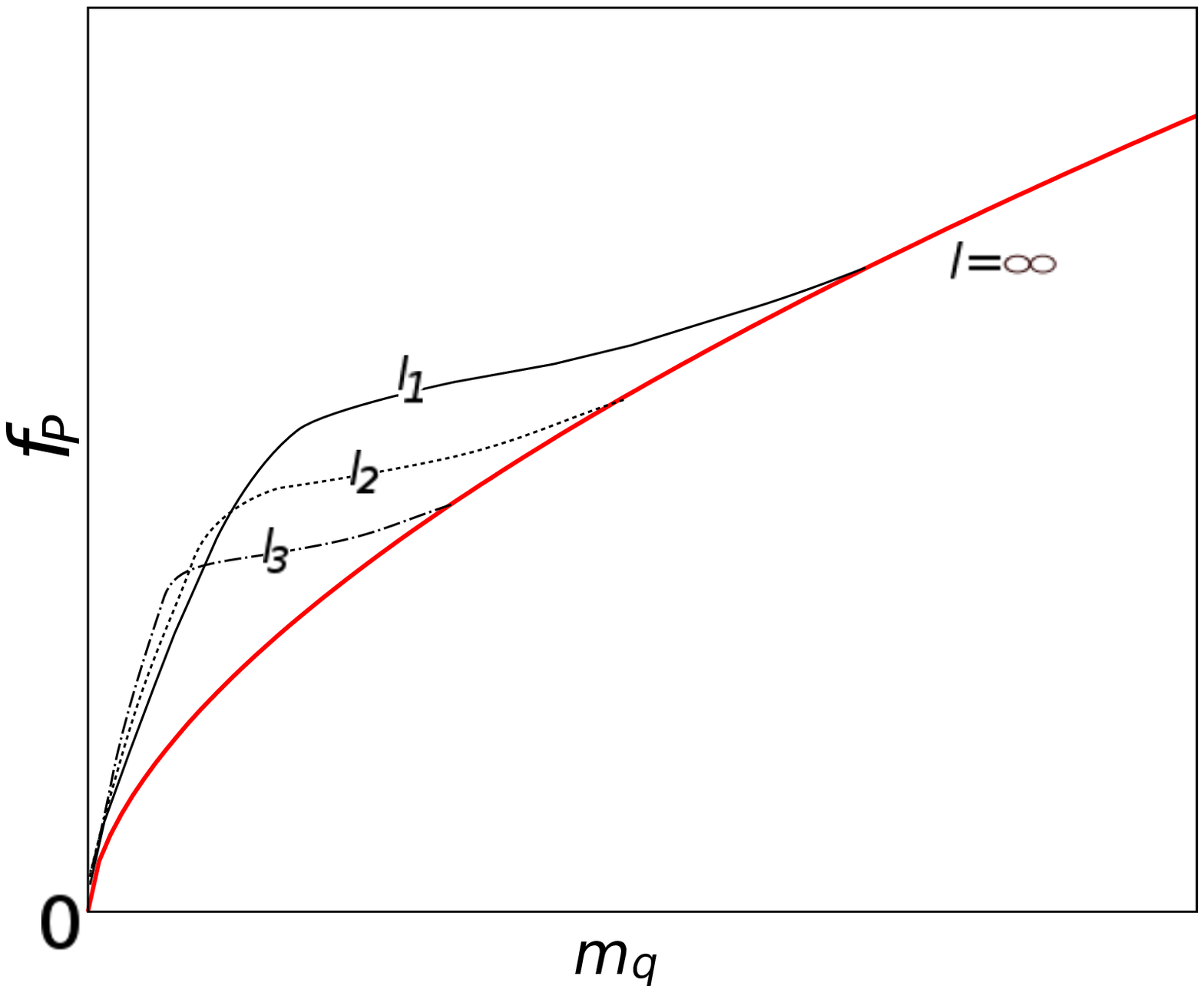}
\caption{
 Example $4$ of the possibility
for finite size effect on the decay constant $f_P(m_q)$
versus the quark mass $m_q$ in the IR-conformal theory, 
in which $f_P$ becomes zero in the chiral limit. 
}
\label{fig:fP_finiteVolume_IRconformal_4}

\end{center}
\end{minipage} 
\end{tabular}
\end{figure}

 Now, we question how the finiteness of the system size
affects to $f_P(m_q)$ in the IR-conformal theory.
 The analysis in the mass-deformed conformal gauge theory
together with the hyperscaling hypothesis 
\cite{DelDebbio:2010hu,DelDebbio:2010ze}
shows that a scaling variable in the finite volume method
is given by the combination $L m_q^{1/(1 + \gamma_\star)}$, 
where $\gamma_\star$ is the mass anomalous dimension
at the infrared fixed point.
 However, it does not predict the form of the scaling function. 
 In particular, it is uncertain 
whether the finite size effect
tends to increase or decrease $f_P(m_q)$.
 Such a qualitative dependence of $f_P(m_q)$ on $L$ 
will reflect the dynamics of the system at long distance.
 Here, we try to speculate various possibilities for 
the finite size effect on $f_P(m_q)$ 
in the theory without chiral symmetry breaking, 
in order to compare the measured data with them at the stage of the analysis.
 The hyperscaling hypothesis predicts that
the universal part of the decay constant is given
by the term $\propto (m_q)^{\alpha_\star}$
with $\alpha_\star = 1/(1 + \gamma_\star)$
in the IR-conformal theory
at infinite volume \cite{DelDebbio:2010hu,DelDebbio:2010ze},
which is expected to dominate $f_P(m_q)$ if $\gamma_\star > 0$.
 It is plausible that the finite size effect will show up at smaller $m_q$
at larger volume.
 With these ``boundary conditions'',
we list four candidates 
of finite size effects in Figs.~\ref{fig:fP_finiteVolume_IRconformal_1}, 
\ref{fig:fP_finiteVolume_IRconformal_2}, 
\ref{fig:fP_finiteVolume_IRconformal_3}
and \ref{fig:fP_finiteVolume_IRconformal_4}.

 In the example $1$ ($2$) in Fig.~\ref{fig:fP_finiteVolume_IRconformal_1} 
(\ref{fig:fP_finiteVolume_IRconformal_2}),
the finite size effect is presumed to increase (decrease) $f_P$ 
with non-vanishing intercept in the chiral limit.
 Note that the example $1$ can be clearly distinguished from 
the behavior in the ${\chi\hspace{-5pt}/}$-theory as shown in 
Fig.~\ref{fig:fP_finiteVolume_chibreak}.
 In ${\rm SU(2)_C}$ gauge theory with two adjoint Dirac fermions,
the finite size effect appears to act on $f_P$ as in this example $1$,
although it is not obvious that the intercept at $m_q \rightarrow 0$
is non-vanishing.
(See Fig.~5 in Ref.~\cite{DelDebbio:2010hu}.)

 The case of the example $2$ can be distinguished from 
the ${\chi\hspace{-5pt}/}$-theory,
because $f_P(m_q)$ is then expected to approach to zero
at a certain non-zero $m_q$ in the finite volume theory.
 However, from Eq.~(\ref{eq:fP}) we can see that $A_{PP}$ 
vanishes at some quark mass in the example $2$, 
implying the disappearance of contribution of the ground state
in the $PP$ contribution.
 We are thus inclined to consider example $3$ 
with vanishing intercept in the chiral limit
as in Fig.~\ref{fig:fP_finiteVolume_IRconformal_3},
where the curves are further far away from the curve at $l = \infty$
for smaller volumes.
 If this case is realized, the finite size effect on $f_P$ 
may not help to determine whether the system
is a ${\chi\hspace{-5pt}/}$-theory or not. 
 The example $4$ in Fig~\ref{fig:fP_finiteVolume_IRconformal_4} 
illustrates the case in which
the decay constant $f_P$ is enhanced by the finite size effect 
but approaches to zero in the chiral limit.


\section{${\rm SU(2)_C}$ gauge theory with two-flavors}
\label{sec:two-flavor}

\begin{table}[htb]
\caption{PCAC mass, chiral condensate, decay constant
 in two-flavor theory on $L/a=16$ and $\beta = 2.0$ lattices.}
\label{tab:NF2.lattparams.L16.b2.0}
\begin{tabular}{cccc}
\hline
\hline
$\kappa$ & 
 $2 a m_{\rm PCAC}$ & 
 $a^3 \left<\overline{\psi} \psi\right>_{\rm subt}$ & $a f_P$ \\
\hline
$0.1540$ & 
 $+0.6657\ (13)$ & 
 $0.8861\ (18)$ & 
 $0.2566\ (38)$ \\

$0.1580$ & 
 $+0.4271\ (10)$ & 
 $0.6346\ (14)$ & 
 $0.2158\ (30)$ \\

$0.1620$ & 
 $+0.1800\ (21)$ & 
 $0.3055\ (37)$ & 
 $0.1583\ (73)$ \\

$0.1625$ & 
 $+0.1505\ (10)$ & 
 $0.2624\ (18)$ & 
 $0.1474\ (28)$ \\

$0.1630$ & 
 $+0.1183\ (12)$ & 
 $0.2116\ (21)$ & 
 $0.1344\ (30)$ \\
\hline
\hline
\end{tabular}
\end{table}

\begin{table}[htb]
\caption{PCAC mass, chiral condensate, decay constant
 in two-flavor theory on $L/a=24$ and $\beta = 2.0$ lattices.}
\label{tab:NF2.lattparams.L24.b2.0}
\begin{tabular}{cccc}
\hline
\hline
$\kappa$ & 
 $2 a m_{\rm PCAC}$ & 
 $a^3 \left<\overline{\psi} \psi\right>_{\rm subt}$ & $a f_P$ \\
\hline
$0.1540$ &
 $+0.6708\ (17)$ &
 $0.8935\ (23)$ & 
 $0.2607\ (23)$ \\

$0.1620$ &
 $+0.18347\ (97)$ &
 $0.3113\ (19)$ & 
 $0.1648\ (27)$ \\

$0.1625$ &
 $+0.15266\ (93)$ &
 $0.2656\ (21)$ & 
 $0.1513\ (14)$ \\

$0.1630$ &
 $+0.11949\ (68)$ &
 $0.2140\ (13)$ & 
 $0.1379\ (16)$ \\

$0.1637$ &
 $+0.06946\ (67)$ &
 $0.1317\ (13)$ & 
 $0.1135\ (14)$ \\
\hline
\hline
\end{tabular}
\end{table}

\begin{table}[htb]
\caption{Meson masses in two-flavor theory
 on $L/a=16$ and $\beta = 2.0$ lattices.}
\label{tab:NF2.mesons.L16.b2.0}
\begin{tabular}{ccccc}
\hline
\hline
$\kappa$ &
 $a M_P$ & 
 $a M_V$ & 
 $a M_S$ & 
 $a M_A$ 
\\
\hline
$0.1540$ & 
 $1.1797\ (17)$ &
 $1.2446\ (28)$ &
 $2.083\  (82)$ &
 $2.048\  (44)$ 
\\
$0.1580$ &
 $0.9637\ (16)$ &
 $1.0525\ (29)$ &
 $1.856\  (81)$ &
 $1.833\  (53)$ 
\\
$0.1620$ &
 $0.6321\ (62)$ &
 $0.770\ (16)$  &
 $1.44\ (28)$ &
 $1.42\ (13)$ 
\\
$0.1625$ &
 $0.5797\ (32)$ &
 $0.7147\ (42)$ &
 $1.10\ (22)$ &
 $1.25\ (13)$ 
\\
 $0.1630$ &
 $0.5155\ (33)$ &
 $0.6647\ (67)$ &
 $1.12\ (22)$ &
 $0.96\ (12)$ 
\\
\hline
\hline
\end{tabular}
\end{table}

\begin{table}[htb]
\caption{Meson masses in two-flavor theory
 on $L/a=24$ and $\beta = 2.0$ lattices.}
\label{tab:NF2.mesons.L24.b2.0}
\begin{tabular}{ccccc}
\hline
\hline
$\kappa$ &
 $a M_P$ &
 $a M_V$ &
 $a M_S$ &
 $a M_A$
\\
\hline

$0.1540$ &%
 $1.1841\ (26)$ &
 $1.2452\ (29)$ &
 $2.20\ (14)$ &
 $2.091\ (82)$
\\

$0.1620$ &%
 $0.6413\ (18)$ &
 $0.7647\ (39)$ &
 $1.53\ (11)$ &
 $1.41\ (20)$ 
\\

$0.1625$ &
 $0.5792\ (31)$ &
 $0.7082\ (63)$ &
 $1.132\ (93)$ &
 $1.329\ (70)$ 
\\

$0.1630$ &
 $0.5176\ (22)$ &
 $0.6543\ (37)$ &
 $1.08\ (18)$ &
 $1.060\ (93)$ 
\\
$0.1637$ & 
 $0.3930\ (25)$ &
 $0.5519\ (48)$ &
 $1.192\ (74)$ &
 $1.137\ (42)$ 
\\
\hline
\hline
\end{tabular}
\end{table}

 In this section, we observe
the qualitative features of the finite size effects
on the various hadronic quantities simulated 
for the ${\chi\hspace{-5pt}/}$-theory
at very weak gauge coupling.
 The simulation is performed in the 
${\rm SU(2)_C}$ gauge theory with two Dirac fermions
in the fundamental representation, 
referred to as the two-flavor or $N_F = 2$ theory.
 The data to be obtained for our target system, the six-flavor theory, 
will be confronted with the knowledge and experiences
obtained in this section.
 We also examine the utility of the subtracted chiral condensate
(\ref{eq:def:scond}) by looking at the data
in the two-flavor theory.


 To highlight the finite size effects,
the simulation is first performed in the
two-flavor theory at $\beta \equiv 4 /g_0^2 = 2.0$
with lattice sizes, $16^3 \times 64$ and $24^3 \times 64$.
 Tables \ref{tab:NF2.lattparams.L16.b2.0}
and \ref{tab:NF2.lattparams.L24.b2.0}
list the simulation parameters, 
and the results for
PCAC masses, subtracted chiral condensate 
$\left<\overline{\psi} \psi\right>_{\rm subt}$ and decay constant $f_P$, 
for $L/a=16$ and $L/a=24$, respectively.
 The gauge configurations 
are stored once every $10$ trajectories. 
 The number of configurations used in the measurement
is $100 \sim 300$, depending on the parameters.
 Statistical errors are
estimated by the single elimination jack-knife method 
throughout this work. 
 The auto-correlation among the configurations
is checked by binning data and varying the bin size.

 The masses of lightest mesons in the various flavor-nonsinglet channels
are shown in Tabs.~\ref{tab:NF2.mesons.L16.b2.0} and 
\ref{tab:NF2.mesons.L24.b2.0}.
 Note that the effective mass plots in the scalar and axial-vector
channels exhibit large fluctuation at $t$ greater than $8 \sim 10$
\footnote{
 Such large fluctuation does not appear 
for small quark both at weak coupling in $N_F=2$ theory
and at $\beta = 2.0$ in $N_F=6$ theory.
}.
 Thus, the masses of these channels 
are obtained by fit at smaller $t$ and the limited range if required.

\begin{figure}[htb]
\begin{center}
\includegraphics[width=11.0cm,clip]{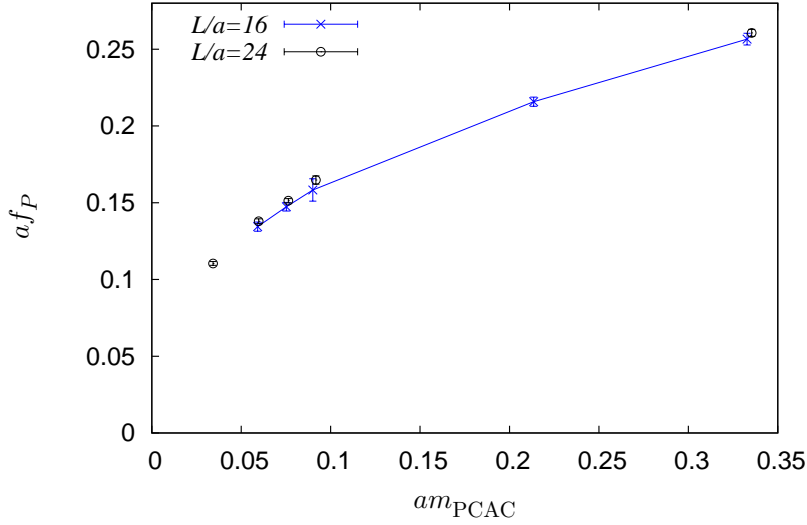}
\caption{
 Decay constant $f_P$ in two-flavor theory at $\beta = 2.0$
and two different volumes, 
$16^3$ (blue cross dots) and $24^3$ (black circle dots).
 The line is written just to guide eyes
to the plots of $L/a=16$.
}
\label{fig:NF2.fP.b2.0}
\end{center} 
\end{figure}
\begin{figure}[htb]
\begin{center}
\includegraphics[width=11.0cm,clip]{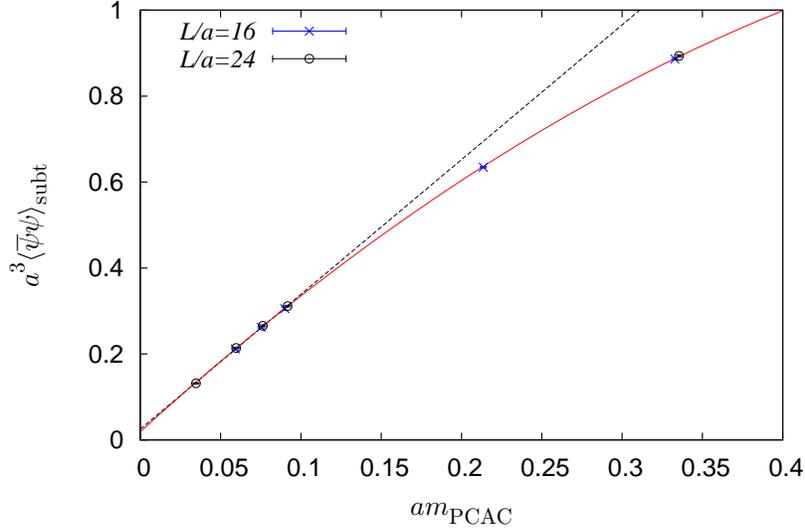}
\caption{
 Subtracted chiral condensate in two-flavor theory at $\beta = 2.0$
and two different volumes, 
$16^3$ (blue cross dots) and $24^3$ (black circle dots).
 No detectable finite size effect is observed, 
and the linear fit (dotted line)
to the data of quark masses $a m_{\rm PCAC} < 0.1$
and the quadratic fit to all data
are performed.
 Only the fit curves with the central values for the coefficients 
are drawn.
}
\label{fig:NF2.scond.b2.0}
\end{center} 
\end{figure}

\begin{figure}[thb]
\begin{center}
\includegraphics[width=10.0cm,clip]{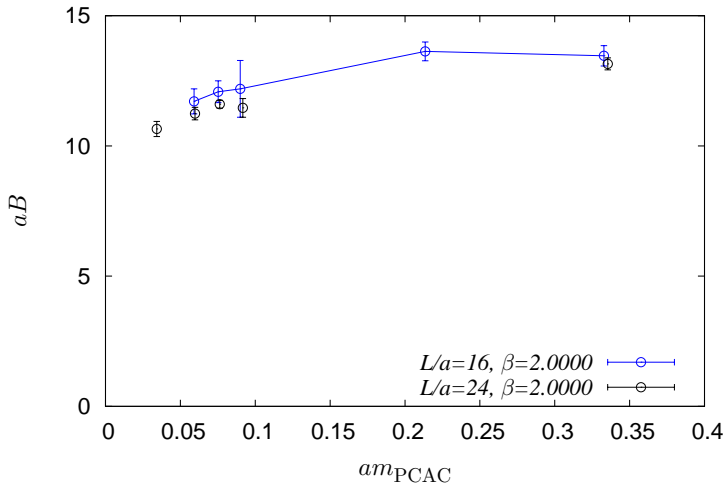}
\caption{
 $B$ defined by Eq.~(\ref{eq:def:bchi}) in two-flavor theory at $\beta = 2.0$.
}
\label{fig:NF2.bchi.b2.0}
\end{center} 
\end{figure}

 Figures \ref{fig:NF2.fP.b2.0} and \ref{fig:NF2.scond.b2.0}
show the quark mass dependence of 
decay constant $f_P$ and subtracted chiral condensate, respectively.
 At present, no explicit analysis for the finite size effect
on $f_P$ is available according to the chiral perturbation theory 
for the breaking pattern ${\rm SU}(2 N_F) \rightarrow {\rm Sp}(2 N_F)$
anticipated in the ${\rm SU(2)_C}$ gauge theory.
 In Fig.~\ref{fig:NF2.fP.b2.0}
the finite size effect is small but seems to tend to decrease $f_P$
\footnote{
 The tendency of decrease of $f_P$ can be more clearly observed in 
Fig.~\ref{fig:NF2.bchi.b2.0} for $B$ which is proportional to $1/f_P^2$, 
together with the fact that 
the subtracted chiral condensate 
$\left<\overline{\psi}\psi\right>_{\rm subt}$ does not 
suffer visible finite size effect.
}.
 This is compatible with 
the schematic behavior in Fig.~\ref{fig:fP_finiteVolume_chibreak}
predicted from the chiral perturbation theory 
for ${\rm SU}(N_F)_{\rm L} \times {\rm SU}(N_F)_{\rm R} 
\rightarrow {\rm SU}(N_F)_{\rm V}$ in the ${\rm SU(3)_C}$ gauge theory with
chiral symmetry breaking.

 As mentioned in Sec.~\ref{subsec:scond},
we will use 
the subtracted chiral condensate
$\left<\overline{\psi}\psi\right>_{\rm subt}$ in Eq.~(\ref{eq:def:scond}) 
as one of the quantities to examine the occurrence of
spontaneous chiral symmetry breaking in the foregoing analysis.
 Here, we discuss 
the utility of $\left<\overline{\psi}\psi\right>_{\rm subt}$.

 The subtracted chiral condensate
$\left<\overline{\psi} \psi\right>_{\rm subt}$ is 
dominated by the contribution proportional to $m_{\rm PCAC}$
and its chiral limit seems to be well below $O(1)$ in the lattice unit.
 The result for $\left<\overline{\psi} \psi\right>_{\rm subt}$
at $\beta = 2.0$ in the two-flavor theory 
is shown in Fig.~\ref{fig:NF2.scond.b2.0}, 
which implies that it is actually the case. 
 This is the first indication of the utility 
of $\left<\overline{\psi} \psi\right>_{\rm subt}$
as the counterpart of the chiral condensate in the continuum.
 Contrastingly, the chiral limit of 
$\left<\overline{\psi}\psi\right>$ 
directly calculated as in Eq.~(\ref{eq:condNiave})
is $O(1)$ in the lattice unit,
reflecting the dominance of cubic UV divergence.
 
 From Eq.~(\ref{eq:def:scond}), 
the subtracted chiral condensate 
is proportional to $m_{\rm PCAC}$. 
 Therefore, in order for $\left<\overline{\psi} \psi\right>_{\rm subt}$ 
to produce a finite and non-vanishing VEV in the thermodynamic limit, 
the four-volume sum of pseudoscalar correlator must 
diverge linearly with respect to $m_{\rm PCAC}$.
 The four-volume sum is regular with respect to the quark mass
at finite volume,
and a singularity is possibly developed only in the infinite volume limit, 
compatible with Fig.~\ref{fig:conds_finiteVolume}. 
 Practically, what we can do at best is
to examine if the data are not incompatible with 
the non-vanishing VEV in the chiral limit as follows.
 First, no finite size effect can be 
seen in the existing data shown in Fig.~\ref{fig:NF2.scond.b2.0}.
 It is thus plausible to consider that they approximate 
the mass dependence in the infinite volume very well
(Unexpectedly, the size is too small to take the thermodynamic limit.). 
 Meanwhile, Fig.~\ref{fig:NF2.scond.b2.0}
indicates that
the shape of the mass dependence of the subtracted chiral condensate
is convex upward.
 The linear extrapolation to such data will thus tend to overestimate 
the chiral limit.
 Despite this fact, we first try to fit a linear function  
\begin{equation}
 f_2\left(x = a m_{\rm PCAC}\right) = a_0 + a_1 x \,,
  \label{eq:linearFunc}
\end{equation}
to the data of $a m_{\rm PCAC} < 0.1$, and get the result
\begin{equation}
 a_0 = 0.0250\ (28)\,,\quad a_1 = 3.136\ (38)\,.
\label{eq:scond_fit_b2.000_linear}
\end{equation}
 We also fit a quadratic function 
\begin{equation}
 f_3\left(x = a m_{\rm PCAC}\right) = b_0 + b_1 x + b_2 x^2\,,
  \label{eq:quadraticFunc}
\end{equation}
to all the data, and find 
\begin{equation}
 b_0 = 0.0190\ (19)\,,\quad b_1 = 3.397\ (29)\,,\quad
 b_2 = -2.368\ (74)\,.
 \label{eq:scond_fit_b2.000}
\end{equation}
 Both are not incompatible
with the non-vanishing VEV in the thermodynamic limit.

 For a given quark mass and volume, we define $a B$ by
\begin{equation}
 a B \equiv a \frac{\left<\overline{\psi} \psi\right>_{\rm subt}}{f_P^2}\,.
 \label{eq:def:bchi}
\end{equation}
 The parameter in the chiral perturbation theory, $B_0$ 
\cite{Gasser:1984gg}, 
is given by the same equation but with 
the renormalized condensate and decay constant in the chiral limit.
 $B_0$ will thus be obtained by multiplicative renormalization
of $\left.B\right|_{m_{\rm PCAC} \rightarrow 0}$.
 As our main purpose is to explore
qualitative dynamics at long distance,
e.g. finiteness and non-vanishment in the chiral limit, 
use of $B$ will suffice.
 Figure~\ref{fig:NF2.bchi.b2.0} shows $B$ 
calculated by Eq.~(\ref{eq:def:bchi}) 
with use of the subtracted chiral condensate
$\left<\overline{\psi} \psi\right>_{\rm subt}$
at $\beta = 2.0$ in the two-flavor theory.
 It indicates that $B$
is finite and non-vanishing in the chiral limit.
 Those features give
another support for the utility of the subtracted chiral condensate.

\begin{figure}[thb]
\begin{tabular}{cc}
\begin{minipage}{0.47\hsize}
\begin{center}
\includegraphics[width=\hsize,clip]{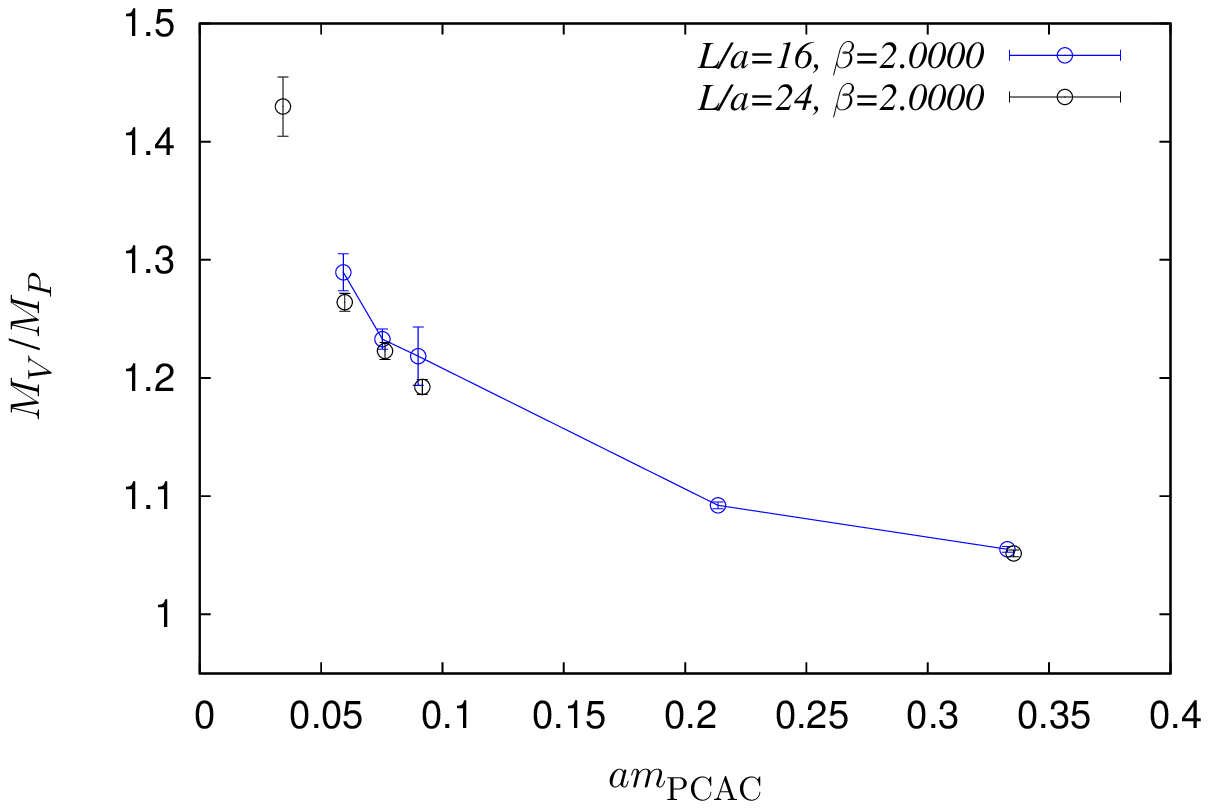}
\caption{
 $M_V /M_P$ in two-flavor theory at $\beta = 2.0$.
}
\label{fig:NF2.mVmP.b2.0}
\end{center} 
\end{minipage}
\quad 
\begin{minipage}{0.47\hsize}
\begin{center}
\includegraphics[width=\hsize,clip]{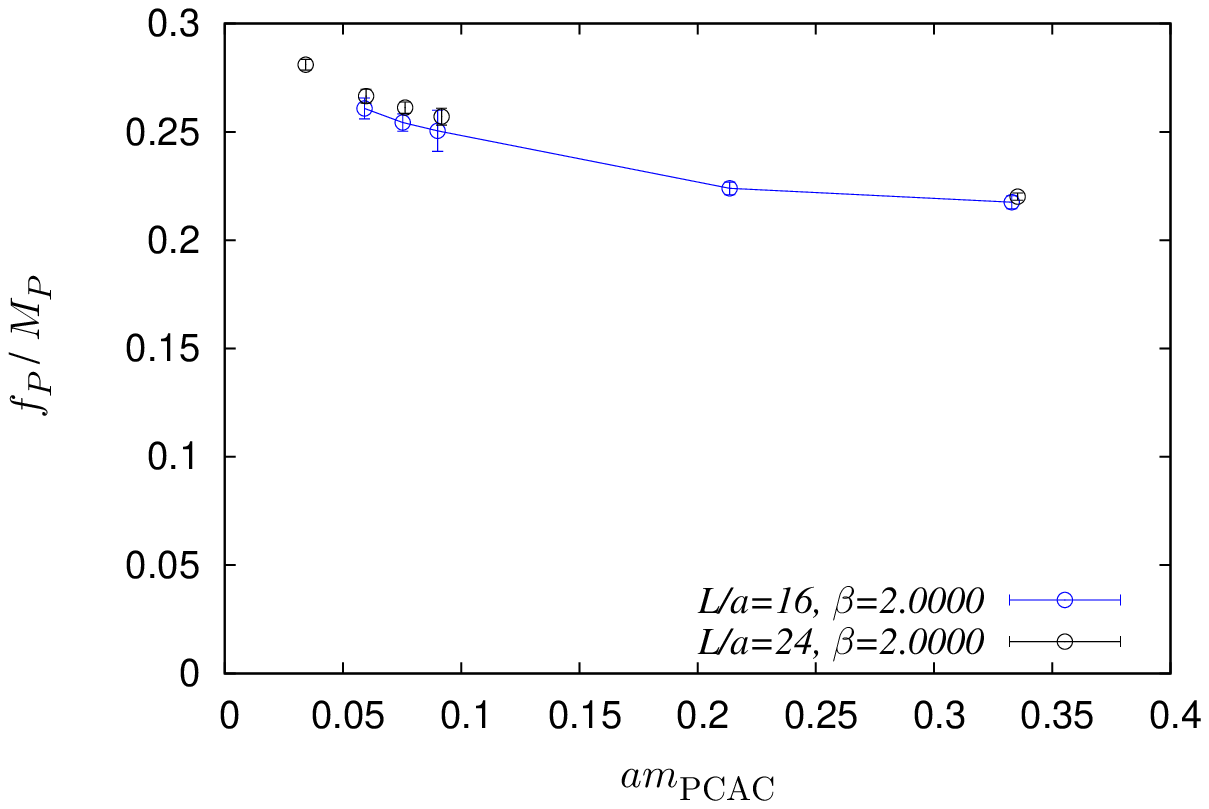}
\caption{
 $f_P /M_P$ in two-flavor theory at $\beta = 2.0$.
}
\label{fig:NF2.fPmP.b2.0}
\end{center} 
\end{minipage}
\end{tabular} 
\end{figure}

 Figures \ref{fig:NF2.mVmP.b2.0} and \ref{fig:NF2.fPmP.b2.0} show
the ratios $M_V /M_P$ and $f_P /M_P$, respectively, 
both of which increase for smaller quark mass as expected
in the ${\chi\hspace{-5pt}/}$-theory such as the two-flavor ${\rm SU(2)_C}$
gauge theory.

\begin{table}[htb]
\caption{PCAC mass, chiral condensate, decay constant
 in two-flavor theory on $L/a=16$ and $\beta = 4.0$ lattices.
}
\label{tab:NF2.lattparams.L16.b4.0}
\begin{tabular}{cccc}
\hline
\hline
$\kappa$ & 
 $2 a m_{\rm PCAC}$ & 
 $a^3 \left<\overline{\psi} \psi\right>_{\rm subt}$ & $a f_P$ \\
\hline
$0.1280$ & 
 $+0.39760\ (38)$ & 
 $0.29195\ (27)$ & 
 $0.04980\ (75)$ \\

$0.1320$ & 
 $+0.17976\ (19)$ & 
 $0.14372\ (16)$ & 
 $0.03591\ (38)$ \\

$0.1340$ & 
 $+0.06456\ (32)$ &
 $0.05377\ (27)$ & 
 $0.01551\ (16)$ \\
\hline
\hline
\end{tabular}
\end{table}

\begin{table}[htb]
\caption{PCAC mass, chiral condensate, decay constant
 in two-flavor theory on $L/a=24$ and $\beta = 4.0$ lattices.}
\label{tab:NF2.lattparams.L24.b4.0}
\begin{tabular}{cccc}
\hline
\hline
$\kappa$ & 
 $2 a m_{\rm PCAC}$ & 
 $a^3 \left<\overline{\psi} \psi\right>_{\rm subt}$ & $a f_P$ \\
\hline
$0.1280$ & 
 $+0.3926\ (12)$ & 
 $0.28830\ (9)$ & $-$ \\

$0.1320$ & 
 $+0.16795\ (10)$ & 
 $0.13451\ (8)$ &
 $0.03457\ (18)$ \\

$0.1340$ & 
 $+0.05630\ (11)$ & 
 $0.047016\ (9)$ &
 $0.01581\ (16)$ \\
\hline
\hline
\end{tabular}
\end{table}

\begin{table}[htb]
\caption{Meson masses in two-flavor theory
 on $L/a=16$ and $\beta = 4.0$ lattices.
 The column is left in blank unless
the effective mass plots obtained from the point-source exhibit plateau.}
\label{tab:NF2.mesons.L16.b4.0}
\begin{tabular}{ccccc}
\hline
\hline
$\kappa$ & 
 $a M_P$ & 
 $a M_V$ & 
 $a M_S$ & 
 $a M_A$ 
\\
\hline
$0.1280$ & 
 $0.5547\ (35)$ &
 $0.5582\ (34)$ &
 $0.5710\ (43)$ &
 $0.5745\ (44)$ 
\\
$0.1320$ &
 $0.5970\ (33)$ &
 $-$ & 
 $0.6019\ (32)$ &
 $-$ 
\\
  $0.1340$ &
 $0.5987\ (33)$ &
 $0.6096\ (34)$ &
 $0.6003\ (34)$ &
 $0.6112\ (34)$ 
\\
\hline
\hline
\end{tabular}
\end{table}

\begin{table}[htb]
\caption{Meson masses in two-flavor theory
 on $L/a=24$ and $\beta = 4.0$ lattices.}
\label{tab:NF2.mesons.L24.b4.0}
\begin{tabular}{ccccc}
\hline
\hline
$\kappa$ &
 $a M_P$ &
 $a M_V$ &
 $a M_S$ &
 $a M_A$
\\
\hline

$0.1280$ & $-$ & 
 $-$ & 
 $-$ & 
 $-$ 
\\
$0.1320$ & $0.3085\ (37)$ & 
 $0.3179\ (21)$ & 
 $0.3231\ (31)$ & 
 $0.3312\ (41)$ 
\\
 $0.1340$ & $0.2717\ (22)$ & 
 $-$ & 
 $0.2720\ (40)$ & 
 $-$ 
\\

\hline
\hline
\end{tabular}
\end{table}


 Next, the simulation is performed at relatively weak coupling, 
$\beta = 4.0$, and sizes of lattices, $16^3 \times 64$ and $24^3 \times 64$,
in which the average plaquette, $\left<W\right>$, 
where
\begin{equation}
 W 
 \equiv
 \frac{1}{n_U}
   \sum_{n \in \Gamma_4} \sum_{\mu < \nu}
   \frac{1}{N_C}\,{\rm tr}\,U(n;\,\mu,\nu)\,,
 \label{eq:def:plaquette}
\end{equation}
with $n_U$ the total number of the summed plaquettes, is around $0.8$.
 The simulation parameters are listed
in Tabs.~\ref{tab:NF2.lattparams.L16.b4.0}
and \ref{tab:NF2.lattparams.L24.b4.0}.

\begin{figure}[thb]
\begin{center}
\includegraphics[width=11.0cm,clip]{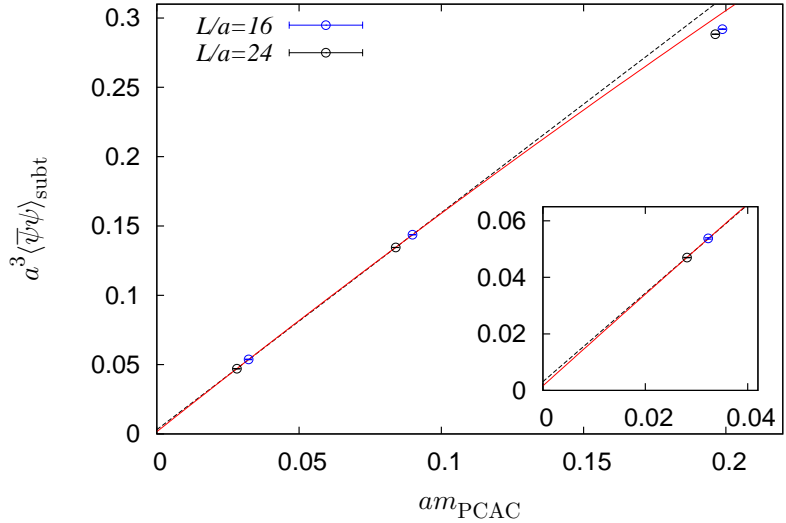}
\caption{
 Subtracted chiral condensate in two-flavor theory at $\beta = 4.0$
in two different volumes, 
$16^3$ (blue cross dots) and $24^3$ (black circle dots).
 The linear fit (black dotted line) and the quadratic fit (red curve)
to such data that satisfy $a m_{\rm PCAC} \le 0.1$ are performed.
}
\label{fig:NF2.scond.b4.0}
\end{center} 
\end{figure}

\begin{figure}[thb]
\begin{center}
\includegraphics[width=11.0cm,clip]{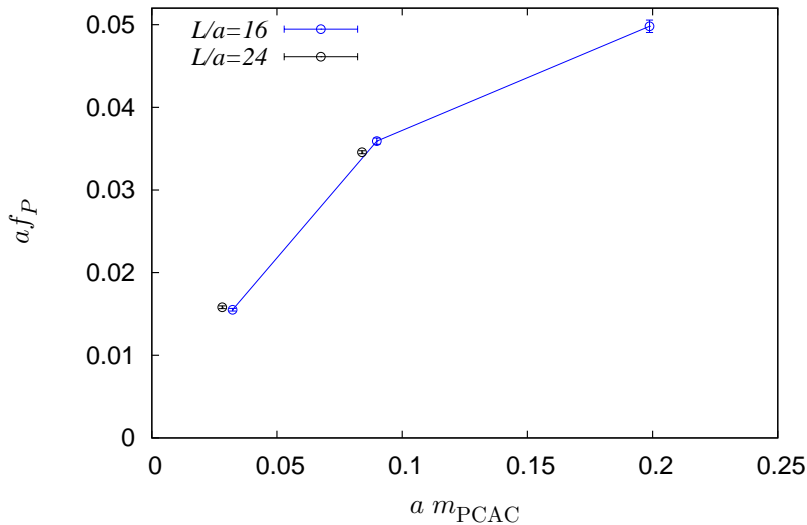}
\caption{
 Decay constant $f_P$ in two-flavor theory at $\beta = 4.0$
in two different volumes, 
$16^3$ (blue cross dots) and $24^3$ (black circle dots).
}
\label{fig:NF2.fP.b4.0}
\end{center} 
\end{figure}

 Figure~\ref{fig:NF2.scond.b4.0} shows 
$a^3 \left<\overline{\psi} \psi\right>_{\rm subt}$ 
measured at $\beta =4.0$ 
together with the result for the fit to the data
with $a m_{\rm PCAC} \le 0.1$
as they show inappreciable finite volume correction.
 The value in the chiral limit is found to be
\begin{equation}
 \lim_{m_{\rm PCAC} \rightarrow 0}
 \left.
  a^3 \left<\overline{\psi} \psi\right>_{\rm subt}
 \right|_{\beta = 4.0} 
 = 
 \left\{
  \begin{array}{ll}
   0.00310\  (18) & {\rm linear\ fit} \\
   0.0016\ (14) & {\rm quadratic\ fit} \\
  \end{array}
 \right.
 \,,
  \label{eq:scond_fit_b4.000_linear}
\end{equation}
in this case. 

\begin{figure}[thb]
\begin{tabular}{cc}
\begin{minipage}{0.47\hsize}
\begin{center}
\includegraphics[width=\hsize,clip]{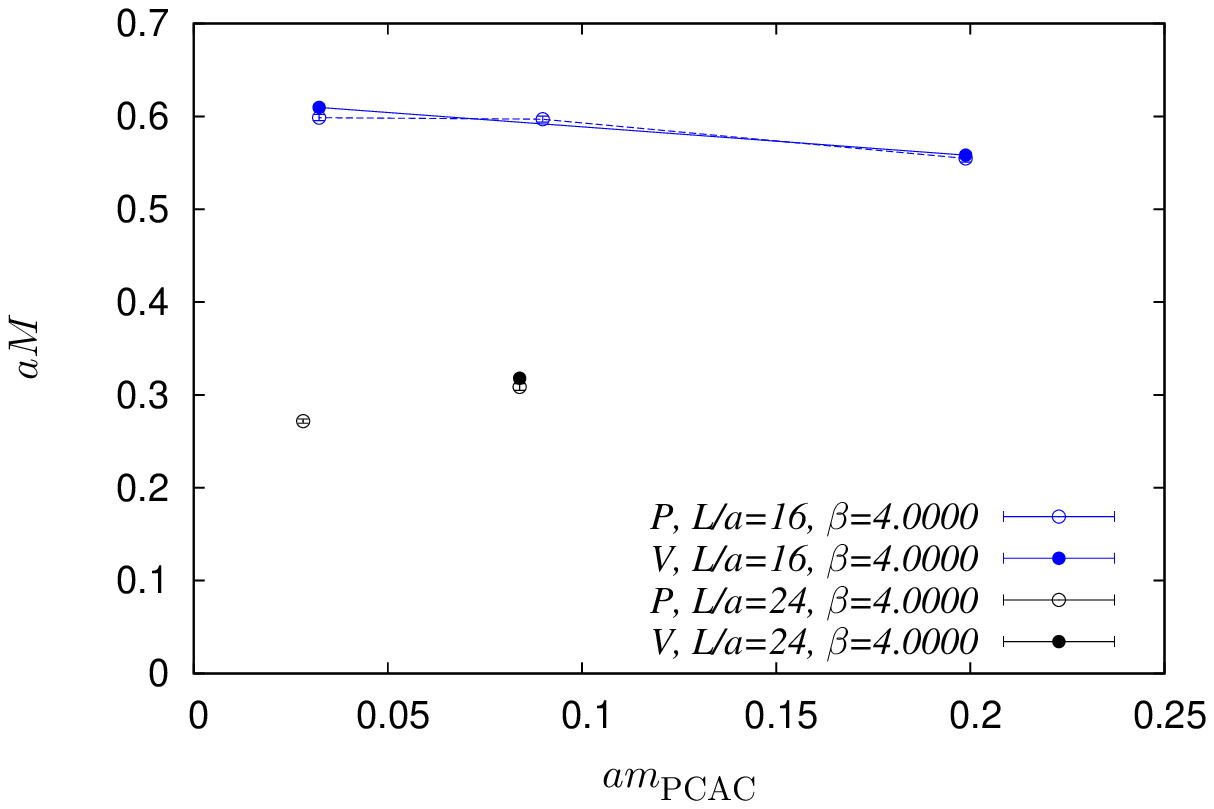}
\caption{
 $M_P$ (dashed line) and 
the lightest vector meson mass $M_V$
in two-flavor theory at $\beta = 4.0$.
 The difference between $M_V$ and $M_P$ 
is negligibly small, 
and both of them are bounded from below at small quark masses.
}
\label{fig:NF2.mP_mV.b4.0}
\end{center}
\end{minipage} 
\quad
\begin{minipage}{0.47\hsize}
\begin{center}
\includegraphics[width=\hsize,clip]{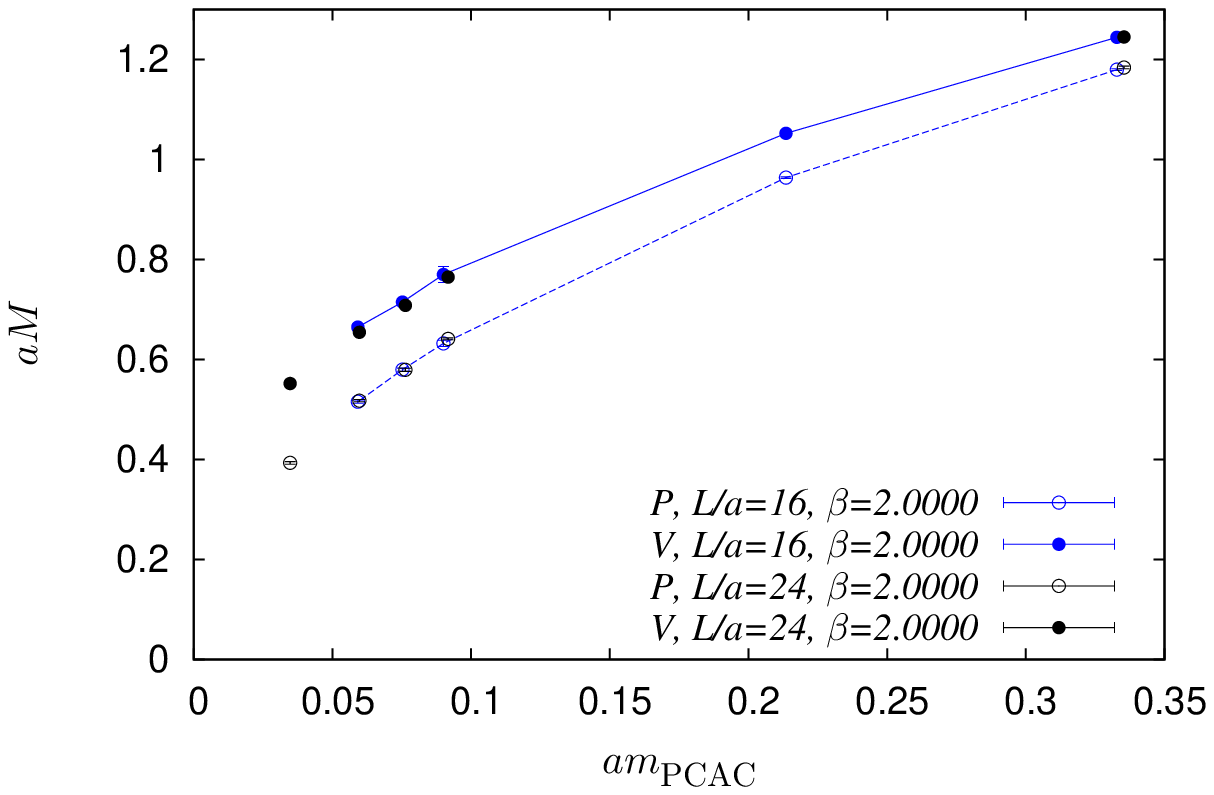}
\caption{
 $M_P$ (dashed line) and 
the lightest vector meson mass $M_V$
in two-flavor theory at $\beta = 2.0$.
}
\label{fig:NF2.mP_mV.b2.0}
\end{center} 
\vspace{1.0cm} 
\end{minipage}
\end{tabular}
\end{figure}

 Next, we turn our attention to $a f_P$ at $\beta = 4.0$. 
 Figure~\ref{fig:NF2.fP.b4.0} indicates 
that $a f_P$ approaches to $0$
in the limit $m_{\rm PCAC} \rightarrow 0$ at finite volume.
 To find the reason of this behavior, 
we plot the mass $M_P$ of the pseudoscalar meson for $\beta = 4.0$
in Fig.~\ref{fig:NF2.mP_mV.b4.0}.
 In contract to Fig.~\ref{fig:NF2.mP_mV.b2.0} at $\beta = 2.0$,
$M_P$ is bounded from below 
and does not seem to approach to $0$ in the limit $m_{\rm PCAC} \rightarrow 0$.
 The simulated lattices with $L/a=16$ and $a m_{\rm PCAC} < 0.1$ 
belong to the region where $M_P$ no longer decreases.
 Figure~\ref{fig:NF2.mP_mV.b4.0} also shows
that the size of the lower bound depends on the volume.
 Hence, the lower bound originates from the finite size effect.
 Since $a f_P$ is calculated by Eq.~(\ref{eq:fP}), 
the chiral extrapolation of the data of $f_P$ at finite volume
in the region where $M_P$ almost ceases to decrease
eventually yields zero in the massless quark limit,
even in the ${\chi\hspace{-5pt}/}$-theory.
 In Figure~\ref{fig:NF2.fP.b4.0}, 
visible finite size effect is observed
in $f_P$, and thus the chiral limit should be taken 
for a set of values that are obtained from those raw data 
in the infinite volume extrapolation. 
 We recall that the result of the finite volume correction shown
in Fig.~\ref{fig:fP_finiteVolume_chibreak}
assumes $4 \pi F_P \cdot L \gg 1$ for 
$F_P \equiv
\left.f_P\right|_{L/a\rightarrow\infty,\,m_{\rm PCAC}\rightarrow 0}$,
which is presumably not satisfied for all of our lattices
at $\beta = 4.0$.

\begin{figure}[thb]
\begin{tabular}{cc}
\begin{minipage}{0.47\hsize}
\begin{center}
\includegraphics[width=\hsize,clip]{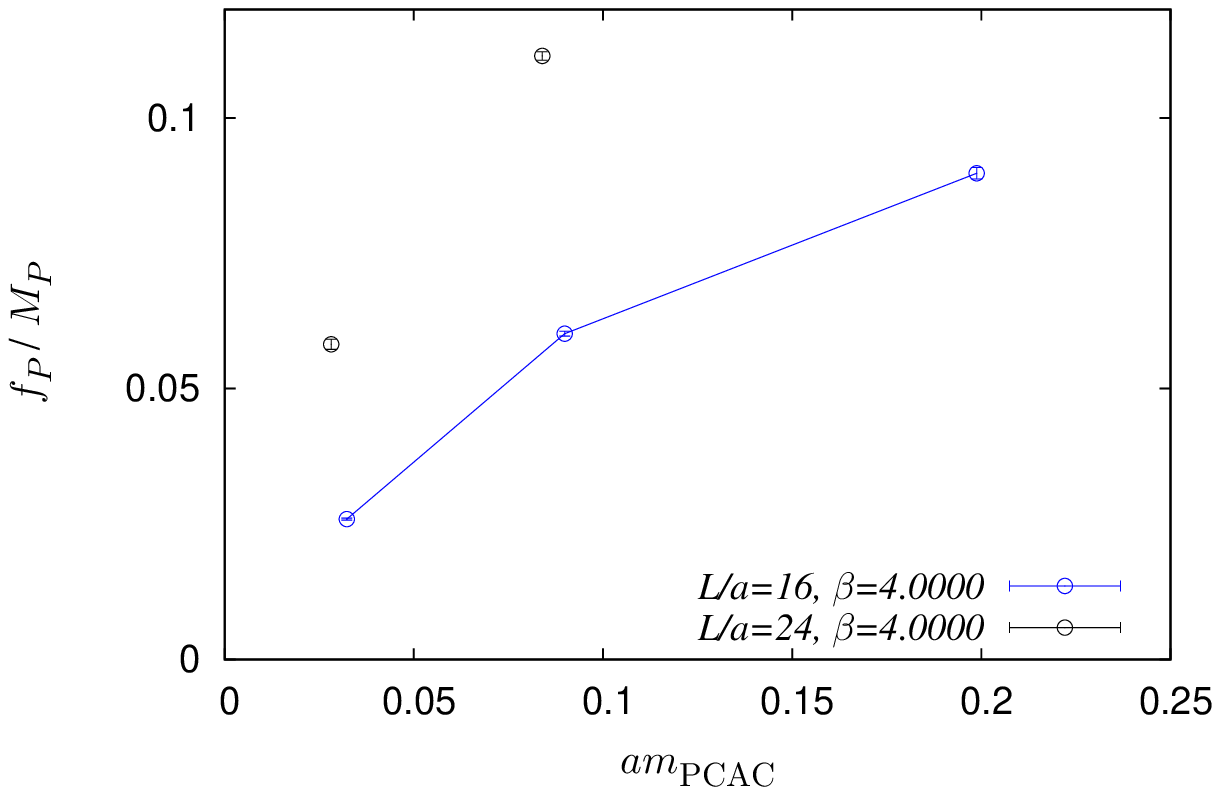}
\caption{
 $f_P /M_P$ in two-flavor theory at $\beta = 4.0$.
}
\label{fig:NF2.fPmP.b4.0}
\end{center}
\end{minipage}
\quad
\begin{minipage}{0.47\hsize}
\begin{center}
\includegraphics[width=\hsize,clip]{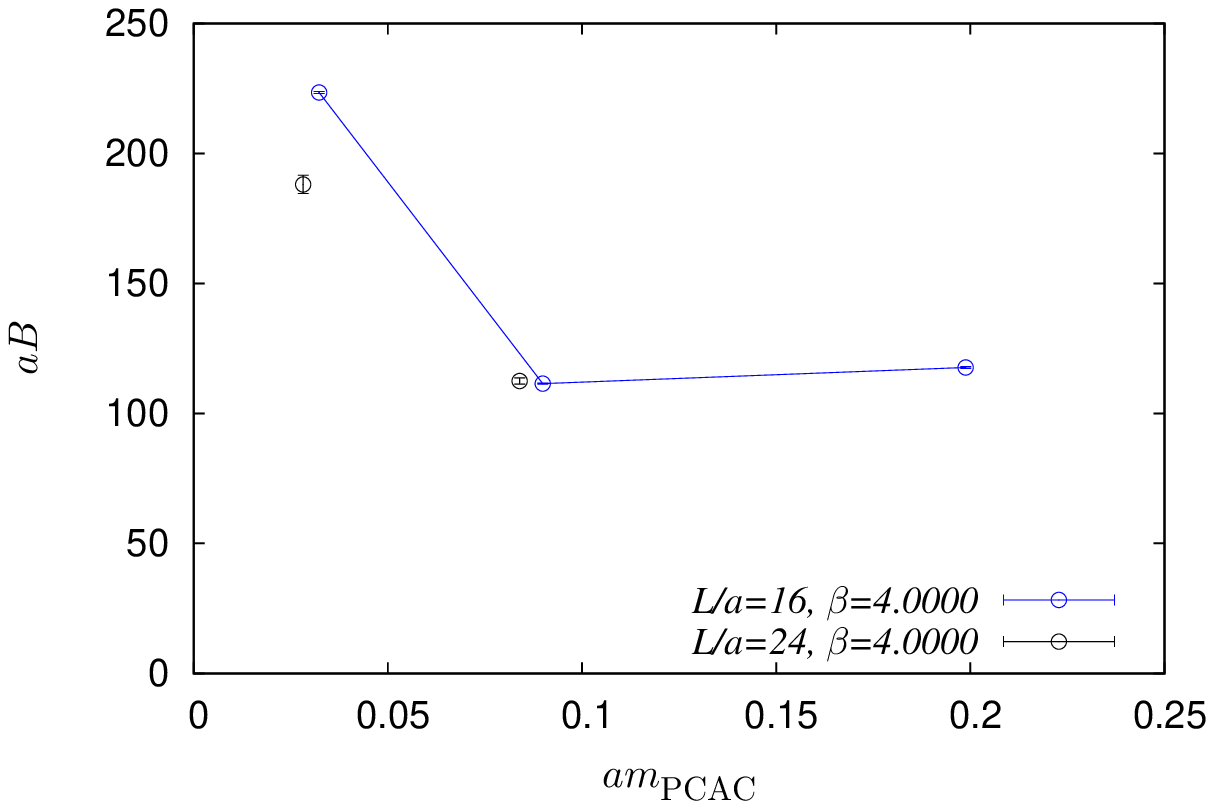}
\caption{
$B$ in two-flavor theory at $\beta = 4.0$.
}
\label{fig:NF2.bchi.b4.0}
\end{center} 
\vspace{0.2cm} 
\end{minipage}
\end{tabular}
\end{figure}

 Since $f_P$ decreases while $M_P$ is bounded from below,
the ratio $f_P /M_P$ at each lattice size also decreases to zero 
in the limit $m_{\rm PCAC} \rightarrow 0$,
as in Fig.~\ref{fig:NF2.fPmP.b4.0}.
 We have seen in Fig.~\ref{fig:NF2.scond.b4.0} that 
the chiral condensate possibly approaches to a non-vanishing value 
in the chiral limit.
 Since $f_P$ approaches to zero,
$B$ defined by Eq.~(\ref{eq:def:bchi}) will diverge
in the chiral limit at finite volume. 
 This explains the behavior observed in Fig.~\ref{fig:NF2.bchi.b4.0}.

 For future reference, we summarize our findings 
in the weak coupling simulation for the system with chiral symmetry breaking: 
\begin{enumerate}
 \item 
  Meson masses stop to decrease in the chiral limit at finite volume, 
with the size-dependent lower bound.
  Pseudoscalar meson tends to form a pair with scalar meson after 
the lower bound is reached. (See Tabs.~\ref{tab:NF2.mesons.L16.b4.0}
and \ref{tab:NF2.mesons.L24.b4.0}.)
 \item
  The decay constant $f_P$ calculated by Eq.~(\ref{eq:fP})
approaches to zero in the chiral limit at finite volume.
 \item
  $f_P /M_P$ approaches to zero in the chiral limit at finite volume
even in the ${\chi\hspace{-5pt}/}$-theory.
 \item
  $B(m_{\rm PCAC})$ defined by Eq.~(\ref{eq:def:bchi}), 
which is seen to approach
to a finite value at the modest coupling constant $\beta = 2.0$, 
blows up in limit $m_{\rm PCAC} \rightarrow 0$ at finite volume.
\end{enumerate}

\clearpage

\section{Brief look at phase structure
of two-color six-flavor Wilson fermions}
\label{sec:phaseStructure}

 This work is intended to provide the 
information complementary to that
found by the study of the running coupling constant in the Schr\"{o}dinger 
functional scheme 
\cite{Hayakawa:2013yfa}.
 In the latter, the calculation is performed at 
the massless points $\kappa_c(\beta)$ for a number of $\beta$, 
and the continuum limit is taken explicitly.
 Rather, this work focuses on the dependence 
on quark masses of the various quantities,
which requires the computation at various $\kappa$.
 As the first exploratory study, 
we perform the simulation at a fixed value of $\beta$, 
and forgive taking the explicit continuum limit.
 In this section, we examine the choice for its appropriate value.

 As seen in Sec.~\ref{subsec:gg}, 
no dimensionful parameter exists in the the lattice action.
 The lattice spacing $a$ and the physical size $L$ of the system 
is controlled mostly by 
the bare coupling constant $g_0^2 = 4/\beta$.
 The lattice with geometry $l^3 \times N_T$ 
represents smaller $a$ and physical size $L = l a$ at the weaker coupling
in the asymptotically free gauge theory.
 As the number of flavors is increased,
the effective coupling runs more slowly.
 Thus, to capture the strong coupling dynamics of the ${\rm SU}(2)_{\rm C}$
gauge theory with six flavors,
the simulation with fine lattices 
requires very large size of lattice 
and vast statistics to sample gauge fields with non-trivial topology, 
which cannot be done with the available computational resources.
 In this work, we attempt to study the dynamics of 
the six-flavor theory by carrying out the simulation 
on coarse lattices.
 The results will be checked by making a comparison with 
the results found in the Schr\"{o}dinger functional scheme
in which the continuum limit is explicitly taken.
 We thus start with choosing an appropriate value of $\beta$ 
which does not belong to the strong coupling phase 
peculiar to the lattice action (\ref{eq:latticeAction}) with Wilson fermions.
 For that purpose, it is inevitable to study more or less 
the phase structure of the system with six Wilson fermions. 

\begin{figure}[thb]
\begin{tabular}{cc}
\begin{minipage}{0.47\hsize}
\begin{center}
\includegraphics[width=\hsize,clip]{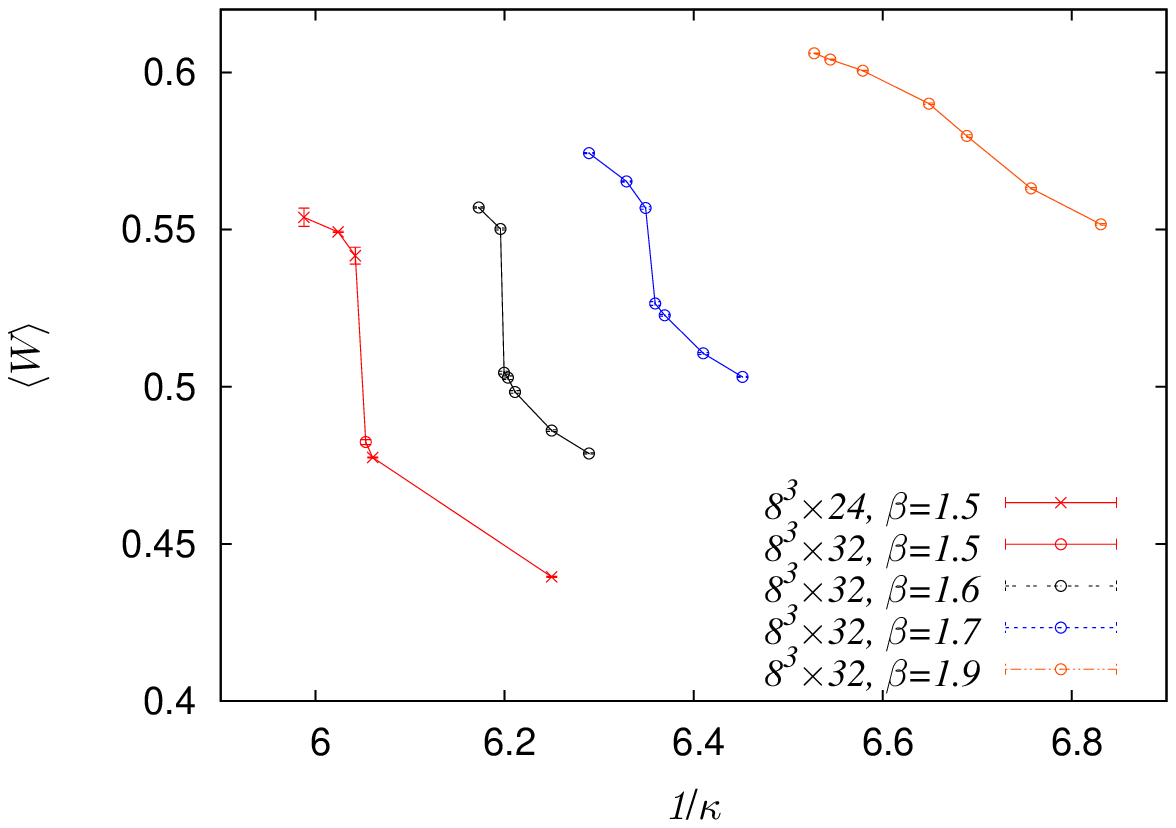}
\caption{
 $\left<W\right>$ as a function of $1 / \kappa$.
Lines are added just to guide eyes.
}
\label{fig:plaq_smallBeta_8x8x8}
\end{center}
\vspace{1.0cm}
\end{minipage}
\quad
\begin{minipage}{0.47\hsize}
\begin{center}
\includegraphics[width=\hsize,clip]{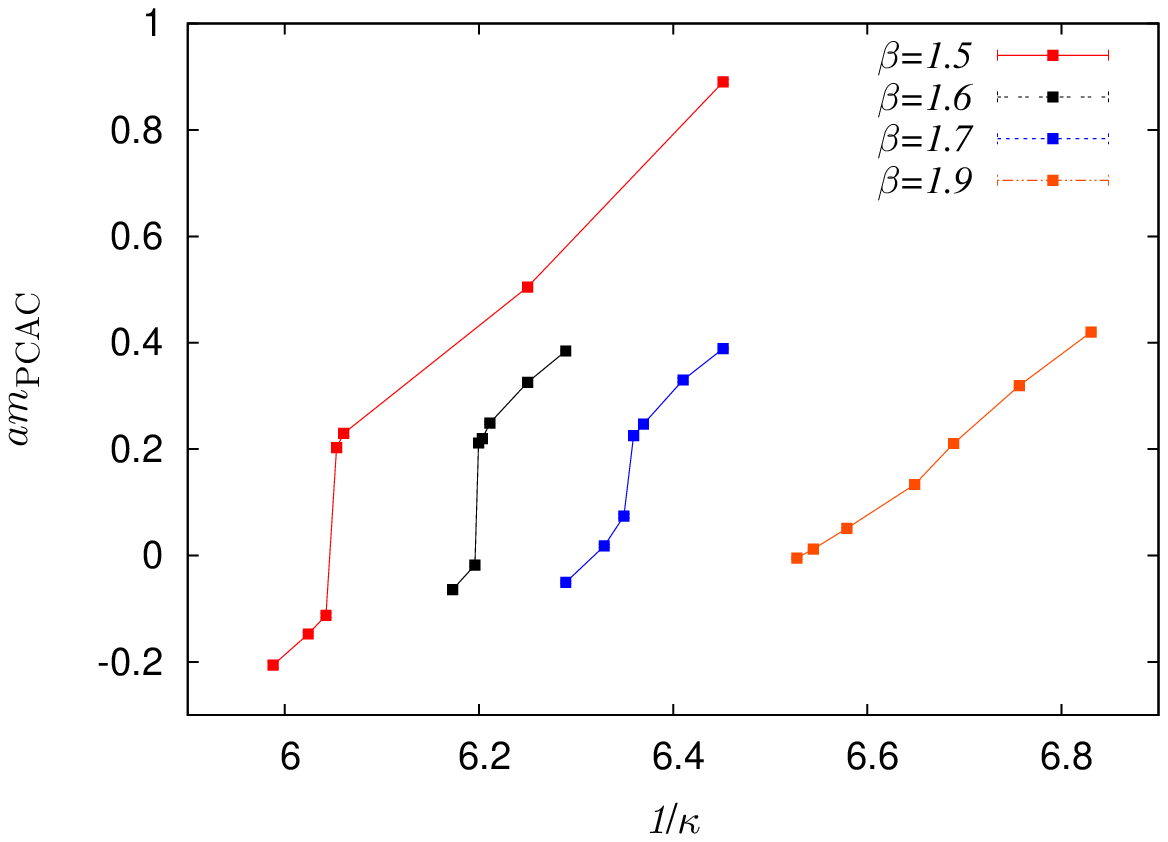}
\caption{
 PCAC mass $a m_{\rm PCAC}$ as a function of $1 / \kappa$.
 The results of $8^3 \times 24$ and $8^3 \times 32$ are plotted 
on the same footing for $m_{\rm PCAC}$ at $\beta=1.5$.
}
\label{fig:pcac_smallBeta_8x8x8}
\end{center}
\end{minipage}
\end{tabular}
\end{figure}

\begin{figure}[ht]
\includegraphics[width=11.0cm,clip]{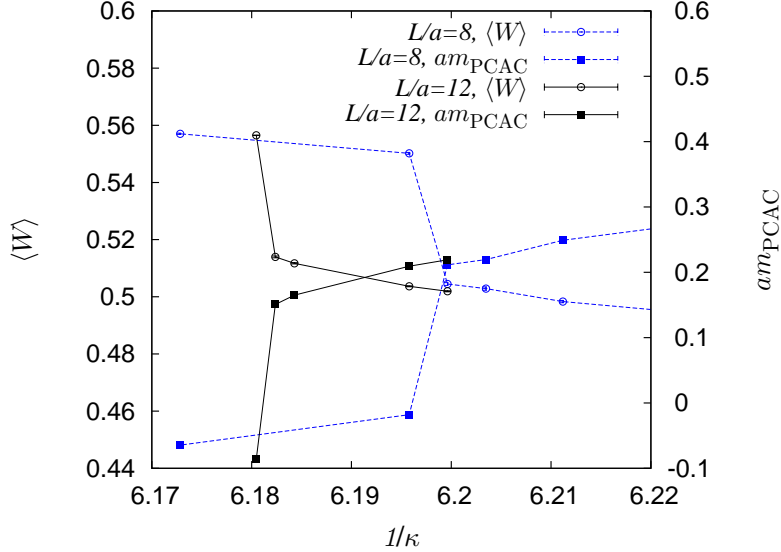}
\caption{
$\left<W\right>$
(empty circle dot)
and PCAC mass $a m_{\rm PCAC}$ (filled square dot)
versus $1 / \kappa$ 
on $l=8$ (dashed blue curve) and $l=12$ (black curve).
}
\label{fig:plaqANDpcac.8x8x8.12x12x12.b1.6}
\end{figure}

 Contrary to the work \cite{Iwasaki:2003de},
which surveys the phase structure of many flavor Wilson lattice gauge theories 
to explore the aspects of the systems, 
our aim is very restrictive here; 
we circumvent the parameters corresponding to 
the strong coupling phase which is separated from 
the weak coupling phase by the first-order bulk phase transition or 
the sharp crossover.
 Since such an unwanted strong coupling phase are attributed 
to the structure at short distance, we monitor 
the transition by the VEV of the averaged plaquettes,
$\left<W\right>$,
where 
and the PCAC mass $a m_{\rm PCAC}$ defined
in Sec.~\ref{subsec:PCACmass_decayConstat}, 
using the lattices of sizes, $8^3 \times 32$ and $8^3 \times 24$.
 The result is shown in
Figs.~\ref{fig:plaq_smallBeta_8x8x8} and \ref{fig:pcac_smallBeta_8x8x8}.
 Discontinuous or abrupt change of
$\left<W\right>$ for the plaquette (\ref{eq:def:plaquette})
and $a m_{\rm PCAC}$ at some $\kappa$
is observed for $\beta \lesssim 1.7$.

 In order to examine the property of the transition, 
simulation is performed on larger lattices, $12^3 \times 64$,
at $\beta = 1.6$, with a focus put on the vicinity of the transition point.
 Figure~\ref{fig:plaqANDpcac.8x8x8.12x12x12.b1.6} shows
that the values for the hopping parameter, $\kappa_t(L/a)$, 
at which the transition occurs are the same within $1$ \%
between two volumes, 
indicating that the transition 
is not related to any long distance physics.
 With these results, we choose $\beta = 2.0$ as a simulation parameter.
 It will then turn out that $a m_{\rm PCAC}$ depends nearly linearly
on the inverse of the hopping parameter, $1 / \kappa$, at $\beta = 2.0$.
(See Fig.~\ref{fig:kinvVSmPCAC.b2.0}.) 


\clearpage

\section{Results}
\label{sec:results}
 
 In this section, we present the results of the lattice simulation
carried out for the ${\rm SU(2)}_{\rm C}$ gauge theory 
with six-flavors. 
 We give three signatures supporting that the
six-flavor theory is not the one with chiral symmetry breaking.

\begin{table}[thb]
\caption{PCAC mass, chiral condensate, decay constant
 in six-flavor theory on $L/a=16$ and $\beta = 2.0$ lattices.
 The latticee geometry is $16^3 \times 32$ for $\kappa \le 0.14300$, 
and $16^3 \times 64$ for $\kappa \ge 0.14500$.
}
 \label{tab:NF6.lattparams.L16.b2.0}
\begin{tabular}{cccc}
\hline
\hline
$\kappa$ & 
 $2 a m_{\rm PCAC}$ & 
 $a^3 \left<\overline{\psi} \psi\right>_{\rm subt}$ & $a f_P$ \\
\hline
$0.13800$ & 
 $+1.3073\ (69)$ &
 $1.1942\ (63)$ & 
 $0.2268\ (68)$\\

$0.14100$ & 
 $+0.9848\ (64)$ &
 $0.9644\ (63)$ & 
 $0.2015\ (65)$\\

$0.14300$ & 
 $+0.7686\ (30)$ &
 $0.7878\ (29)$ & 
 $0.1859\ (46)$\\

$0.14500$ & 
 $+0.5423\ (10)$ &
 $0.5830\ (11)$ & 
 $0.1424\ (14)$\\

$0.14600$ & 
 $+0.4335\ (10)$ &
 $0.4772\ (12)$ & 
 $0.12697\ (78)$\\

$0.14700$ & 
 $+0.3197\ (10)$ &
 $0.3603\ (11)$ & 
 $0.1024\ (17)$\\

$0.14750$ & 
 $+0.2663\  (8)$ &
 $0.3034\  (9)$ & 
 $0.08803\ (14)$\\

$0.14760$ & 
 $+0.2572\  (8)$ &
 $0.2936\  (10)$ & 
 $0.0892\ (13)$\\

$0.14770$ & 
 $+0.2470\  (8)$ &
 $0.2828\   (9)$ & 
 $0.09006\  (67)$\\

$0.14785$ & 
 $+0.2285\ (10)$ &
 $0.2623\  (12)$ & 
 $0.0858\ (19)$\\

$0.14800$ & 
 $+0.2133\ (10)$ &
 $0.2455\ (12)$ & 
 $0.0799\ (15)$\\

$0.14900$ & 
 $+0.1102\  (5)$ &
 $0.12917\ (57)$ & 
 $0.04425\ (74)$\\

$0.14950$ & 
 $+0.06524\ (54)$ &
 $0.07711\ (64)$ & 
 $0.02673\ (32)$\\
\hline
\hline
\end{tabular}
\end{table}

\begin{table}[thb]
\caption{PCAC mass, chiral condensate, decay constant
 in six-flavor theory on $L/a=24$ and $\beta = 2.0$ lattices.
 All lattices has the same geometry $24^3 \times 48$.
}
 \label{tab:NF6.lattparams.L24.b2.0}
\begin{tabular}{cccc}
\hline
\hline
$\kappa$ & 
 $2 a m_{\rm PCAC}$ & 
 $a^3 \left<\overline{\psi} \psi\right>_{\rm subt}$ & $a f_P$ \\
\hline

$0.14100$ & 
 $+0.9833\ (21)$ &
 $0.9623\ (20)$ & 
 $0.2054\ (34)$ 
 \\
$0.14300$ & 
 $+0.7629\ (16)$ & 
 $0.7825\ (16)$ & 
 $0.1816\ (27)$ 
 \\
$0.14400$ & 
 $+0.6541\ (21)$ & 
 $0.6867\ (22)$ & 
 $0.1626\ (18)$ 
 \\

$0.14500$ & 
 $+0.5429\ (17)$ & 
 $0.5839\ (18)$ & 
 $0.1491\ (21)$ 
 \\

$0.14600$ & 
 $+0.4345\ (17)$ & 
 $0.4787\ (19)$ & 
 $0.1258\ (14)$ 
 \\

$0.14700$ & 
 $+0.3229\ (9)$ & 
 $0.3639\ (10)$ & 
 $0.1012\  (7)$ 
 \\

$0.14800$ & 
 $+0.21313\ (45)$  &
 $0.24551\  (54)$ & 
 $0.07749\ (54)$ 
 \\

$0.14850$ & 
 $+0.16341\ (58)$ &
 $0.19043\  (71)$ & 
 $0.06549\ (63)$ 
 \\

$0.14900$ & 
 $+0.10884\ (71)$  & 
 $0.12802\  (83)$ & 
 $0.04702\ (83)$ 
 \\

$0.14930$ & 
 $+0.07927\ (54)$ & 
 $0.09371\  (65)$ & 
 $0.03569\ (79)$ 
 \\

$0.14950$ & 
 $+0.06317\ (70)$ &
 $0.07495\  (83)$ & 
 $0.02923\ (50)$ 
 \\

$0.14980$ & 
 $+0.03460\ (63)$ & 
 $0.04120\  (74)$ & 
 $0.01552\ (35)$ 
 \\
$0.15000$ & 
 $+0.01702\ (34)$ & 
 $0.02036\ (41)$ & 
 $0.00776\ (17)$ 
 \\

\hline
\hline
\end{tabular}
\end{table}

\begin{table}[thb]
\caption{PCAC mass, chiral condensate, decay constant
 in six-flavor theory on $L/a=32$ and $\beta = 2.0$ lattices.
 All lattices has the same geometry $32^3 \times 64$.
}
 \label{tab:NF6.lattparams.L32.b2.0}
\begin{tabular}{cccc}
\hline
\hline
$\kappa$ & 
 $2 a m_{\rm PCAC}$ & 
 $a^3 \left<\overline{\psi} \psi\right>_{\rm subt}$ & $a f_P$ \\
\hline

$0.14850$ & 
 $+0.16108\ (20)$ & 
 $0.18773\ (23)$ & 
 $0.06225\ (45)$ 
 \\

$0.14920$ & 
 $+0.09165\ (25)$ & 
 $0.10835\ (29)$ & 
 $0.04138\ (69)$ 
 \\

$0.14965$ & 
 $+0.04672\ (35)$ & 
 $0.05567\ (41)$ & 
 $0.02326\ (32)$ 
 \\

$0.14980$ & 
 $+0.03239\ (19)$ & 
 $0.03868\ (22)$ & 
 $0.01582\ (23)$ 
 \\

\hline
\hline
\end{tabular}
\end{table}

\begin{table}[htb]
\caption{Meson masses in six-flavor theory
 on $L/a=16$ and $\beta = 2.0$ lattices.}
\label{tab:NF6.mesons.L16.b2.0}
\begin{tabular}{ccccc}
\hline
\hline
$\kappa$ &
 $a M_P$ & 
 $a M_V$ & 
 $a M_S$ & 
 $a M_A$ 
\\
\hline

$0.13800$ &
 $1.5152\ (51)$ & 
 $1.5388\ (53)$ & 
 $2.270\ (60)$ & 
 $2.237\ (48)$ 
 \\

$0.14100$ &
 $1.3213\ (52)$ & 
 $1.3543\ (75)$ & 
 $1.908\ (61)$  & 
 $1.926\ (50)$ 
 \\

$0.14300$ &
 $1.1551\ (34)$ & 
 $1.1845\  (51)$ & 
 $1.701\  (69)$ & 
 $1.718\  (65)$ 
 \\

$0.14500$ &
 $0.9347\ (16)$ & 
 $0.9412\ (23)$ & 
 $1.321\ (17)$ & 
 $1.323\ (13)$ 
 \\

$0.14600$ &
 $0.8086\ (16)$  & 
 $0.8391\ (19)$ & 
 $1.111\ (26)$ & 
 $1.145\ (23)$ 
 \\

$0.14700$ &
 $0.6587\ (32)$ & 
 $0.6871\ (46)$ & 
 $0.871\  (34)$ & 
 $0.868\ (22)$ 
 \\

$0.14750$ &
 $0.5813\ (29)$ & 
 $0.6110\ (34)$ & 
 $0.765\ (18)$ & 
 $0.777\ (21)$ 
 \\

$0.14760$ &
 $0.5755\  (27)$ & 
 $0.6079\  (35)$ & 
 $0.745\ (11)$ & 
 $0.754\ (11)$ 
 \\

$0.14770$ &
 $0.5674\  (17)$ & 
 $0.6045\  (19)$ & 
 $0.752\ (27)$ & 
 $0.763\ (22)$ 
 \\

$0.14785$ &
 $0.5679\  (29)$ & 
 $0.6034\  (31)$ & 
 $0.6792\  (92)$ & 
 $0.707\ (12)$ 
 \\

$0.14800$ &
 $0.5574\  (27)$ & 
 $0.5950\  (33)$ & 
 $0.645\ (12)$ & 
 $0.6777\  (72)$ 
 \\

$0.14900$ &
 $0.5404\ (26)$ & 
 $0.5761\ (28)$ & 
 $0.5571\ (35)$ & 
 $0.5913\ (40)$ 
 \\

$0.14950$ &
 $0.5427\ (29)$ & 
 $0.5812\ (30)$ & 
 $0.5539\ (31)$ & 
 $0.5909\ (31)$ 
 \\
\hline
\hline
\end{tabular}
\end{table}

\begin{table}[htb]
\caption{Meson masses in six-flavor theory
 on $L/a=24$ and $\beta = 2.0$ lattices.}
\label{tab:NF6.mesons.L24.b2.0}
\begin{tabular}{ccccc}
\hline
\hline
$\kappa$ &
 $a M_P$ & 
 $a M_V$ & 
 $a M_S$ & 
 $a M_A$ 
\\
\hline

$0.14100$ &
 $1.319\  (3)$ & 
 $1.346\  (3)$ & 
 $1.814\ (74)$ & 
 $1.789\ (51)$ 
 \\

$0.14300$ &
 $1.154\  (3)$ & 
 $1.184\  (3)$ & 
 $1.628\ (44)$ & 
 $1.675\ (39)$ 
 \\

$0.14400_{\rm P}$ &
 $1.050\  (3)$ & 
 $1.079\  (4)$ & 
 $1.476\ (28)$ & 
 $1.489\ (37)$ 
 \\

$0.14500$ &
 $0.9400\ (34)$ & 
 $0.9702\ (44)$ & 
 $1.273\ (21)$ & 
 $1.228\ (65)$ 
 \\

$0.14600$ &
 $0.8105\ (44)$ & 
 $0.8401\ (58)$ & 
 $1.005\ (54)$ & 
 $1.067\ (59)$ 
 \\

$0.14700$ &
 $0.6554\ (19)$ & 
 $0.6847\ (26)$ & 
 $0.926\ (18)$ & 
 $0.939\ (12)$ 
 \\

$0.14800$ &
 $0.4801\ (16)$ & 
 $0.5057\ (19)$ & 
 $0.646\ (20)$ & 
 $0.659\ (17)$ 
 \\

$0.14850$ &
 $0.4035\  (21)$ & 
 $0.4334\  (29)$ & 
 $0.5857\  (77)$ & 
 $0.5907\  (54)$ 
 \\

$0.14900$ &
 $0.3472\ (52)$  & 
 $0.3660\ (58)$  & 
 $0.376\ (11)$ & 
 $0.423\ (13)$ 
 \\

$0.14930$ &
 $0.3592\ (49)$ & 
 $0.3799\ (73)$ & 
 $0.3756\ (46)$ & 
 $0.4003\ (91)$ 
 \\

$0.14950$ &
 $0.3631\ (28)$ & 
 $0.3973\ (28)$ & 
 $0.3754\ (35)$ & 
 $0.4088\ (28)$ 
 \\

$0.14980$ &
 $0.3657\ (30)$ & 
 $0.4064\ (38)$ & 
 $0.3711\ (40)$ & 
 $0.4129\ (53)$ 
 \\

$0.15000$ &
 $0.3588\ (37)$ & 
 $0.3960\ (28)$ & 
 $0.3612\ (37)$ & 
 $0.3990\ (30)$ 
 \\
\hline
\hline
\end{tabular}
\end{table}

\begin{table}[htb]
\caption{Meson masses in six-flavor theory
 on $L/a=32$ and $\beta = 2.0$ lattices.}
\label{tab:NF6.mesons.L32.b2.0}
\begin{tabular}{ccccc}
\hline
\hline
$\kappa$ &
 $a M_P$ & 
 $a M_V$ & 
 $a M_S$ & 
 $a M_A$ 
\\
\hline

$0.14850$ &
 $0.3808\ (10)$ & 
 $0.4044\ (11)$ & 
 $0.4977\ (56)$ & 
 $0.5333\ (76)$ 
 \\

$0.14920$ &
 $0.2755\ (23)$ & 
 $0.3019\ (21)$ & 
 $0.3542\ (47)$ & 
 $0.3698\ (50)$ 
 \\

$0.14965$ &
 $0.2510\ (26)$ & 
 $0.2783\ (30)$ & 
 $0.2658\ (26)$ & 
 $0.2933\ (29)$ 
 \\

$0.14980$ &
 $0.2719\ (17)$ & 
 $0.2990\ (24)$ & 
 $0.2759\ (17)$ & 
 $0.3040\ (28)$ 
 \\
\hline
\hline
\end{tabular}
\end{table}

 The lattices used 
in this work are all homogeneous in the spatial directions.
 The number of the sites along each direction,  $l = L/a$, 
is either $l = 16$, $24$ or $32$ in this work.
 As explained above, 
the inverse of the gauge coupling constant $\beta$ is set to $2.0$
for all lattices.
 The hopping parameters are summarized 
in Tab.~\ref{tab:NF6.lattparams.L16.b2.0} for $l=16$, 
Tab.~\ref{tab:NF6.lattparams.L24.b2.0} for $l=24$, 
and 
Tab.~\ref{tab:NF6.lattparams.L32.b2.0} for $l=32$, respectively.
 The gauge configurations 
are stored once every $10$ trajectories for $L/a=16$ and $24$, 
and the measurements 
are performed for  $100 \sim 300$ configurations 
depending on parameters.
 For $L/a=32$, 
the configurations are stored once
every $5$ trajectories for $L/a=32$ 
owing to the upper limit on the time available to a single job.
 The measurements are carried out for more than $400$ configurations.

\subsection{meson masses}
\label{subsec:result:mesonMasses}

\begin{figure}[thb]
\begin{center}
\includegraphics[width=14.0cm,clip]{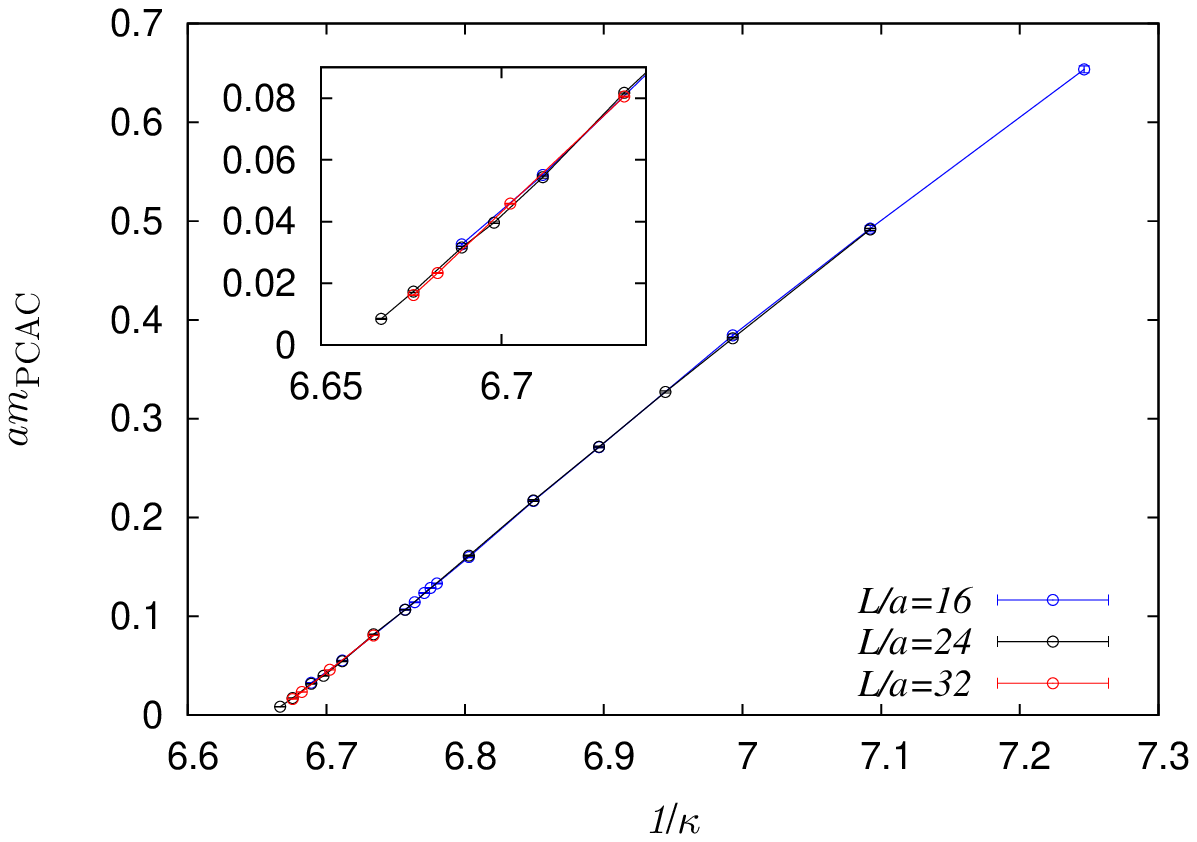}
\caption{
 $a m_{\rm PCAC}$ versus the inverse of the hopping parameter 
in six-flavor theory.}
\label{fig:kinvVSmPCAC.b2.0}
\end{center} 
\end{figure}

 The lattice action employed in this work
depends on three parameters, 
the number $l = L/a$ of sites along spatial direction, 
the inverse of the bare gauge coupling constant $\beta = 4/ g_0^2$,
and the hopping parameter $\kappa$. 
 For each set of lattice parameters,
the (bare) PCAC quark mass in unit of the lattice spacing $a$, 
$a m_{\rm PCAC}$, 
is obtained as described in Sec.~\ref{subsec:PCACmass_decayConstat}.
 Figure~\ref{fig:kinvVSmPCAC.b2.0} summarizes
$a m_{\rm PCAC}$ as a function of $1 /\kappa$ 
at each lattice size.
 It shows that $a m_{\rm PCAC}$ does not sustain 
visible finite size effect and is thus a short-distance quantity.
 Moreover, it depends on $1 / \kappa$ monotonically.
 It thus enables to convert 
the dependence of various observables on $\kappa$
to that on $a m_{\rm PCAC}$, 
which is often referred to
as the ``quark mass dependence'' somewhat loosely.

\begin{figure}[thb]
\begin{center}
\includegraphics[width=14.0cm,clip]{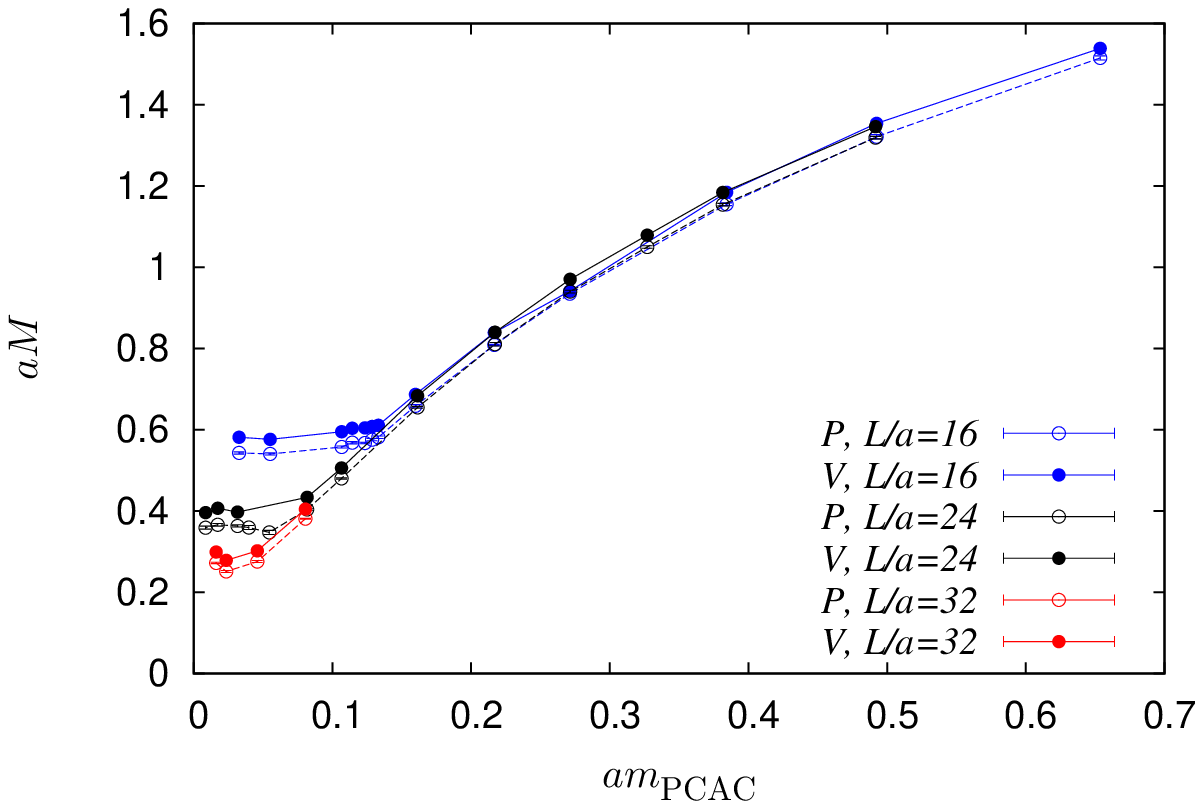}
\caption{
 The lightest pseudoscalar meson mass $a M_P$ (open dot)
and the lightest vector meson mass $a M_V$ (filled dot)
versus $a m_{\rm PCAC}$ in six-flavor theory.
 The individual lines are drawn just to guide eyes.}
\label{fig:pcacVSmP_mV.b2.0}
\end{center} 
\end{figure}

 Figure \ref{fig:pcacVSmP_mV.b2.0} 
summarizes the masses of the lightest mesons 
in the (flavor-nonsinglet) pseudoscalar and vector channels.
 The numerical values of meson masses are tabulated in
Tabs.~\ref{tab:NF6.mesons.L16.b2.0}, 
\ref{tab:NF6.mesons.L24.b2.0} and \ref{tab:NF6.mesons.L32.b2.0}.
 For a fixed lattice size,
the meson masses are seen to be bounded from below, 
and cannot approach to zero even for $m_{\rm PCAC} \rightarrow 0$.
 The bounds on meson masses depend on the sizes of lattices and 
are smaller for larger size of lattice.
 Moreover, the meson mass at $l_1$ 
is compatible with that at $l_2 > l_1$ for the quark mass
above some $a m_{\rm PCAC}$, 
but below it the change of the former slows down and branches off from 
those of larger sizes of lattices.
 This phenomenon is hence considered to be caused by the finite size effect.
 That is, the finite size effect acts to increase the meson masses, 
and it eventually prevents them to decrease further
for quark mass smaller than some value.

 The finite size effect observed above in the meson masses 
in our target system is qualitatively different from that
in ${\rm SU}(2)_{\rm C}$ gauge theory with two adjoint Dirac fermions
\cite{DelDebbio:2010hu,Bursa:2011ru}, 
where the pseudoscalar meson mass is smaller for smaller size of lattice.
 Tables \ref{tab:NF6.mesons.L16.b2.0}, 
\ref{tab:NF6.mesons.L24.b2.0} and \ref{tab:NF6.mesons.L32.b2.0}
also show that the pseudoscalar meson becomes degenerate in mass
with the scalar meson once the decrease of the meson masses
almost stops.
 The finite size effect
on the meson masses in our target system
resembles that obtained in the results of the simulation for $N_F = 2$ system
at very weakly coupling ($\beta = 4.0$) 
and finite volume in Sec.~\ref{sec:two-flavor}.

\begin{figure}[thb]
\begin{center}
\includegraphics[width=14.0cm,clip]{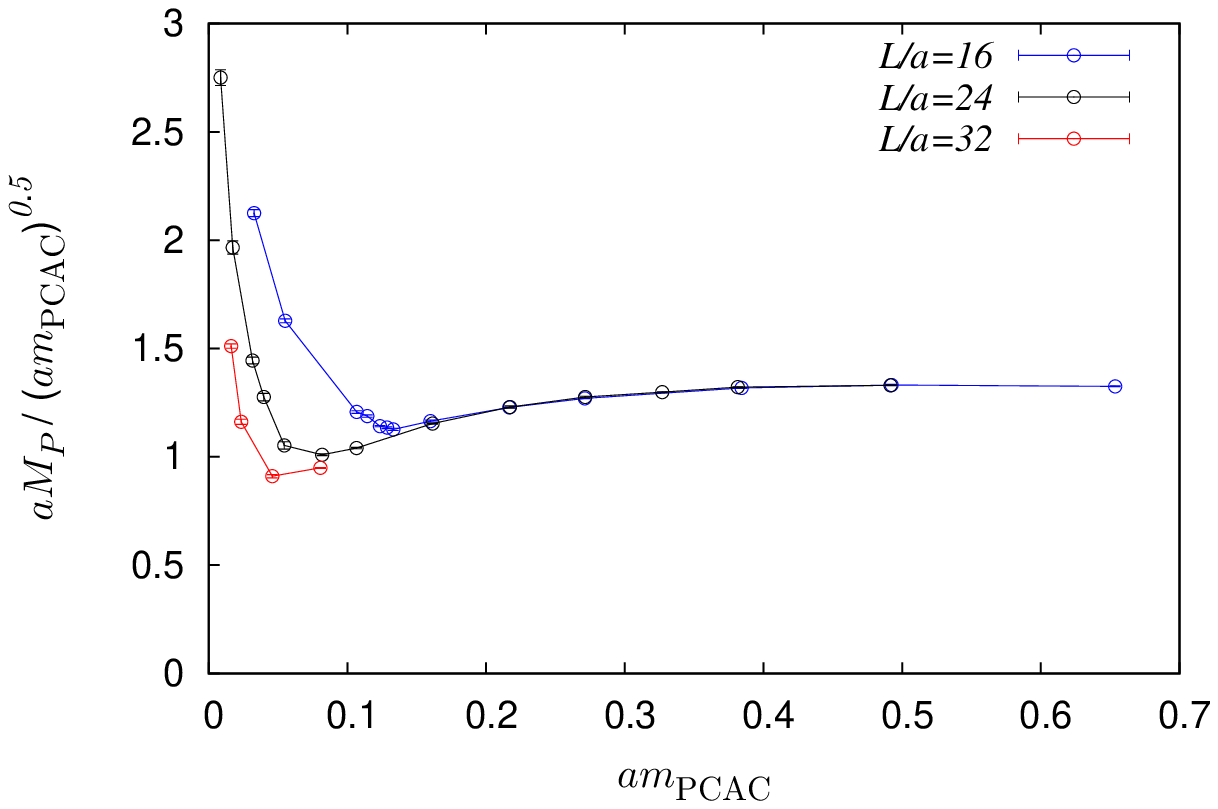}
\caption{
 The quark mass dependence of 
$X_{1/2} = a M_P /(a m_{\rm PCAC})^{1/2}$ in six-flavor theory.
 $X_{1/2}$ decreases below $a m_{\rm PCAC} \lesssim 0.3$ 
until the finite size effect shows up.
}
\label{fig:pcacVSmP_mPCAC0.5.b2.0}
\end{center} 
\end{figure}

\begin{figure}[thb]
\begin{center}
\includegraphics[width=14.0cm,clip]{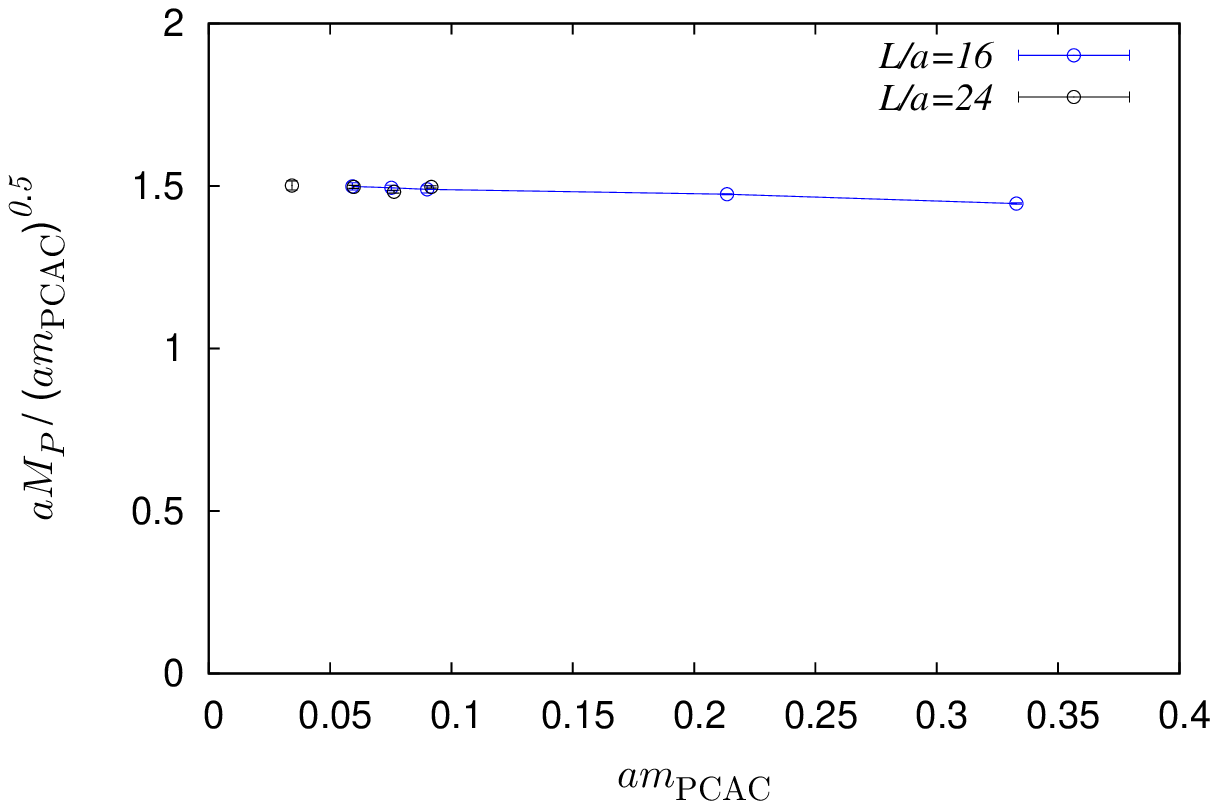}
\caption{
 The quark mass dependence of 
$a M_P /(a m_{\rm PCAC})^{1/2}$ in two-flavor theory. 
}
\label{fig:pcacVSmP_mPCAC0.5.b2.0.NF2}
\end{center} 
\end{figure}

 Another noteworthy feature found in Fig.~\ref{fig:pcacVSmP_mV.b2.0}
is the approximate degeneracy of
the vector meson with the pseudoscalar meson
in mass.
 This fact prevents us to try to fit the data using chiral perturbation theory.
 The approximate degeneracy in mass between pseudoscalar meson
and vector meson at first sight reminds us of the heavy quark limit
in which they form a multiplet.
 If the quark is heavy enough, 
the meson mass should vary almost linearly.
 However, Fig.~\ref{fig:pcacVSmP_mV.b2.0} shows that 
this possibility is unlikely.
 To examine this point more closely, the ratio 
\begin{equation}
 X_{1/2} \equiv \frac{a M_P}{(a m_{\rm PCAC})^{1/2}} \,,
\end{equation}
is plotted in Fig.~\ref{fig:pcacVSmP_mPCAC0.5.b2.0}.
 In what follows, the dependence of $a$ on the quark mass 
is assumed to be small enough that it does not to affect to the observation.
 From Fig.~\ref{fig:pcacVSmP_mPCAC0.5.b2.0},
$a M_P$ seems to vary with the power $0.5$
in the region $a m_{\rm PCAC} \gtrsim 0.35$.
 The ratio $X_{1/2}$ in that figure
starts to increase at some quark mass 
depending on the size of lattice, 
and then diverges for $m_{\rm PCAC} \rightarrow 0$,
since $M_P$ eventually stops to decrease at finite volume. 
 We consider that the results for such parameters offer no 
knowledge on the dynamics of the theory.

 Rather, the most important point to notice is as follows;  
if we look at the data toward smaller quark mass,
the part of $X_{1/2}$
that do not suffer finite volume correction so much
{\it decrease} in the region $a m_{\rm PCAC} \lesssim 0.3$.
 If the chiral symmetry breaking is realized in the system, 
$M_P$ should be better approximated
by $(m_{\rm PCAC})^{1/2}$ for lighter quarks.
 As a reference, we show $X_{1/2}$ simulated for the two-flavor theory
at $\beta = 2.0$ in Fig.~\ref{fig:pcacVSmP_mPCAC0.5.b2.0.NF2} .
 In contrast to it, in the six-flavor theory, 
$a M_P \propto (a m_{\rm PCAC})^{1/2}$
for $0.35 < a m_{\rm PCAC} < 0.65$, but 
it changes with different exponent, 
$a M_P \propto (a m_{\rm PCAC})^\alpha$ with $\alpha > 0.5$, 
at smaller quark mass.
 This is the first evidence supporting 
the absence of chiral symmetry breaking in the six-flavor theory.

%

\begin{figure}[thb]
\begin{tabular}{cc}
\begin{minipage}{0.47\hsize}
\begin{center}
\includegraphics[width=\hsize,clip]{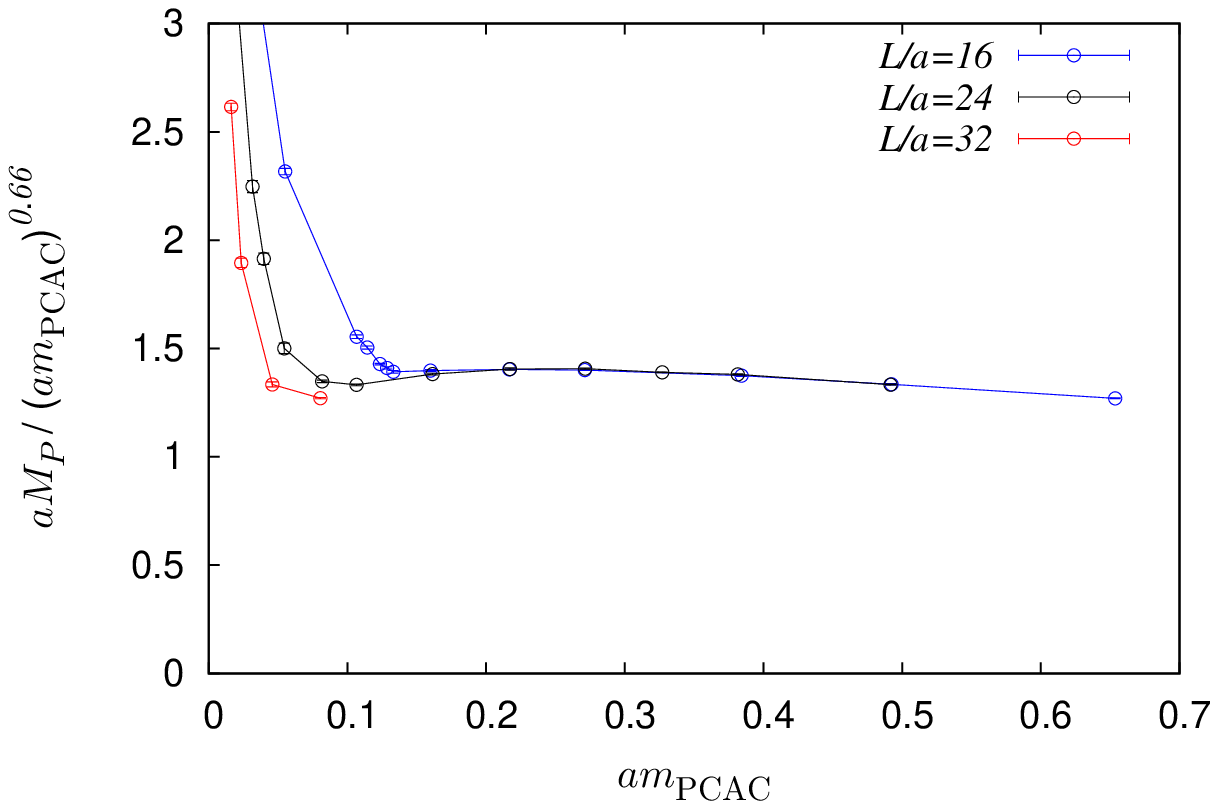}
\caption{
 The quark mass dependence of 
$a M_P /(a m_{\rm PCAC})^{\alpha}$ with $\alpha = 0.66$
in six-flavor theory.}
\label{fig:pcacVSmP_mPCAC0.66.b2.0}
\end{center} 
\end{minipage}
\quad 
\begin{minipage}{0.47\hsize}
\begin{center}
\includegraphics[width=\hsize,clip]{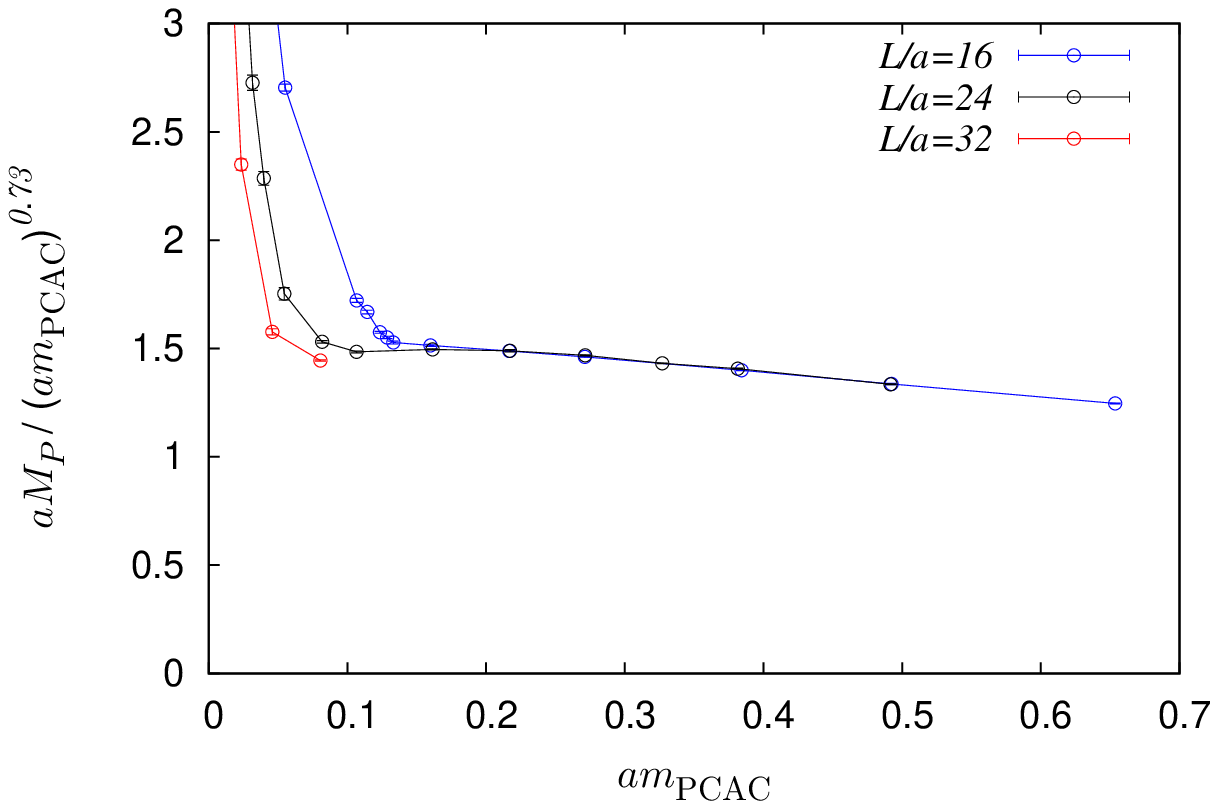}
\caption{
 The quark mass dependence of 
$a M_P /(a m_{\rm PCAC})^{\alpha}$ with $\alpha = 0.73$
in six-flavor theory.}
\label{fig:pcacVSmP_mPCAC0.73.b2.0}
\end{center} 
\end{minipage}
\end{tabular}
\end{figure}

 Figure~\ref{fig:pcacVSmP_mPCAC0.5.b2.0} shows that
available data are all in the transition region 
over which the exponent $\alpha_{M_P}$ changes 
from $0.5$ to that in the basin of the IR-fixed point.
 Moreover,
the scaling $M_P \propto (m_{\rm PCAC})^{\alpha_{M_P},\,\star}$ 
with $\alpha_{M_P,\,\star} = 1 /(1 + \gamma_\star)$
predicted according to the hyperscaling hypothesis 
\cite{DelDebbio:2010hu,DelDebbio:2010ze}
implies that $1 > \alpha_{M_P,\,\star} > 0.5$ for $0 < \gamma_\star < 1$. 
 Unless the data at very small quark mass $\lesssim 0.01$ are available,
the fit to determine the term $c_{M_P} (a m_{\rm PCAC})^{\alpha_{M_P}}$ 
just looks at the height of the data, 
but poorly captures the bent of the curve, 
implying no hope for reliable estimate of $\alpha_{M_P}$.
 Here, we are just content with observing
$X_{\alpha} \equiv a M_P /(a m_{\rm PCAC})^{\alpha}$
for $\alpha = 0.66$
($\gamma \equiv 1/\alpha - 1 \simeq 0.52$) 
and $\alpha = 0.73$ ($\gamma \simeq 0.3$)
in Figs.~\ref{fig:pcacVSmP_mPCAC0.66.b2.0} 
and \ref{fig:pcacVSmP_mPCAC0.73.b2.0}, respectively.
 Those figures indicate that 
the exponent $\alpha_{M_P}$ suggested in the present spectroscopy study  
at $\beta = 2.0$ is compatible with 
that found in the study of the running coupling constant
defined in the Schr\"{o}dinger functional scheme
\begin{equation}
 0.26 \le \gamma_{\star,\,{\rm SF}} \le 0.74\,.
 \label{eq:gammaStar_SF}
\end{equation}  

 As discussed in Sec.~\ref{subsec:FV}, 
the spectrum and the dynamics at low energy, 
which could also be affected by the representations of fermions,
determine the qualitative behavior of the finite size effect.
 It is thus necessary to scrutinize the dependence of other observables, 
such as $f_P$ and the subtracted chiral condensate,
on quark masses and lattice sizes.

\subsection{decay constant}

\begin{figure}[thb]
\begin{center}
\includegraphics[width=14.0cm,clip]{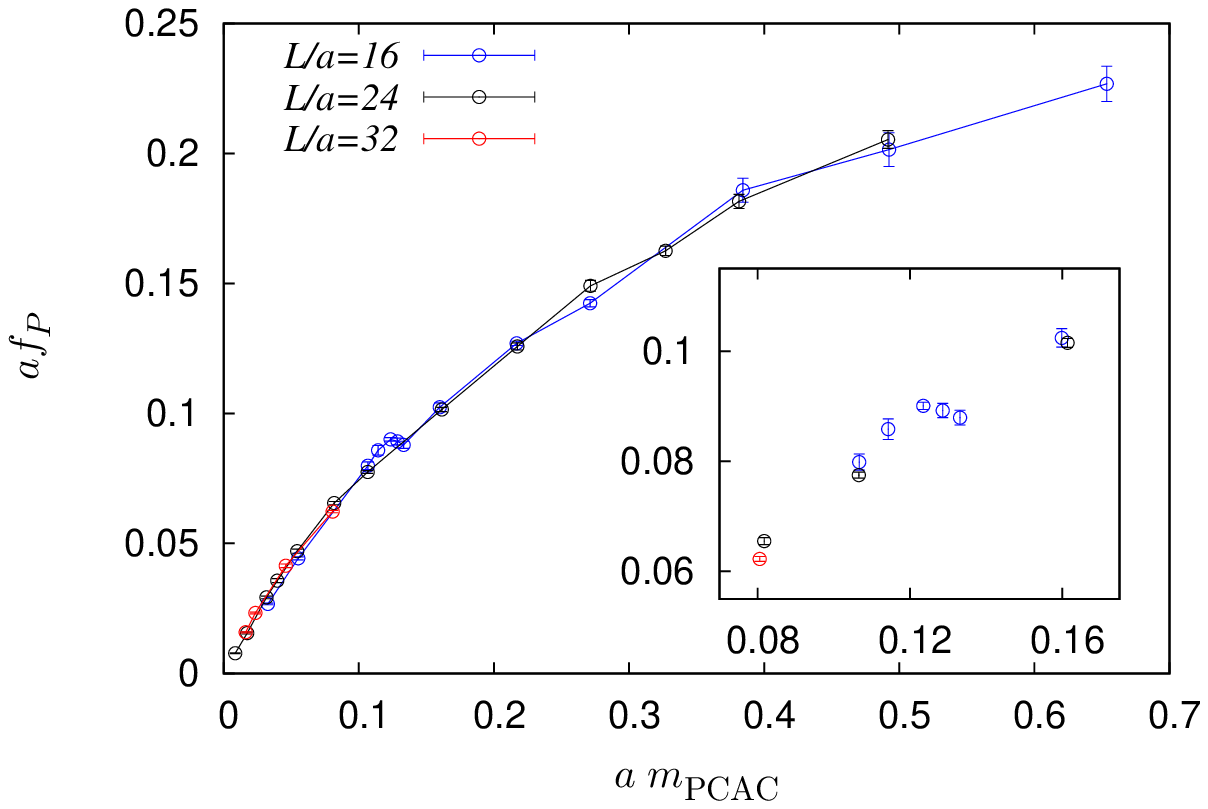}
\caption{
 The quark mass dependence of $a f_P$ in six-flavor theory.
}
\label{fig:pcacVSfP.b2.0}
\end{center} 
\end{figure}

 Figure \ref{fig:pcacVSfP.b2.0} shows 
the quark mass dependence of the decay constant $f_P$ 
of the lightest pseudoscalar meson $P$ in unit of the lattice spacing $a$
obtained through Eq.~(\ref{eq:fP}).
 There, $f_P$ appears to vanish in the chiral limit,
$m_{\rm PCAC} \rightarrow 0$.
 However, this seems to originate from 
the bounded behavior of the pseudoscalar meson mass $M_P$ 
in the chiral limit at finite volume:
Eq.~(\ref{eq:fP}) implies that $f_P$ will vanish
unless $M_P$ approaches to zero for $m_{\rm PCAC} \rightarrow 0$.
 The analysis of the data shows
that the factor $A_{PP}$ in Eq.~(\ref{eq:asymTwoPoint}) 
related to the wave function of $P$
is slightly more sensitive to the finite size effect than $M_P$.
 Once $M_P$ almost ceases to decrease,
$a f_P$ starts to vanish almost linearly in $a m_{\rm PCAC}$.
 Actually, we saw in Sec.~\ref{sec:two-flavor} that 
the same phenomenon happens 
even in the two-flavor theory at very weakly coupling
($\beta = 4.0$).
 It is caused by the finite size effect.
 These imply that we cannot find anything about 
dynamics in the quark mass region where meson masses cease to decrease at
each lattice size
($L \ll 1/M_P$ in our case).

 We rather focus on the behavior of $f_P$ 
around the quark mass where the finite size effect begins to 
show up but the meson masses do not cease to decrease.
 For the lattice size $L/a=16$,
Fig.~\ref{fig:pcacVSmP_mV.b2.0} implies that
$a M_P$ begins to suffer visible finite size effect
around $a m_{\rm PCAC} = 0.12 \sim 0.14$. 
 The simulation has thus been done for many parameters in this region.
 Fig.~\ref{fig:pcacVSfP.b2.0} shows that 
the finite size effect acts to increase $a f_P$ in this region.
 The difference between $L/a=24$ and $L/a=32$ 
found at $a m_{\rm PCAC} \simeq 0.08$ is also compatible with 
such a tendency. 
 This behavior is opposite to that
in Fig.~\ref{fig:fP_finiteVolume_chibreak} expected for 
the finite size effect in the theory with chiral symmetry breaking.
 The behavior of $L/a=16$ data suggests 
the example $4$ in Fig.~\ref{fig:fP_finiteVolume_IRconformal_4} 
for the finite size effect on $a f_P$.
 This difference in the features of the finite size effect
implies that the six-flavor is unlikely a ${\chi\hspace{-5pt}/}$-theory.


\begin{figure}[thb]
\begin{center}
\includegraphics[width=14.0cm,clip]{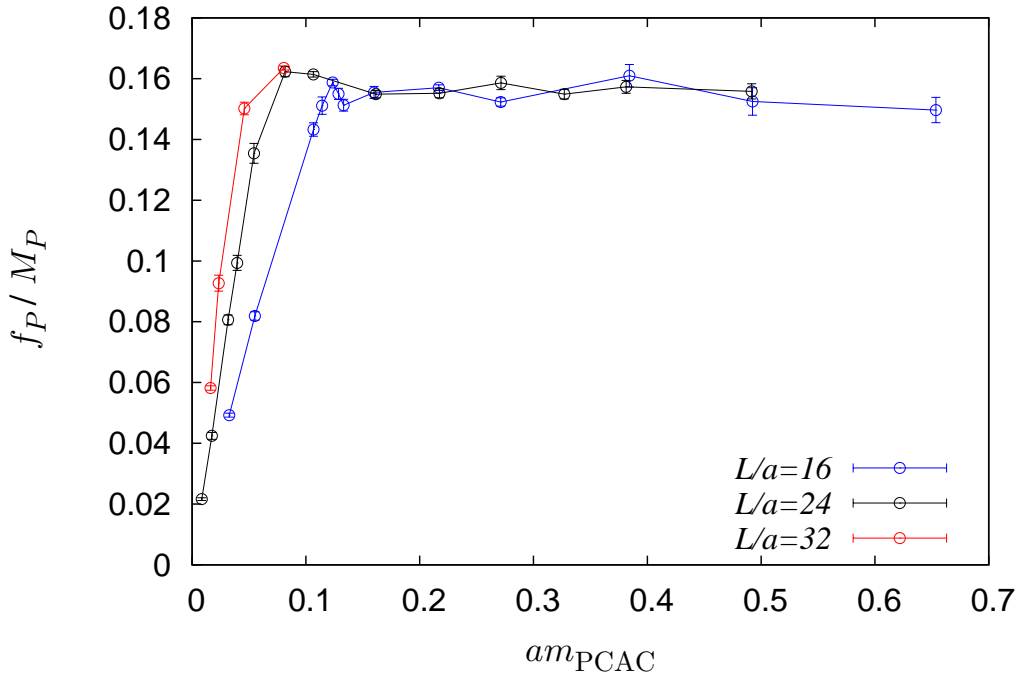}
\caption{
 The quark mass dependence of 
the ratio $f_P /M_P$ in six-flavor theory.}
\label{fig:pcacVSfpmp.b2.0}
\end{center} 
\end{figure}

 We next focus on the ratio $f_P / M_P$.
 In the ${\chi\hspace{-5pt}/}$-theory,
this ratio should blow up in the chiral limit.
 In the IR-conformal theory, 
if the hyperscaling hypothesis is valid, 
$f_P$ depends on the quark mass with the same power 
as $M_P$ \cite{DelDebbio:2010hu,DelDebbio:2010ze}
so that the dependence of their ratio $f_P /M_P$ on the quark mass 
will fade away in the scaling region. 
 Figure \ref{fig:pcacVSfpmp.b2.0}
shows the result in the six-flavor theory.
 The behavior of $f_P$ and $M_P$ implies that
$f_P /M_P$ starts to decrease once $M_P$ almost stops to decrease
at finite volume.
 We recall that, 
even in two-flavor theory, 
the ratio $f_P /M_P$ seems to start to blow up 
at far smaller quark mass (Figure \ref{fig:NF2.fPmP.b2.0}).
 The data for $0.13 \lesssim a m_{\rm PCAC} \lesssim 0.25 $ 
are almost flat,
but the data with smaller $a m_{\rm PCAC}$ leave
the possibility for the ratio to increase in the chiral limit.
 However, the behavior of the data $L/a=16$
around $a m_{\rm PCAC} \sim 0.13$ suggests the possibility that 
$f_P /M_P$ jumps once before starting to drop steeply.
 To settle this issue, simulation with larger lattices, 
say, of size $L/a=48$, is required.

\subsection{chiral condensate}

\begin{figure}[thb]
\begin{center}
\includegraphics[width=14.0cm,clip]{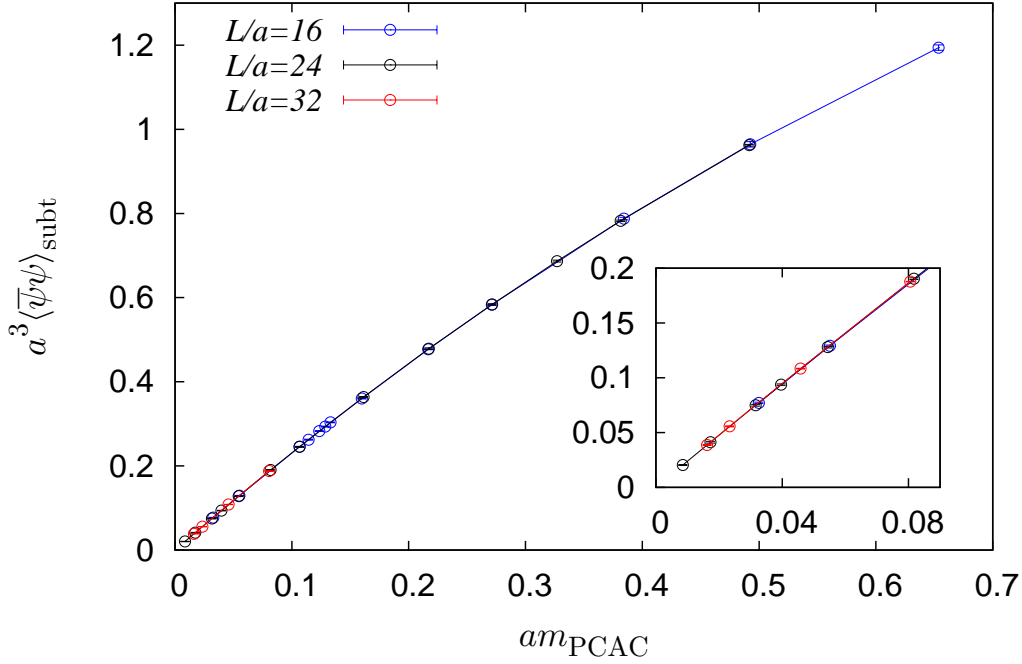}
\caption{
 The subtracted chiral condensate versus 
$a m_{\rm PCAC}$ in six-flavor theory.
 The lines are drawn for the individual sizes of lattices 
just to guide the eyes.}
\label{fig:pcacVSconds.b2.0}
\end{center} 
\end{figure}

 We next focus on the chiral condensate.
 We recall the result of the hyperscaling analysis 
applied to the chiral condensate \cite{DelDebbio:2010ze}.
 In the attractive basin of the IR fixed point, 
the non-analytic and universal term with
the exponent
$\alpha_{\overline{q} q} = \frac{3 - \gamma_\star}{1 + \gamma_\star}$
for the mass anomalous dimension $\gamma_\star$
is expected to emerge.
 The analysis thus far indicates that 
$\gamma_\star < 1$ provided that
the six-flavor theory turns out to be IR-conformal.
 Then, $\alpha_{\overline{q} q} > 1$.
 Therefore, the analytic term linear in the quark mass dominates
the chiral condensate at small quark mass. 

 Figure \ref{fig:pcacVSconds.b2.0} shows 
the result for the subtracted chiral condensate 
$\left<\overline{\psi} \psi\right>_{\rm subt}$ 
in the six-flavor theory.
 The lines in that figure are drawn to guide the eyes
to the plots of the individual sizes of lattices.
 The three lines overlap with each other.
 We stress the non-triviality of this fact, 
and then make an important observation 
on the properties of the data of the subtracted chiral condensate, 
which is crucial for our succeeding analysis.
 The subtracted chiral condensate
$\left<\overline{\psi} \psi\right>_{\rm subt}\left(m_{\rm PCAC},\,L/a\right)$
defined in Eq.~(\ref{eq:def:scond}) 
will diminish more rapidly once the finite size effect almost 
stops decrease of the pseudoscalar meson mass $M_P$.
 With this in our mind, we examine the quark mass and volume dependence 
of meson masses in Fig.~\ref{fig:pcacVSmP_mV.b2.0} more closely.
 On one hand, at $L/a=16$, meson masses start to suffer 
the finite size effect around $a m_{\rm PCAC} \sim 0.14$, 
and has already ceased to decrease around $a m_{\rm PCAC} \sim 0.10$.
 On the other hand, meson masses at $L/a=32$ and $a m_{\rm PCAC} \sim 0.08$ 
do not seem to suffer visible finite size effect.
 Figure~\ref{fig:pcacVSconds.b2.0} shows that
no visible finite size effect is observed 
for $\left<\overline{\psi} \psi\right>_{\rm subt}$
even for the quark mass in this region. 
 This indicates that 
{\it the finite size effect appears to be negligibly small
in $\left<\overline{\psi} \psi\right>_{\rm subt}$
over the entire region of available data}
\footnote{
 We will examine 
the effect of the data with very small quark masses 
on the result deliberately.
}. 
 Based on this observation, we perform various fits 
and ask whether the chiral limit
is consistent with $0$ within available precision.

 We consider the following three types of target data to be fit;
\begin{description}
\item[(S1)]
 data with $a m_{\rm PCAC} \le 0.2$\,,
\item[(S2)]
 data with $a m_{\rm PCAC} \le 0.1$\,,
\item[(S3)]
 data for which $a m_{\rm PCAC} \le 0.2$
 and $f_P /M_P$ does not belong to the sharply falling region
 in Fig.~\ref{fig:pcacVSfpmp.b2.0}\,.
\end{description}
 No restrictions on lattice sizes are imposed.
 As the fit functions, the following three will be examined;
\begin{description}
\item[(F1)]
 linear function $f_2(x = a m_{\rm PCAC})$ in Eq.~(\ref{eq:linearFunc})\,,
\item[(F2)]
 quadratic function $f_3(x = a m_{\rm PCAC})$ in Eq.~(\ref{eq:quadraticFunc})\,,
\item[(F3)]
 function containing a term with indefinite exponent
\begin{equation}
 f_4(x = a m_{\rm PCAC}) = 
 c_0 + c_1 x + c_2 x^{\alpha}\,.
 \label{eq:varexpFunc}
\end{equation}
\end{description}

\begin{table}[thb]
\caption{Result for the linear fit to the subtracted chiral condensate, 
where $a_0$ is the value in chiral limit.
 See the text for the definition of three data data, 
${\bf S1}$, ${\bf S2}$ and ${\bf S3}$.} 
\label{tab:NF6.conds.lin.b2.0}
\begin{tabular}{ccc}
\hline
\hline
 data set & $a_0$ & $a_1$ \\ 
\hline
 {\bf S1} & $0.00406\ (78)$ & $2.2499\ (86)$ \\
 {\bf S2} & $0.00137\ (27)$ & $2.3191\ (58)$ \\
 {\bf S3} & $0.0131\ (11)$  & $2.1766\ (88)$ \\
\hline
\hline
\end{tabular}
\end{table}

\begin{table}[thb]
\caption{Result for the fit to the subtracted chiral condensate
by the quadratic fit,
where $b_0$ is the value in chiral limit.
${\bf S1}^\prime$ is obtained by discarding the data
with $a m_{\rm PACA} \le 0.03$ in {\bf S1}. } 
\label{tab:NF6.conds.quadratic.b2.0}
\begin{tabular}{cccc}
\hline
\hline
 data set & $b_0$ & $b_1$ & $b_2$ \\ 
\hline
 {\bf S1} & $-0.00003\ (12)$ & $2.4030\ (35)$ & $-0.927\ (21)$\\
 {\bf S2} & $ 0.00017\ (20)$ & $2.3897\ (97)$ & $-0.76\ (10)$\\
 {\bf S3} & $ 0.00142\ (76)$ & $2.382\ (13)$ & $-0.849\ (53)$ \\
\hline
 ${\bf S1}^\prime$ & $-0.00021\ (24)$ & $2.4068\ (59)$ & $-0.947\ (31)$\\
\hline
\hline
\end{tabular}
\end{table}

 As seen in Fig.~\ref{fig:pcacVSconds.b2.0},
the shape of the chiral condensate
versus quark mass is convex upward.
 Therefore, the chiral extrapolation by linear fit to
the data with negligible finite size effect will tend to overestimate
the value in the massless limit. 
 If it turns out to vanish within the available precision,
the result will support absence of chiral symmetry breaking.
 This motivates us to do the linear fit first, but 
Tab.~\ref{tab:NF6.conds.lin.b2.0} shows that the answer is not affirmative.

\begin{figure}[thb]
\begin{tabular}{cc}
\begin{minipage}{0.47\hsize}
\begin{center}
\includegraphics[width=\hsize,clip]
{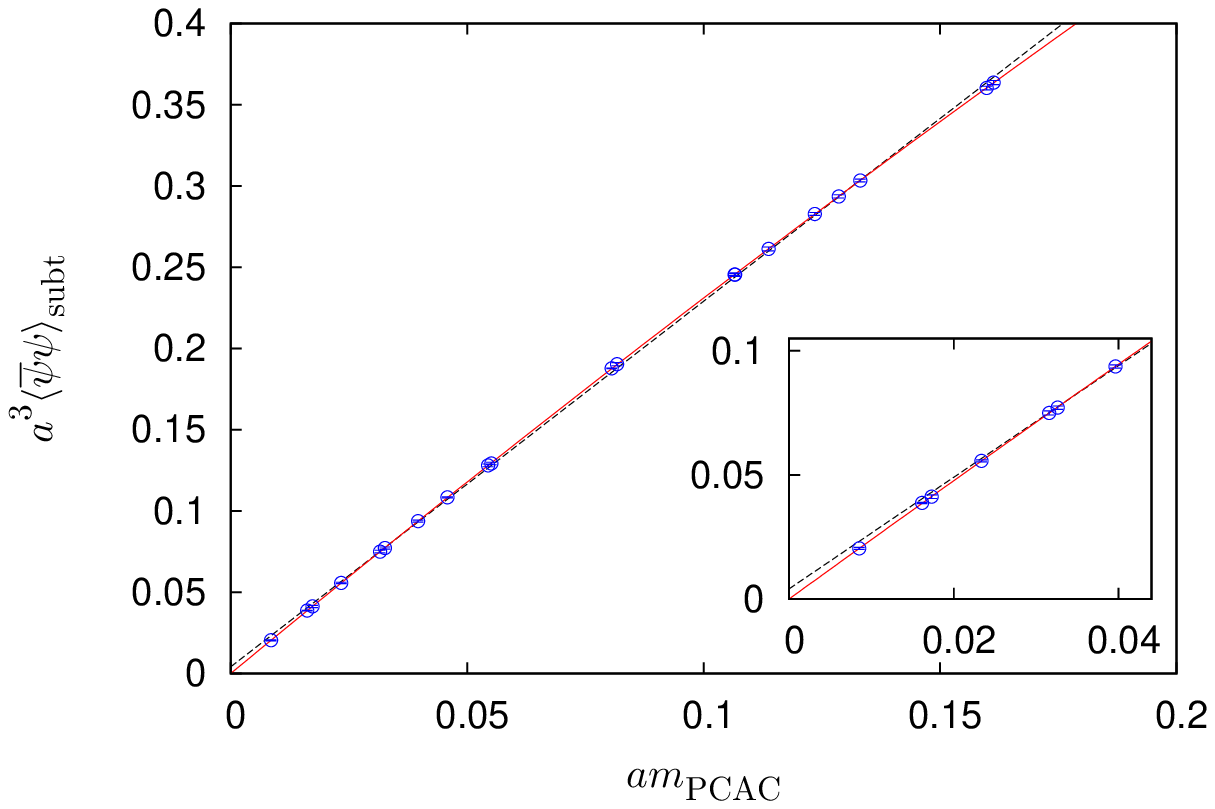}
\caption{
 Result for the fit to the data set {\bf S1}
by the linear function (dotted line) and quadratic function (red curve). 
 Only the fit curves with the central values for the coefficients 
are drawn.
}
\label{fig:fit.pcacVSconds.S1.b2.0}
\end{center}
\end{minipage}
\quad 
\begin{minipage}{0.47\hsize}
\begin{center}
\includegraphics[width=\hsize,clip]
{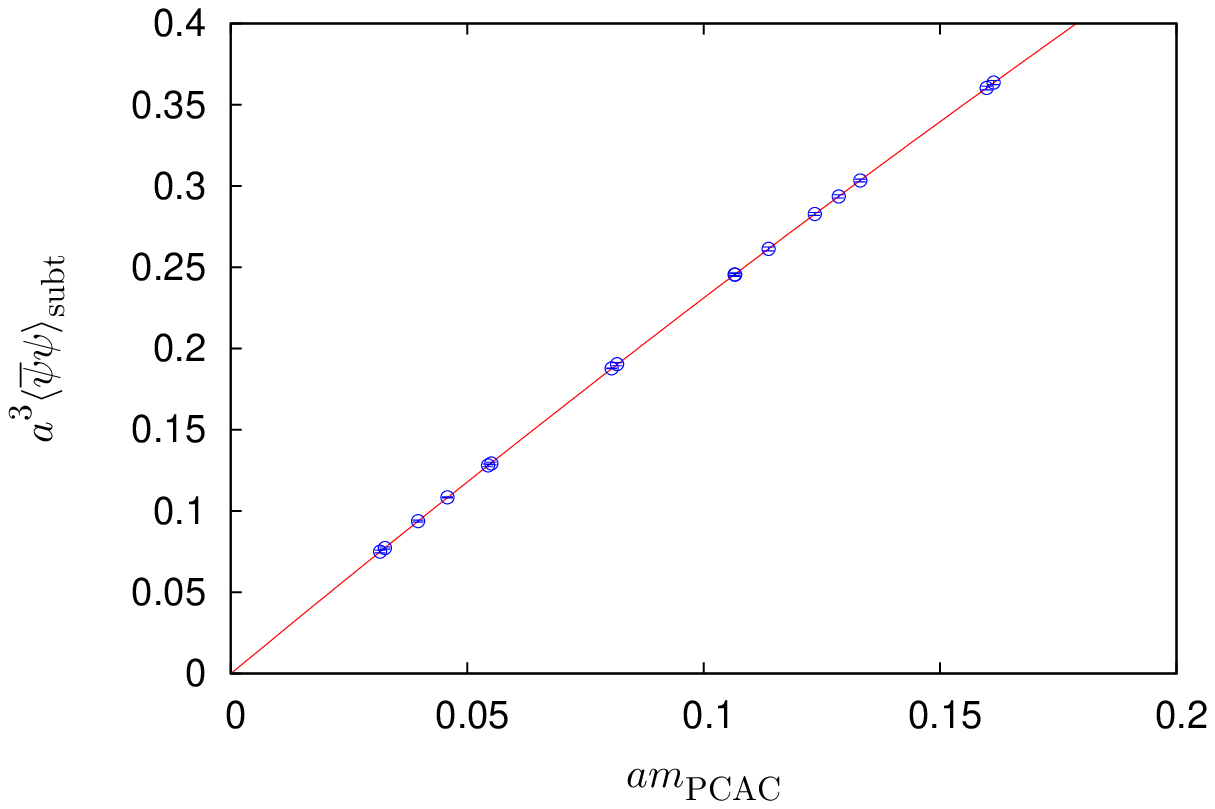}
\caption{
 Result for the quadratic fit to 
the data set ${\bf S1}^\prime$
obtained from the data set ${\bf S1}$
by discarding the data with $a m_{\rm PCAC} \le 0.03$.
}
\label{fig:fit.pcacVSconds.S1p.b2.0}
\vspace{1.47cm} 
\end{center} 
\end{minipage}
\end{tabular} 
\end{figure}

 Table \ref{tab:NF6.conds.quadratic.b2.0} shows
the result of the fit with the quadratic function (\ref{eq:quadraticFunc}).
 The quadratic function consists
of the leading and the next-to-leading (NLO) analytic 
corrections, but with
no non-analytic term such as chiral logarithm. 
 We regard it as being one of the trial functions
to examine 
{\it the consistency of the presence of the chiral symmetry breaking}.
 The intercept $b_0$ at $m_{\rm PCAC} = 0$ 
is non-vanishing for the fit to the data set ${\bf S3}$, 
but the smallest $a M_P$ in ${\bf S3}$ is about $0.38$ there.
 From the above observation about the smallness of the finite size effect
on the data of the subtracted chiral condensate 
$\left<\overline{\psi} \psi\right>_{\rm subt}$, 
there is no reason to disregard the results
for the data sets, ${\bf S1}$ and ${\bf S2}$.
  Figure~\ref{fig:fit.pcacVSconds.S1.b2.0} shows
how the fit curves pass through the data in ${\bf S1}$.
 One notices that the data with small quark masses
are slightly separated from the line determined by the linear fit. 
 In order to examine the significance of the data with 
relatively small quark mass
in the quadratic fit, we intentionally remove 
the data with $a m_{\rm PCAC} \le 0.03$ from ${\bf S1}$, 
and carry out the quadratic fit.
 As shown in Figure~\ref{fig:fit.pcacVSconds.S1p.b2.0},
the result does not exhibit essential change.
(The coefficients are listed 
as ${\bf S1}^\prime$ in Tab.~\ref{tab:NF6.conds.quadratic.b2.0}.)
 This indicates that the difference in the chiral limit 
between the two functions is not attributed to 
those data with relatively small quark mass.
 Rather, Tab.~\ref{tab:NF6.conds.quadratic.b2.0} shows
the difference in the coefficient of
the NLO term, $b_2$, between ${\bf S1}$ 
and ${\bf S2}$, 
indicating that $b_2$ is mostly determined by the subset of data
with larger quark mass in each set, 
which is thought to suffer far smaller finite size effect.
 With $b_2$ determined as such, 
$\left<\overline{\psi} \psi\right>_{\rm subt}$
in the limit $m_{\rm PCAC} \rightarrow 0$
is found to be consistent with zero
both for ${\bf S1}$ and ${\bf S2}$
within the precision of data available at present.

\begin{table}[thb]
\caption{Result for fit to the subtracted chiral condensate by the function 
(\ref{eq:varexpFunc}) with the exponent $\alpha$ to be determined.
$c_0$ is the value in chiral limit.} 
\label{tab:NF6.conds.power.b2.0}
\begin{tabular}{ccccc}
\hline
\hline
 data set & $c_0$ & $c_1$ & $c_2$ & $\alpha$ \\ 
\hline
 {\bf S1} & $0.00002\ (22)$ & $2.398\ (15)$ & $-0.96\ (12)$ & $2.04\ (12)$\\
 {\bf S2} & $-0.00023\ (79)$ & $2.48\ (31)$ & $-0.401\ (55)$ & $1.39\ (80)$\\
 {\bf S3} & $0.001\  (11)$   & $2.38\ (38)$ & $-0.845\ (93)$ & $2.0\ (1.8)$\\
\hline
\hline
\end{tabular}
\end{table}

 Table \ref{tab:NF6.conds.power.b2.0} is the result for 
the fit with the indefinite exponent (\ref{eq:varexpFunc}). 
 For the data set {\bf S1}, the exponent is found
to be nearly equal to $2$ 
and the values of the other parameters coincide with
those in Tab.~\ref{tab:NF6.conds.quadratic.b2.0} 
obtained by the quadratic fit.
 The fit with the function (\ref{eq:varexpFunc})
contains four parameters to be determined, 
and the values of the parameters are contaminated by the uncertainty larger
than that in the quadratic fit.
 Nevertheless, the intercept $c_0$ is consistent with zero.
 This motivates us
to carry out the fit with the function (\ref{eq:varexpFunc}) 
but now with $c_0$ constrained to $0$, i.e. assuming IR-conformality, 
and try to examine if the resulting exponent is consistent with 
that (\ref{eq:gammaStar_SF}) found in Schr\"{o}dinger functional scheme.

 As remarked above, 
the chiral condensate has a non-analytic and universal piece 
$(m_q)^{\alpha_{\overline{q} q}}$
with the exponent 
$\alpha_{\overline{q} q} = \frac{3 - \gamma_\star}{1 + \gamma_\star}$
in the mass-deformed IR-conformal theory with hyperscaling hypothesis, 
and it is sub-dominant
as long as the inequality $1 < \alpha_{\overline{q} q} < 2$ is satisfied
(Recall the presence of the next-leading-order analytic term
$\propto (m_{\rm PCAC})^2$.).
 As pointed out in Sec.~\ref{subsec:result:mesonMasses},
the exponent $\alpha_{M_P,\,\star} = 1 /(1 + \gamma_\star)$ 
appearing in the scaling of the pseudoscalar meson mass $M_P$ is smaller
than $1$ if $0 < \gamma_\star < 1$, 
and the determination of $\alpha_{M_P}$ requires 
the data at very small quark masses.
 In contrast, it may be possible
to give very rough estimation on $\alpha_{\overline{q} q} (> 1)$ 
by performing the fit with the function (\ref{eq:varexpFunc}) 
the set including the data at the quark mass
that is modest but belongs to the basin of IRFP.
 Meanwhile, we observed in Fig.~\ref{fig:pcacVSmP_mPCAC0.5.b2.0}
that the exponent may be still
in the process of changing towards the value at the infrared fixed point.
 Thus, we perform the fit with the function 
(\ref{eq:varexpFunc}) under the constraint $c_0 = 0$
only for the set ${\bf S2}$ consisting of data with $a m_{\rm PCAC} < 0.1$,
and get
\begin{equation}
 c_1 = 2.420\ (31)\,,\quad c_2 = -0.46\ (20)\,,\quad
 \alpha = 1.65\ (30)\,.
\end{equation}
 The above value of the exponent $\alpha$ is not inconsistent with 
$1.3 
\le \alpha_{\overline{q} q,\,{\rm SF}} 
\le 2.2$ corresponding to 
$\gamma_\star$ in Eq.~(\ref{eq:gammaStar_SF}), 
which was obtained by the calculation
in the Schr\"{o}dinger functional scheme.


\begin{figure}[thb]
\begin{center}
\includegraphics[width=14.0cm,clip]{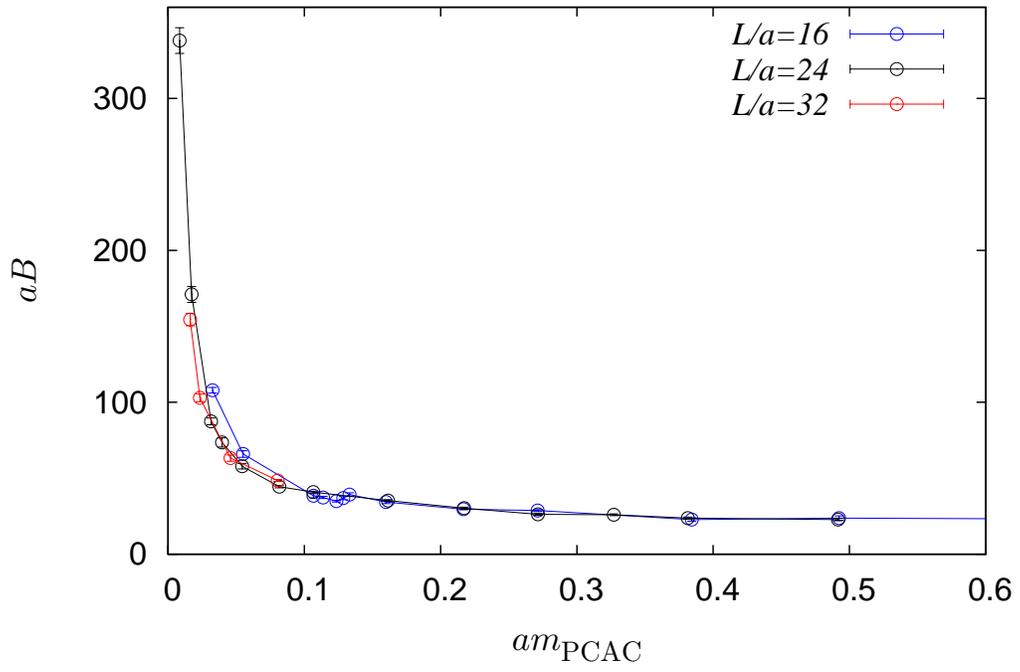}
\caption{
 $B$ versus 
$a m_{\rm PCAC}$ in six-flavor theory.}
\label{fig:pcacVSbchi.b2.0}
\end{center} 
\end{figure}

 Finally, Fig.~\ref{fig:pcacVSbchi.b2.0} shows
the quark mass dependence of  
$B$ defined by Eq.~(\ref{eq:def:bchi}) in the six-flavor theory.
 We have seen that it decreases and approaches to a non-vanishing
value in the chiral limit in 
the two-flavor theory (Fig.~\ref{fig:NF2.bchi.b2.0}). 
 In the six-flavor theory, 
it increases as quark mass decreases,
even if the focus is restricted to the range, 
$a m_{\rm PCAC} \gtrsim 0.08$,
in which the finite size effect is not substantial.
 This is discriminated from 
the behavior of $B$ in Fig.~\ref{fig:NF2.bchi.b4.0} 
obtained for the two-flavor theory at weak coupling, 
where the increase is caused by the large finite size effect.
 Since $\left<\overline{\psi} \psi\right> \propto m_q$
due to dominance of analytic term, 
and $f_P^2 \propto m_q^{2 /(1+\gamma_\star)}$ 
if the hyperscaling hypothesis holds in this system, 
$B \propto m_q^{-(1-\gamma_\star) /(1+\gamma_\star)}$.
 Thus, the qualitative behavior found in Fig.~\ref{fig:pcacVSbchi.b2.0} 
is compatible with the scaling with $\gamma_\star < 1$.

\clearpage

\section{Summary and discussion}
\label{sec:summary}

 To study the quantum-mechanical dynamics of 
the ${\rm SU}(2)_{\rm C}$ gauge theory with six Dirac fermions
in the fundamental representation, 
we perform simulation on the lattices 
with a fixed bare gauge coupling constant ($\beta = 2.0$)
and linear size up to $L/a = 32$, 
and present the first results.
 In this work we use Wilson fermion with no $O(a)$ improvement,
leaving the influence 
of the lattice fermions on the chiral property 
as a subject to be investigated in the future work.

 We first observe that 
the finite size effect on the meson masses turns out to be substantial
and to put the lower bound on masses at each size of lattice.
 To know the quantitative dependence of 
the lightest pseudoscalar meson mass $M_P$ on the quark mass, 
it is inevitable to carry out larger lattice simulation
such as $L/a = 48$.
 Nevertheless, we have seen that the data 
with quark mass $a m_{\rm PCAC} \gtrsim 0.35$ exhibit
the dependence as $M_P \propto (m_{\rm PCAC})^{0.5}$, 
but tends to change more rapidly for smaller quark mass.
 In the theory with chiral symmetry breaking, 
the exponent $\alpha_{M_P}$ in $M_P \propto (m_{\rm PCAC})^{\alpha_{M_P}}$ 
should be closer and closer to $0.5$ for smaller 
and smaller quark mass.
 This point supports 
that the theory is an IR-conformal theory with 
the exponent $\alpha_{M_P}$ larger than $0.5$.

 With help of the explicit simulation of the two-flavor theory,
we demonstrate the utility of the subtracted chiral condensate 
$\left<\overline{\psi}\psi\right>_{\rm subt}$ defined 
by Eq.~(\ref{eq:def:scond}) as a quantity
which may help to examine the occurrence of chiral symmetry breaking
in the Wilson fermion simulation.
 We discuss that 
the finite size effect is negligibly small 
in all available data of $\left<\overline{\psi}\psi\right>_{\rm subt}$. 
 With this observation, the chiral extrapolation is performed 
and its massless limit is seen to be compatible with $0$
within the precision of the available data.

 We focus on the qualitative feature of 
finite size effect on the decay constant $f_P$, 
and search its possible difference
from the theory with chiral symmetry breaking.
 From close examination of data simulated for the six-flavor theory, 
the finite size effect seems to increase $f_P$, 
opposite to that in the ${\chi\hspace{-5pt}/}$-theory.
 We conjecture example $4$
shown in Fig~\ref{fig:fP_finiteVolume_IRconformal_4}
is realized in the $N_F = 6$ theory.

 At present, the proposed utility 
of the finite size effect on $f_P$ 
has a loophole.
 We recall that the qualitative feature seen
in Fig.~\ref{fig:fP_finiteVolume_chibreak} 
for the finite size effect on $f_P$ is obtained 
according to the chiral perturbation theory.
 At present, it is uncertain if 
the decreasing tendency of $f_P$ 
persists even in the circumstance outside 
the applicability of chiral perturbation, 
i.e. on the space with too small volume
compared with the dynamical length scale.
 Naively speaking, it is plausible because, 
if the finite size effect is assumed to tend to increase $f_P$ 
at such weak coupling 
that chiral perturbation is not applicable, 
there must be some transition region of parameters
where the finite size effect accidentally disappears. 
 The issue could be checked by carrying out the simulation
at weak coupling. 
 We attempted to do that at $\beta = 4.0$ in the two-flavor theory, 
but failed to observe a plateau in the effective mass plot 
in the analysis of $M_P$,
which is necessary to get $f_P$ through the PCAC relation,
at $\kappa = 0.128$ and $L/a=24$.
 On the other hand, $M_P$ could be determined on the lattice 
with $\kappa = 0.128$ and $L/a=16$.
 Thus, $\beta = 4.0$ may be too weak 
for the contribution of the excited states to decouple
unless the quark mass is so small that 
the meson masses reach the lower bounds caused by the finite size effect.
 This issue is left as one of the questions to be settled 
in order for the finite size effect 
on $f_P$ to become a device to judge the occurrence 
of chiral symmetry breaking.
 Meanwhile, we note that, 
in the six-flavor, the plateaus can be observed 
in the effective mass plots in all range of measured quark mass, 
which indicate 
the distinction
from the two-flavor theory,
i.e. the system with the chiral symmetry breaking, at very weak coupling. 

\begin{figure}[thb]
\begin{center}
\includegraphics[width=14.0cm,clip]{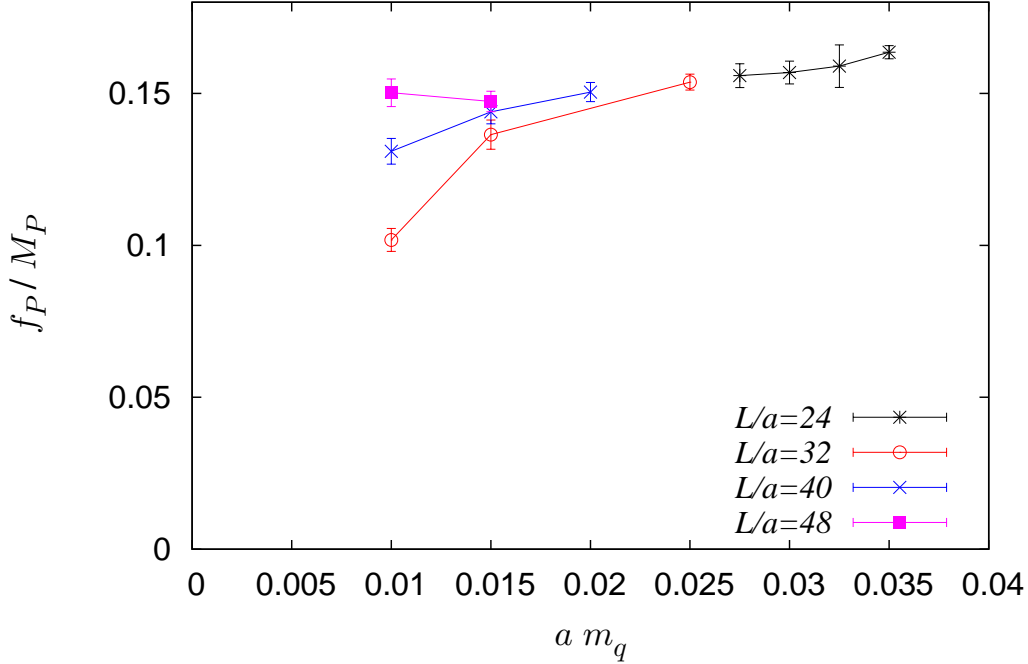}
\caption{
 Quark mass and volume dependence of 
the ratio $f_P /M_P$ in ${\rm SU(3)_C}$ gauge theory
with twelve-flavors, 
calculated from the data found in Ref.~\cite{Fodor:2011tu}.
}
\label{fig:mqVSfpmp.NF12}
\end{center} 
\end{figure}

 The seemingly flatness of the ratio $f_P /M_P$ is also 
compatible with the IR-conformal theory and 
the scaling predicted from the hyperscaling hypothesis.
 At present, we leave the possibility that
this ratio may start to blow up at smaller $M_P$
for the following two reasons. 
 In Fig.~\ref{fig:mqVSfpmp.NF12}, we show $f_P /M_P$ derived  
from the data found in the work by Fodor {\it et~al}.~\cite{Fodor:2011tu}
for ${\rm SU(3)_C}$ gauge theory with $N_F=12$ Dirac fermions.
 The finite size effect appears there. 
 In particular, at the smallest quark mass, $a m_q = 0.01$, 
the data with size $L/a = 48$ can be seen to play an important role.
 However, it is still uncertain if $f_P /M_P$ starts to blow up
as in Fig.~\ref{fig:NF2.fPmP.b2.0}, 
while the finite size effect 
acts to decrease $f_P$ in Ref.~\cite{Fodor:2011tu},
which is the same tendency as in the ${\chi\hspace{-5pt}/}$-theory.
 We are expecting to report the result for $f_P /M_P$ 
obtained using larger lattices in the future.

 Another reason is that, 
although we show here that the chiral symmetry breaking 
does not seem to occur in the six-flavor theory, 
it was demonstrated in our preliminary report \cite{Hayakawa:2012gf} 
that the string tension unlikely vanishes in the chiral limit, 
in contradiction to the IR-conformality. 
 To resolve the issue, 
it may be necessary to study the theory further 
by taking the possibility of realization of confinement
without chiral symmetry breaking taken into account.

\begin{acknowledgments}
 The numerical simulations with large lattices, $L/a \ge 16$, were 
carried out on the computer system $\varphi$ at Nagoya University,
and the servers equipped with GPU cards
at High Energy Accelerator Research Organization (KEK).
 This work is supported partly by JSPS Grands-in-Aid
for Scientific Research 20540261, 22224003, 22740183, 23740177.
 M.~T. is supported in part by LGS (leading graduate school) program.
\end{acknowledgments}

\appendix

\section{Implication on electroweak symmetry breaking}
\label{sec:EWSB}

 It is often said that
${\rm SU}(2)_{\rm C}$ gauge theory is similar to
${\rm SU}(3)_{\rm C}$ gauge theory 
and thus no additional insight is obtained 
by performing separate lattice simulation.
 That is actually not the case, 
because ${\rm SU}(2)_{\rm C}$ gauge theory is one of
${\rm Sp}(2 N)_{\rm C}$ gauge theories
in that its fundamental representation is pseudo-real
due to the existence of a group-invariant symplectic form.
 As is explained below, the chiral symmetry of the 
system with $N_F$ Dirac fermions in the fundamental representation
is thus enhanced from the usual 
${\rm SU}(N_F)_{\rm L} \times {\rm SU}(N_F)_{\rm R} \times {\rm U}(1)_{\rm B}$
to ${\rm SU}(2 N_F)$, 
which is though to be broken
to ${\rm Sp}(2 N_F)$ if the spontaneous breakdown occurs.
 Thus, it is plausible that the spectra as well as the chiral dynamics 
differ from those in ${\rm SU}(3)_{\rm C}$ gauge theories.
 Moreover, from the standpoint of the application 
of the gauge dynamics to the realization of 
the electroweak symmetry breaking, 
this fact serves the effective composite Higgs sector 
quite different from the ${\rm SU}(N_C)_{\rm C}$ gauge theories 
with $N_C \ge 3$.
 The purpose of this appendix is
to summarize these basic kinematic features
of ${\rm SU(2)}_{\rm C}$ gauge theory. 
 We take up ${\rm SU}(2)_{\rm C}$ from a series of 
${\rm Sp}(2 N)_{\rm C}$ as a structure group of the gauge theory, 
but the discussion in what follows persists if 
the symplectic form $\varepsilon_{\rm C}$ of ${\rm SU}(2)_{\rm C}$ 
is replaced with that $\mathcal{J}_{\rm C}$ of ${\rm Sp}(2 N)_{\rm C}$.

 In the chiral representation of gamma matrices,
\begin{equation}
 \gamma^\mu
 =
 \left(
  \begin{array}{cc}
   0 & \left(\sigma^\mu\right)_{\alpha \dot{\delta}}\\
   \left(\overline{\sigma}^\mu\right)^{\dot{\beta} \gamma} & 0
  \end{array}
 \right)\quad
 \left(\alpha,\,\dot{\beta},\,\gamma,\,\dot{\delta} = 1,\,2\right)\,,
\end{equation}
each of $N_F$ Dirac fermions, 
$\psi_{r,\,i}$ ($r = 1,\,2$; $i=1,\,\cdots,\,N_F$),  
is decomposed into a pair of 
two-component spinors (Weyl fermions)
$\left(\xi^+_{r,\,i}\right)_\alpha$ ($\alpha = 1,\,2$), 
$\left(\xi^-_{r,\,i}\right)^{\dot{\beta}}$ ($\dot{\beta} = 1,\,2$), 
which belong to inequivalent irreducible representations of 
Lorentz group ${\rm SO}(3,\,1)$, as
\begin{equation}
 \psi_{r,\,i}
 =
 \left(
  \begin{array}{c}
   \xi^+_{r,\,i} \\ \xi^-_{r,\,i}
  \end{array}
 \right)\,.
\end{equation}
 We recall that the anti-symmetric tensor $(\varepsilon_{\rm C})_{rs}$
($r,\,s = 1,\,2$, $(\varepsilon_{\rm C})_{12} = 1$) is the symplectic form
of ${\rm SU}(2)_{\rm C}$.
 Just as done for a Higgs doublet in the standard model,
$\varepsilon_{rs}\,\left(\xi^-_{s,\,i}\right)^*$, 
where the summation over repeated indices is understood, 
is shown to transform in the same way as $\xi^-_{r,\,i}$
with respect to ${\rm SU}(2)_{\rm C}$.
 Further multiplication of an ${\rm SL}(2)$-invariant anti-symmetric tensor, 
$(\varepsilon_{\rm L})_{\alpha \beta}$,
allows to convert
$(\varepsilon_{\rm C})_{rs}\,\left(\xi^-_{s,\,i}\right)^*$
to
$\left(\xi^{+}_{r,\,i+N_F}\right)_\alpha \equiv 
(\varepsilon_{\rm L})_{\alpha \beta}\, 
(\varepsilon_{\rm C})_{rs}\,\left(\xi^{-\,\dot{\beta}}_{s,\,i}\right)^*$
that transforms exactly in the same way as $\xi^+_{r,\,i}$
under Lorentz transformations as well as ${\rm SU}(2)_{\rm C}$.
 A simple manipulation shows that
the fermionic part of the action in the chiral limit
contains 
$2 N_F$ Weyl fermions $\xi^+_{r,\,I}$ ($I = 1,\,\cdots,\,2 N_F$)
on the same footing\,:
\begin{equation}
 S_F = \int d^4 x\,\sum_{I=1}^{2 N_F}
 \frac{i}{2} \left(\xi^{+\,\dagger}_I\right)_{\dot{\alpha}}
 \left(
  \overline{\sigma}^{\,\mu}
 \right)^{\dot{\alpha}\,\beta} 
 \left(
  \del_\mu - i g G_\mu
 \right) \xi^+_{I,\,\beta}\,.
\end{equation}
 This action is manifestly invariant under the global symmetry
$G = {\rm SU}(2 N_F)$.

 We suppose that the fermion-bilinear operators
\begin{equation}
 W_{IJ} \equiv
 - \xi_I \left(\varepsilon_{\rm C} \otimes \varepsilon_{\rm L}\right) \xi_J\,,
\end{equation}
($\varepsilon_{\rm C}$ and $\varepsilon_{\rm L}$ are 
${\rm SU}(2)_{\rm C}$-invariant and Lorentz-invariant anti-symmetric tensors, 
respectively), 
which are anti-symmetric with respect to $I,\,J$,
serve an appropriate order parameter of chiral symmetry breaking.
 As long as $N_F$ is below a certain number $N_F^{\rm crtl}$, 
the non-perturbative dynamics are expected to give them
non-zero vacuum expectation values (VEVs) of the form
\begin{equation}
 \left<W_{IJ}\right> 
 = \frac{1}{2}\,\Sigma\,\mathcal{J}_{IJ}\,,
 \label{eq:VEV:ChiralConds}
\end{equation}
where $\mathcal{J}$ denotes the symplectic form  of 
the subgroup $H = {\rm Sp}(2 N_F)$ of $G = {\rm SU}(2 N_F)$ 
and $\Sigma$ can be taken to be real, 
implying that the chiral symmetry $G$ is spontaneously broken down to $H$.
 In the basis such that $\mathcal{J}$ takes the form
\begin{equation}
 \mathcal{J}_{IJ}
 =
 - \delta_{I + N_F,\,J} + \delta_{I,\,J + N_F}\,.
 \label{eq:form1_J}
\end{equation}
VEVs in Eq.~(\ref{eq:VEV:ChiralConds}) imply that
\begin{equation}
 \left<
  \overline{\psi}^{\,i} \psi_j
 \right>
 =
 - \Sigma\,\delta_j^{\ i}\,,
\end{equation}
which is the same form
as in QCD whose structure group is ${\rm SU(3)}_{\rm C}$.
 From this fact, one anticipates the pattern of breaking
${\rm SU}(2 N_F) \rightarrow {\rm Sp}(2 N_F)$.

 Just as in the same manner, the degenerate Dirac mass terms, 
$
m \sum_{i = 1}^{N_F} \overline{\psi}^{\,i} \psi_i
$, 
can be seen to be invariant under ${\rm Sp}(2 N_F)$.
 Thus, we expect that 
the lattice actions and techniques developed
for three-color QCD serve a way to examine
whether the breaking pattern 
${\rm SU}(2 N_F) \rightarrow {\rm Sp}(2 N_F)$ 
occurs by taking the vanishing ``external source'' limit, $m \rightarrow 0$. 

 Next, we observe the implication 
of the application of the dynamics of ${\rm SU}(2)_{\rm C}$ gauge theories
to the electroweak symmetry breaking.
 This needs specification of assignment of charges (representations)
under $G_{\rm EW} = {\rm SU}(2)_{\rm L} \times {\rm U}(1)_{\rm Y}$,
i.e. the way of embedding of $G_{\rm EW}$
into $G = {\rm SU}(2 N_F)$.
 Our dynamical assumption is
that $G \rightarrow H = {\rm Sp}(2 N_F)$
with (\ref{eq:VEV:ChiralConds}) at the zeroth order of
the interactions other than ${\rm SU}(2)_{\rm C}$,
i.e. $G_{\rm EW}$-gauge interactions, ``real'' QCD, 
and the interactions from the structure
responsible to generating masses of quarks and leptons
(extended technicolor, ETC) \cite{Dimopoulos:1979es}.
 Even though $H$ has degeneracy
$G /H$ in $G$ at the zeroth order,
the position will be fixed 
so that the energy due to radiative corrections of those interactions
is minimized \cite{Peskin:1980gc,Preskill:1980mz}.
 The unbroken gauge symmetry is $H \cap G_{\rm EW}$ with $H$ 
determined as such, 
and must be the symmetry of electromagnetism, $U(1)_{\rm em}$.

 Since only relative position in $G$ matters, 
vacuum alignment can be asked
by fixing $H$ in $G$ so that the chiral condensates take the form
(\ref{eq:VEV:ChiralConds}) with $\mathcal{J}$ in Eq.~(\ref{eq:form1_J}), 
and questioning which position of $G_{\rm EW}$ in $G$ 
minimize the vacuum energy.
 Denoting a pair of an irreducible representation ${\bf r}$ of ${\rm SU}(2)_L$ 
and the charge $\frac{Y}{2} $with respect to ${\rm U}(1)_{\rm Y}$ 
by $\left({\bf r},\,\frac{{\rm Y}}{2}\right)$, 
we start with $N_F = 2$ and the following simple example
\begin{equation}
 \left(
  \begin{array}{c}
   \xi_1 \\ \xi_2
  \end{array}
 \right)
 \Leftrightarrow \left({\bf 2},\,0\right),\,
 \quad
 \xi_3 \Leftrightarrow \left(0,\,-\frac{1}{2}\right)\,,
 \quad
 \xi_4 \Leftrightarrow \left(0,\,+\frac{1}{2}\right)\,.
  \label{eq:assignment1}
\end{equation}
 We can see that the condensates of the form (\ref{eq:VEV:ChiralConds})
is invariant under the subgroup ${\rm U}(1)_{\rm em}$
but not under the whole $G_{\rm EW}$. 
 However, there is another embedding way of $G_{\rm EW}$ into $G$ as follows: 
\begin{equation}
 \left(
  \begin{array}{c}
   \xi_1 \\ \xi_3
  \end{array}
 \right)
 \Leftrightarrow \left(\frac{1}{2},\,0\right),\,
 \quad
 \xi_2 \Leftrightarrow \left(0,\,-\frac{1}{2}\right)\,,
 \quad
 \xi_4 \Leftrightarrow \left(0,\,+\frac{1}{2}\right)\,.
  \label{eq:assignment2}
\end{equation}
 In this case, the condensates (\ref{eq:VEV:ChiralConds})
are invariant under the whole $G_{\rm EW}$
so that $G_{\rm EW}$ remains unbroken.


 The above example highlights the essence of a property of 
the ``effective Higgs sector'' of ${\rm Sp}(2 N)_{\rm C}$ gauge theories.
 In order to break ${\rm SU}(2)_{\rm L}$, there must be at least
one ${\rm SU}(2)_{\rm L}$-nonsinglet $\xi_I$ ($I=1,\,\cdots,\,n$) 
with $n \ge 2$.
 The chiral condensates
$W_{IJ} \equiv
- \xi_I \left(\mathcal{J}_{\rm C} \otimes \varepsilon_{\rm L}\right) \xi_J$
defined as in Eq.~(\ref{eq:VEV:ChiralConds})
will contain the multiplet transforming as ${\bf n}$. 
 An even $n$, such as ${\bf 2}$, may be suitable to
inducing the desirable breaking pattern 
$G_{\rm EW} \rightarrow {\rm U}(1)_{\rm em}$.
 Since the chiral condensates are anti-symmetric with respect to flavor indices,
{\it they inevitably contain an $SU(2)_L$-invariant component} ${\bf 1}$; 
$\left({\bf n} \otimes {\bf n}\right)_A
= {\bf 1} +{\bf 5} + \cdots + {\bf 2n - 3}$.
 The presence of ${\rm SU(2)_L}$-singlet composite field is
one characteristic property of the effective composite Higgs sector of 
${\rm Sp}(2 N)_{\rm C}$ gauge theory.
 If there are ${\rm SU(2)_L}$-multiplet with $n$
equal to or greater than $4$,
the VEVs of the composite fields belonging to 
the multiplet of odd dimension $\ge 5$ can shift the $\rho$ parameter 
significantly from $1$, 
unless the dynamics of ${\rm Sp}(2 N)_{\rm C}$ gauge theory
function to suppress them.

 A similar consideration brings out the basic feature 
of the ``effective Higgs sector''
of ${\rm SO}(2 N)_{\rm C}$ gauge theories. 
 The chiral condensates
$N_{IJ} \equiv - \xi_I \varepsilon_L \xi_J$ are symmetric 
with respect to $I,\,J$. 
 For a given $n$, $N_{IJ}$ contains $\left({\bf n} \otimes {\bf n}\right)_S
= {\bf 3} + {\bf 7} + \cdots + {\bf 2n - 1}$.
 Therefore, the effective composite Higgs sector of 
${\rm SO}(2 N)_{\rm C}$ gauge theory 
inevitably contains the ${\rm SU(2)_L}$-triplet composite Higgs field, 
as mentioned above, 
the VEV of which yields inconsistency with the experimental constraint.


\bibliographystyle{utphys}
\bibliography{largeNF}

\end{document}